\documentclass[prl,aps,superscriptaddress,twocolumn]{revtex4-2}
\setcitestyle{super}
\usepackage{amsfonts,amssymb,amscd,amsthm}
\usepackage{graphicx}
\usepackage{threeparttable}
\usepackage{multirow}
\usepackage{mathrsfs}
\usepackage[intlimits]{amsmath}
\usepackage{array}
\usepackage{xcolor}
\usepackage[colorlinks, citecolor=red]{hyperref}
\usepackage{etoolbox}
\usepackage{color,soul}
\usepackage{ulem}

\definecolor{blue}{rgb}{0,0,1}
\definecolor{red}{rgb}{1,0,0}
\definecolor{green}{rgb}{0,1,0}

\begin{document}

\title{Exploring topology via momentum-selective tomography}

\author{Sainan Huai}
\thanks{These authors contributed equally to this work.}
\affiliation{Tencent Quantum Laboratory, Tencent, Shenzhen 518057, China}

\author{Ling Lin}
\thanks{These authors contributed equally to this work.}
\affiliation{Institute of Quantum Precision Measurement, State Key Laboratory of Radio Frequency Heterogeneous Integration, College of Physics and Optoelectronic Engineering, Shenzhen University, Shenzhen 518060, China}

\author{Kunliang Bu}
\thanks{These authors contributed equally to this work.}
\affiliation{Tencent Quantum Laboratory, Tencent, Shenzhen 518057, China}

\author{Tianqi Cai}
\thanks{These authors contributed equally to this work.}
\affiliation{Tencent Quantum Laboratory, Tencent, Shenzhen 518057, China}

\author{Zhiwen Zong}
\affiliation{Tencent Quantum Laboratory, Tencent, Shenzhen 518057, China}

\author{Xiaopei Yang}
\affiliation{Tencent Quantum Laboratory, Tencent, Shenzhen 518057, China}

\author{Yuan Li}
\affiliation{Tencent Quantum Laboratory, Tencent, Shenzhen 518057, China}

\author{Yulei Ye}
\affiliation{Tencent Quantum Laboratory, Tencent, Shenzhen 518057, China}

\author{Zhenxing Zhang}
\email{zzxht3@gmail.com}
\affiliation{Tencent Quantum Laboratory, Tencent, Shenzhen 518057, China}

\author{Shengyu Zhang}
\email{shengyzhang@tencent.com}
\affiliation{Tencent Quantum Laboratory, Tencent, Shenzhen 518057, China}

\author{Chaohong Lee}
\email{chleecn@szu.edu.cn}
\affiliation{Institute of Quantum Precision Measurement, State Key Laboratory of Radio Frequency Heterogeneous Integration, College of Physics and Optoelectronic Engineering, Shenzhen University, Shenzhen 518060, China}

\begin{abstract}
Identifying and measuring topological invariants remains a fundamental challenge due to the non-local nature of momentum eigenstates in real space. 
Established approaches based on Bloch oscillations, quantum walks, charge pumping and quench dynamics, typically fail to provide direct momentum-dependent geometric information of each Bloch state.
Here, we propose a novel framework for implementing direct band-topology tomography via momentum-selective measurements, and experimentally demonstrate it on a superconducting processor featuring a highly connected coupling architecture. 
Utilizing a ``star-like" geometry in which an auxiliary probe couples to a one-dimensional topological chain with spatially modulated coupling strengths, we selectively address the desired momentum states of the chain.
This approach enables the direct and robust extraction of the winding phase that contributes to the associated topological invariant. 
Leveraging non-local couplers bridged by long tantalum airbridges, we realize an extended Su-Schrieffer-Heeger (SSH) model and map its full topological phase diagram. 
Through a nested Fourier analysis over time and space, we directly extract the winding phase across Brillouin zone and resolve quantized winding numbers of $\nu=\{0,\pm1,2\}$ characterized by phase winding $\{0, \pm 2\pi,4\pi\}$.
Our work establishes a versatile tool for probing topological characteristics and opens a new avenue for robust quantum computing architectures based on topological protection.  
\end{abstract}

\maketitle

\section{Introduction}\label{Sec1}
Since the introduction of the concept of topology into condensed matter physics \citep{PhysRevLett.45.494, PhysRevLett.49.405, PhysRevLett.61.2015, PhysRevLett.95.146802}, the identification and measurement of topological invariants have remained a challenging task. 
Topological invariants, such as the Chern number and winding number, are defined over the Brillouin zone in momentum space and characterize the global geometric features of the energy bands \citep{RevModPhys.88.035005}. 
This global property endows topological phases with exceptional robustness against local perturbations and impurities, a characteristic that holds promising applications in quantum computing \citep{KITAEV20032, RevModPhys.80.1083}.

However, direct observation of topological invariants in momentum space faces fundamental physical barriers in experimental settings \citep{Goldman2016}. 
In a spatially periodic system, the eigenstates (known as Bloch states) are extended and characterized by a good quantum number: the quasi-momentum. It is thus extremely difficult to accurately prepare and resolve specific eigenstates in real-space experiments.
For a long time, the bulk-boundary correspondence \citep{science.1148047,Hsieh2008,rechtsman2013photonic,Yang2018ideal,Zhirihin2021,braun2024real} has been a cornerstone for understanding band topology, with topological boundary states providing crucial experimental signatures in numerous systems.
While this correspondence-based approach is highly effective, accessing the topology of underlying bulk bands offers a more comprehensive perspective that boundary observations alone do not directly capture.

The measurement of topological invariants has attracted substantial attention in recent years, including Bloch oscillations \citep{atala2013direct,science.1259052,jotzu2014experimental, aidelsburger2015measuring,science.aad5812}, quantum walks \citep{cardano2017detection,PhysRevLett.127.147401, PhysRevX.7.031023}, charge pumping  \citep{lohse2016thouless,nakajima2016topological, xiang2023simulating} and quench dynamics \citep{Science362,PhysRevLett.113.045303,science.aad4568, PhysRevLett.118.185701,flaschner2018observation,tarnowski2019measuring, PhysRevLett.121.250403,PhysRevLett.123.080501,PhysRevLett.125.183001,science.adp6802}.
Usually, to take into account the contributions of all Bloch states, one can either scan the entire Brillouin zone via Bloch oscillations or completely fill the Bloch band. 
Notably, using ultracold atomic systems, a full tomography of the Bloch states has been achieved by filling the entire lower band and a subsequent sudden quench to flat bands \citep{PhysRevLett.113.045303, science.aad4568}.
However, to determine the topological invariant, can one directly measure the momentum-dependent geometric information of each Bloch state, rather than performing a full tomography of all Bloch states?
In principle, the geometric information can be obtained via momentum-selective measurements for each Bloch momentum.
Up to date, a direct measurement of momentum-dependent geometric information remains a paramount yet unfulfilled challenge.

\begin{figure*}[tb]
\includegraphics{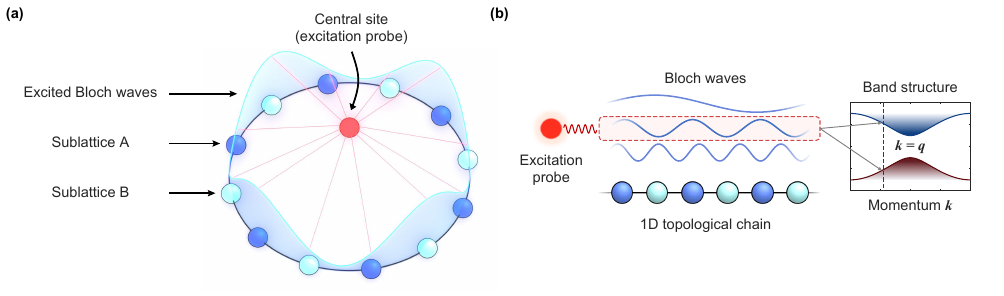}
\caption{\textbf{Schematic illustration for the star-like coupling model and momentum-selective excitation.} (a) Real-space lattice geometry. A one-dimensional (1D) topological chain is configured in a ring geometry under periodic boundary conditions. The sites on the chain (represented by blue spheres) are coupled to an auxiliary central probe site (red sphere), with the coupling strength varying spatially. (b) Mechanism of momentum selection. The central site couples exclusively to the Bloch state of a selected momentum $k$ on the chain when the spatial modulation frequency $q$ of the coupling strengths is matched to it. This creates a probe state for momentum-selective band tomography.}
\label{fig:Fig1}
\end{figure*}

Here we pursue a direct route to band-topology tomography through momentum-selective measurements of geometric information.
The core of the scheme involves the construction of a ``star-like" lattice model, in which an auxiliary probe site couples to every site of a one-dimensional (1D) topological chain through spatially and periodically modulated coupling strengths (Fig.~\ref{fig:Fig1}). 
By precisely tuning the modulation frequency, the probe exhibits high momentum selectivity, coupling only to Bloch states with desired momenta.
This provides a direct momentum-selective access to the winding phase that contributes to the winding number, without requiring complete band filling or full momentum-space imaging.

In this work, we experimentally demonstrate this scheme on a superconducting quantum platform, a versatile architecture for programmable quantum simulation and computation \citep{GoogleQuantumAI2025BelowThreshold,rqkg-He2025AllMicrowaveLeakageSuppression,Lin2026SurfaceCodeLogicalOperations,Zhu2026Combinatorial104Qubits,Wang2026LowOverheadQEC,Zhang2026HighPrecisionCZCalibration,Kim2023QuantumUtility,PhysRevLett.120.130503,PhysRevLett.122.210401,PhysRevLett.131.080401,Yoshioka2025KrylovDiagonalization,McKay2023BenchmarkingScale,Ni2023BreakEvenDiscreteVariable}. 
By leveraging a highly connected network of 9 qubits (Fig.~\ref{fig:Fig2}(b), maximum connectivity $d_0=8$) with individually tunable couplings, we realize an extended SSH model featuring long-range tunneling \citep{PhysRevApplied.19.054028}. 
Using nested Fourier analysis with eigenstate reconstruction, we not only isolate the eigenstate component at a selected energy but also resolve its phase at each momentum, thereby achieving a complete tomography of band topology.
The experimental results show that this method is highly robust against system imperfections and can be potentially extended to multi-particle systems, establishing a universal technical path for probing topological invariants in complex quantum systems. 
The stability of the measured topological invariants suggests potential applications in robust quantum computation, where topological characteristics could be potentially leveraged to protect quantum information.

\section{Theoretical framework}\label{Sec2}

\subsection{Star-like model with momentum-selective coupling} \label{sub:star-like model}
We consider a one-dimensional (1D) chain subject to periodic boundary condition (PBC), coupled to a central auxiliary site (the probe), as illustrated in Fig.~\ref{fig:Fig1} (a).
In this star-like model, each site in the chain couples to the probe site with spatially modulated coupling strength. 
The general Hamiltonian of this system reads
\begin{equation}
\label{eqn:Ham_General}
\hat H = \sum\limits_{x,x'} {\sum\limits_{\sigma ,\sigma '} {t_{x,x'}^{\sigma ,\sigma '}\hat c_{\sigma ,x}^\dag {{\hat c}_{\sigma ',x'}}} }  + \sum\limits_{x = 1}^L {\sum\limits_\sigma  {\eta_\sigma \left( x \right)\hat c_0^\dag {{\hat c}_{\sigma ,x}}} }  + {\rm{H.c.}},
\end{equation}
where $\hat c_{\sigma, x}^\dag$ ($\hat c_{\sigma, x}$) is the creation (annihilation) operator on sublattice (orbital) $\sigma$ of the $x$-th cell, with $x\in [1,L]$ labeling sites on the ring and $x=0$ denoting the central site (the probe). 
In Eq.~\eqref{eqn:Ham_General}, the first term is a general form of tunneling within the lattice, while the second term captures the coupling between the central site and the chain, as schematically demonstrated in Fig.~\ref{fig:Fig1}(a). 
Throughout this work, we consider a system of $L$ unit cells and set the lattice constant to unity for convenience. The parameters $t_{x,x'}^{\sigma, \sigma'}$ and $\eta_\sigma(x)$ denote the intra-chain tunneling amplitudes and the spatially dependent coupling strengths to the central site, respectively. 

Through the Fourier transformation for sites on the ring, ${{\hat c}_{\sigma ,x}} = \frac{1}{{\sqrt L }}\sum\nolimits_k {{e^{\mathrm{i}kx}}} {{\hat c}_{\sigma ,k}}$, the coupling term becomes $\hat{H}_{c}=\sum\nolimits_{k,\sigma} {\;{\eta _\sigma }\left( k \right)\hat c_0^\dag {{\hat c}_{\sigma ,k}}}  + {\rm{H.c.}}$, with ${\eta _\sigma }\left( k \right) = \frac{1}{{\sqrt L }}\sum\nolimits_{x = 1}^L {  {{\eta _\sigma }\left( x \right){e^{\mathrm{i}kx}}} } $ describing the coupling between the central site and Bloch waves of the ring. 
Remarkably, by engineering a spatial modulation in the form of ${\eta _\sigma }\left( x \right) = {\tilde \eta _{\sigma ,q}}{e^{ - \mathrm{i}qx}}$ (with $q$ the modulation frequency), the coupling term then reads ${\eta _\sigma }\left( k \right) = \frac{{{{\tilde \eta }_{\sigma ,q}}}}{{\sqrt L }}{\delta _{q,k}}$, meaning that the central site couples exclusively to the Bloch wave with momentum $k=q$. 
This enables the selective coupling to a chosen momentum state, as depicted in Fig.~\ref{fig:Fig1}(b). 
In fact, by further refining the modulation ${\tilde \eta }_{\sigma ,q}$, the central site can couple to eigenstates of a specified energy quantum number; see the Supplementary Material for details.
This momentum-selective excitation scheme constitutes the cornerstone of our tomography protocol, establishing a direct bridge between real-space experimental control and momentum-space topological invariants.

\subsection{Winding-number tomography }
Now, we apply the momentum-selective coupling to extract the topological invariant of the extended Su-Schrieffer-Heeger (SSH) model \citep{PhysRevLett.42.1698}. 
The SSH model serves as a typical 1D topological model, consisting of two types of sublattices, namely $A$ and $B$ sublattices. 
Introducing cross-cell couplings between two different sublattices extends the SSH model while preserving its chiral symmetry. 
The real-space Hamiltonian of the extended SSH model reads,
\begin{eqnarray}
\label{eqn:Ham_SSH_real_space}
\hat H_{\rm{SSH}} &=&  - \sum\limits_{x = 1}^L {\left( {{t_1}\hat c_{B,x}^\dag {{\hat c}_{A,x}} + {t_2}\hat c_{A, x + 1}^\dag {{\hat c}_{B,x}} } \right)}  \nonumber \\
&& - \sum\limits_{x = 1}^L {\left( { {\gamma _1}\hat c_{A, x - 1}^\dag {{\hat c}_{B,x}} + {\gamma _2}\hat c_{A, x + 2}^\dag {{\hat c}_{B,x}}} \right)}  + \rm{H.c.}, \nonumber \\
\end{eqnarray}
where $t_{1}$, $t_{2}$ and $\gamma_{1,2}$ are the intra-cell, inter-cell and cross-cell tunneling amplitudes, respectively.
The eigenstate of Hamiltonian in Eq.~\eqref{eqn:Ham_SSH_real_space} can be solved as
\begin{eqnarray}
\label{eqn:SSH_eigenstate_solution}
|{\psi _{k, \pm}} \rangle  &=& {{u_{A,k}}|k,A\rangle  \pm {u_{B,k}}|k,B\rangle } \nonumber \\
& =& \sum\limits_{x = 1}^L {{e^{ikx}}\left( {{u_{A,k}}|x,A\rangle  \pm {u_{B,k}}|x,B\rangle } \right)}.
\end{eqnarray}
Here, $u_{A,k}=t_k/|\sqrt{2}t_k|$, $u_{B,k}=1/\sqrt{2}$, and ${t_k} =  - {t_1} - {t_2}{e^{ - \mathrm{i}k}} - {\gamma _1}{e^{\mathrm{i}k}} - {\gamma _2}{e^{ - 2\mathrm{i}k}} $.

Since the parameter space of the chiral-symmetric SSH model is homotopy equivalent to $S^1$, the topological phases of the system are classified by the integer group $\mathbb{Z}$. Therefore its topological invariant is the winding number $\nu \in \mathbb{Z}$ \citep{RevModPhys.88.035005, Ryu2010, PhysRevLett.113.046802}, which can be calculated by
\begin{eqnarray}
\label{eqn:winding_number_formula}
\nu  &=&  \frac{1}{{\pi i}}\int_0^{2\pi } {dk\;\left( { {u_{B,k}^*}{\partial _k}{u_{B,k}} - {u_{A,k}^*}{\partial _k}{u_{A,k}}} \right)} \nonumber \\
& =& \frac{1}{{2\pi}}\int_0^{2\pi } {dk\;{\partial _k}\phi_{k}}.
\end{eqnarray}
Here we define ${\phi_{k}} = \arg \left({u_{B,k}}{u_{A,k}^*}\right) $ as the winding phase for simplicity. 
In a discrete system, the winding number is evaluated through $\nu  = \frac{1}{{2\pi }}\sum_m {\arg } \left[ {{e^{{\rm{i}}\left( {{\phi _{{k_{m + 1}}}} - {\phi _{{k_m}}}} \right)}}} \right]$, with $k_m=2m\pi/L$ being the discrete momentum for an integer $m=\{1,2,\cdots,L\}$. 
This global phase accumulation fundamentally underpins the topological robustness of the system: slight variations in individual phases $\phi_k$ induced by perturbations do not alter the integral related to their derivatives (or the sum related to their differences).

Because the winding number is a global topological invariant defined over the entire Brillouin zone, it is challenging to directly detect the winding number in real space. 
Our strategy is to decompose this global quantity into a set of momentum-dependent winding phases $\{\phi_k\}$ and then measure $\phi_k$ one by one.
Through engineering the probe couplings with designed spatial modulations, the star-like model selectively couples the central probe site to the Bloch states of desired momenta.
Therefore the target winding phase $\phi_q$ is imprinted onto the real-space profile of a zero-energy state in the gap.
In the following, we employ the above star-like model to realize the tomography of band topology and give the winding number directly through Eq.~\eqref{eqn:winding_number_formula}.

To preserve chiral symmetry, the central site is engineered to couple solely to the $B$-sublattice sites of the ring.
The associated Hamiltonian reads
\begin{equation}
\label{eqn:Ham_coupled}
\hat H = {{\hat H}_{{\rm{SSH}}}}  + \sum\limits_k {\eta_B \left( k \right)\hat c_0^\dag {{\hat c}_{B,k}}}  + {\rm{H.c.}}.
\end{equation}
The corresponding chiral operator is given by $\hat \Gamma  = \sum\nolimits_x {\left( {\hat c_{A,x}^\dag {{\hat c}_{A,x}} - \hat c_{B,x}^\dag {{\hat c}_{B,x}} + \hat c_0^\dag {{\hat c}_0}} \right)} $, and the chiral symmetry $\hat{\Gamma}^{-1}\hat{H}\hat{\Gamma} = -\hat{H}$ ensures that the spectrum is symmetric about $E=0$. 
In the superconducting-qubit implementation, the spatial modulation is chosen as $\eta_B(x) = -[\alpha + \beta \cos(qx)]$, $q=2n\pi/L, n=1,2,\cdots, L$ in real space, which can be realized by tuning the coupling between qubits. 
After Fourier transformation, the coupling takes the form of $\eta_B \left( k \right) = - \sqrt L \left[ {\alpha {\delta _{k,0}} + \frac{\beta }{2}\left( {{\delta _{k,q}} + {\delta _{k, - q}}} \right)} \right]$, showing that only the eigenstates with momenta $k \in \{0,\pm q\}$ will be coupled to the central site. 
Rather than isolating a state of single momentum $+q$, which would require complex-valued spatial modulation of both coupler amplitudes and local phases, our protocol employs purely real-valued amplitude modulation. As shown below, this choice remains effective for topological characterization, as the zero-energy mode encodes the target winding phase $\phi_q$, which can be faithfully extracted from the real-space profile.

Within momentum space, the eigenstate of the whole ``star-like'' system can be expanded as superposition of Bloch states at momenta $k=\{0,\pm q\}$ and the central site excitation: $|{\Psi _{q,n}}\rangle  \propto \sum\nolimits_m {\left( {{C_{0,m}}|{\psi _{0,m}}\rangle  + {C_{q,m}}|{\psi _{q,m}}\rangle  + {C_{ - q,m}}|{\psi _{ - q,m}}\rangle } \right)}  + |{x_0}\rangle $, where $|x_0\rangle$ denotes the central site. 
Here, the coefficients denote the corresponding expansion amplitudes. 
Due to the chiral symmetry, there exists a special eigenstate with zero-energy in the middle gap,
\begin{eqnarray}
\label{eqn:zero_energy_Eigenstate}
|{\Psi }_{q, n=0}\rangle  \propto  \sum\limits_{x = 1}^L \left[{C_1{\mathop{\rm Re}} ( {{e^{  \mathrm{i}(qx - \phi_q)}}} ) } + C_2 {e^{-\mathrm{i}\phi_0} } \right] |x,A\rangle  + |{x_0}\rangle ,
\end{eqnarray}
where $C_{1,2}$ are the superposition coefficients. 
Note that this zero-energy state is not an edge state as the topological chain is a ring without boundary. 
Instead, it serves as a probe state whose real-space wavefunction encodes the winding phase ${\phi_q}$. This winding phase can therefore be extracted experimentally from this zero-energy eigenstate via Fourier transformation ${\phi _q} =  - \arg \sum\nolimits_x {|\langle x,A|{\Psi _{q,n = 0}}\rangle| {e^{ - iqx}}} $, where taking the absolute amplitude acts as a simple gauge-fixing step to eliminate arbitrary global phases. As we will see in the next section, this gauge-fixing strictly preserves the topological signature.

\begin{figure*}[tb]
\includegraphics{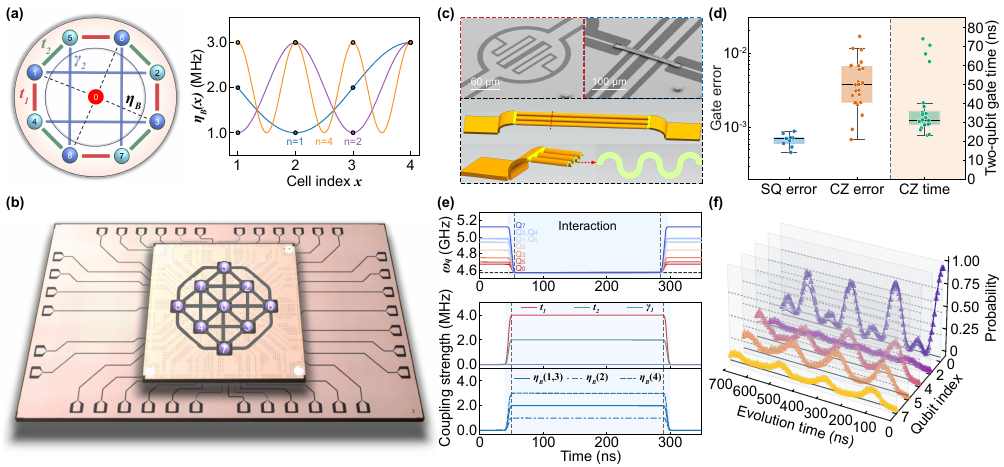}
\caption{\textbf{Device and control overview.} (a) Illustration of experimental model and the spatial modulation of coupling. Left: the extended-SSH lattice realized on the outer ring, with intra-cell ($t_1$) and inter-cell ($t_2$) hoppings, cross-cell coupling ($\gamma_2$), and star-like central coupling ($\eta_B$) to the auxiliary qubit $Q_0$. Right: the programmed central-to-$B$-sublattice couplings $\eta_B(x) = -[\alpha + \beta\cos(qx)]$ across the four unit cells $x$, where the cell indices $x=1,2,3,4$ correspond to couplings from $Q_0$ to $Q_1$, $Q_6$, $Q_3$, $Q_8$, respectively. Spatial modulation profiles are shown for the target momenta $n=1,2,4$ ($n=3$ coincides with $n=1$). (b) Optical micrograph of the flip-chip L9 processor. (c) Top: Scanning electron microscope images of the symmetric floating-coupler crossover architecture, in which a long airbridge spans over the asymmetric floating coupler to establish the cross-diagonal interconnect. Bottom: overall and cross-sectional views of the 3D model of long tantalum airbridge; (d) Benchmarking of chip L9: single-qubit gate error, two-qubit CZ-gate error, and two-qubit CZ gate swap time. Boxes span the interquartile range, central lines mark the medians, and dots denote individual qubits/pairs. (e) Representative pulse sequence realizing the extended-SSH Hamiltonian. Top: qubit frequency trajectories $\omega_q$ steering all qubits into mutual resonance during the interaction window. Middle: chain hoppings $t_1$, $t_2$, and $\gamma_2$. Bottom: the four central couplings $\eta_B(1)$--$\eta_B(4)$ programmed to follow $\eta_B(x)$ for the selected momentum. (f) Coherent population dynamics following a $\pi$ pulse on the central qubit $Q_0$. Site-resolved population distribution across qubits $Q_{0}, Q_{2}, Q_{4}, Q_{5},$ and $Q_{7}$ versus coherent evolution time. Data points (dots) and theory including dissipation and dephasing (lines).}
\label{fig:Fig2}
\end{figure*}

\section{Experimental Demonstration}\label{Sec3}

\begin{figure*}[tb]
\includegraphics{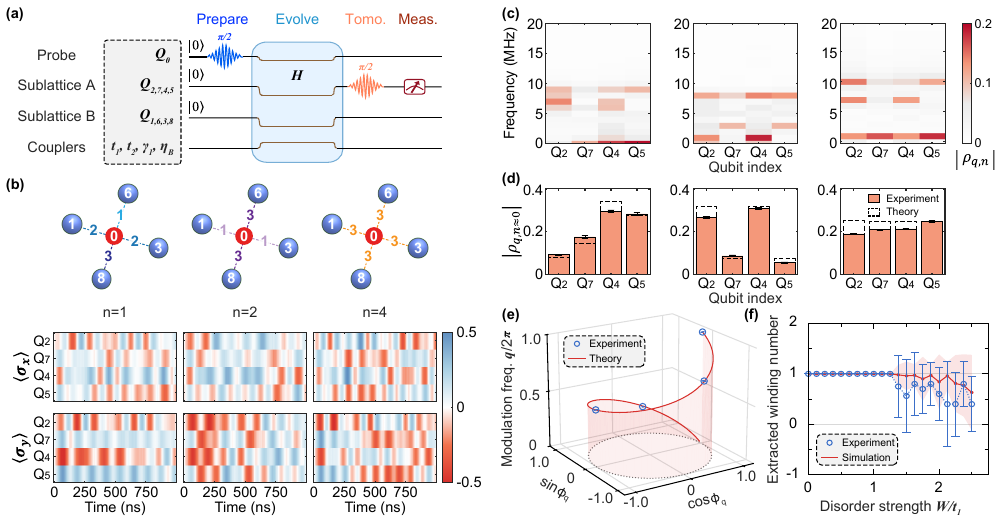}
\caption{\textbf{Band-topology tomography protocol and experimental extraction of the winding number.} (a) Pulse sequence of the protocol, comprising preparation, dynamic evolution and measurement. (b) Top: the three spatial modulations of the couplings $\eta_B$ between the auxiliary qubit $Q_0$ and the $B$-sublattice qubits for the target momenta $n=1,2,4$, numbers give the relative coupling amplitudes on each branch. Bottom: the corresponding measured $\langle \sigma_j^{x} \rangle$ and $\langle \sigma_j^{y} \rangle$ of the $A$-sublattice qubits versus evolution time. (c) Temporal Fourier spectra derived from the dynamics in (b) for each $A$-sublattice qubit, shown for $n=1,2,4$ (left to right). (d) The component extracted from the Fourier spectra near the zero-energy region (bars, experiment; dashed boxes, theory). (e) Extracted phase $\phi_q$ versus the modulation frequency $q$ (circles, experiment; line, theory from $L=100$ system), obtained by a spatial Fourier transform of the zero-energy component in (d). (f) Extracted winding number versus disorder strength $W$ (in units of $t_1$), open circles with error bars, experiment; solid line with shaded band, simulation averaged over disorder realizations. Parameters are $t_1=2.0\;\mathrm{MHz}$ and $t_2=4.0\;\mathrm{MHz}$.}
\label{fig:Fig3}
\end{figure*}

\subsection{Experimental platform and Hamiltonian implementation}

Our experiment is performed on L9, a flip-chip superconducting quantum processor integrating nine frequency-tunable transmon qubits configured in a window-blossoms structure\citep{Sui2026Platonic} [Fig.~\ref{fig:Fig2}(b)]. The device incorporates 24 independently controlled floating couplers \citep{Sete2021FloatingCoupler,Field2024ModularCoupler,Sete2024ErrorBudget} that enable programmable modulation of qubit–qubit interactions. To accommodate non-local connectivity \citep{Bu2025TantalumAirbridges,Wang2026LowOverheadQEC,Marxer2023Long} without introducing line-crossing overhead, the interconnects are bridged by crossover structures embedded with long, stiffener-reinforced tantalum (Ta) airbridges \citep{Bolgar2025StiffenerAirbridges,Chen2026InPrep} [Fig.~\ref{fig:Fig2}(c)]. Under such coupling architecture, the nine qubits exhibit an average isolated energy-relaxation time $T_1$ of $59.3~\mu\mathrm{s}$ and an average spin-echo dephasing time $T_2^{\mathrm{echo}}$ of $29.3~\mu\mathrm{s}$ (see the Supplementary Material II). Benchmarking of the gate set yields an average single-qubit gate fidelity of $99.93\%$ and an average two-qubit CZ-gate fidelity of $99.5\%$~[Fig.~\ref{fig:Fig2}(d)]. 
Notably, the central qubit $Q_0$ can be engineered to enable a single-site degree of up to $d_0 = 8$. This high-connectivity architecture establishes the exact structural prerequisites to host the star-like geometry\citep{Renger2026AllToAll,Song2017TenQubit,Song2019TwentyQubit} and long-range interaction required for our model, allowing the central site to selectively address the outer sublattices (Fig.~\ref{fig:Fig2}(b) and Supplementary Fig. S5). 

To experimentally simulate an extended SSH model, we employ a sub-network comprising the 9 qubits and 16 active couplers shown in Fig.~\ref{fig:Fig2}(a). The eight outer qubits form $L = 4$ unit cells, with $\{Q_2, Q_4, Q_5, Q_7\}$ assigned as the $A$ sublattice and $\{Q_1, Q_3, Q_6, Q_8\}$ as the $B$ sublattice. The cross-cell coupling here only involves the $\gamma_2$ terms. The $\gamma_1$ coupling can be realized in this network scheme by changing the choice of unit cell.
The coupling strengths between the central site and the SSH chain are spatially modulated by programming the four couplers linked to the B-sublattice sites according to $\eta_B(x) = -[\alpha + \beta \cos(qx)], x\in [1,L]$. 
All couplers not actively utilized in a given experimental run are continuously biased to their designated ZZ-free idle positions \citep{Yan2018TunableCoupling,Kandala2021ZZSuppression,Stehlik2021TunableCoupling}. Further device details are provided in the Supplementary Material.

To implement the band-topology tomography protocol, we first realize the target system Hamiltonian in Eq.~(\ref{eqn:Ham_coupled}) through the pulse sequence summarized in Fig.~\ref{fig:Fig2}(e). All qubits are initialized at their designated idle frequencies, with all floating couplers biased at their designated $ZZ$-free positions. Fast flux pulses then simultaneously steer the qubits into mutual resonance and tune the couplers along the extended-SSH chain to the target hoppings $t_1$, $t_2$, and $\gamma_2$ \citep{Deng2024HighOrderPumping,zhang2024emergence,Wang2025TwoParticleCorrelation,Tao2025Emulating}. Concurrently, the four central couplers $\{\mathrm{C_{01}}, \mathrm{C_{06}}, \mathrm{C_{03}}, \mathrm{C_{08}}\}$ connecting $Q_0$ to the {$B$} sublattice are programmed to the spatial profile $\eta_B(x)$ for the target momentum $q$ [Fig.~\ref{fig:Fig2}(a)], while all other unused couplers remain deactivated. 

As a characterization step, we verify the experimental realization of the Hamiltonian in Eq.~(\ref{eqn:Ham_coupled}) by preparing $Q_0$ in the single-excitation state $|1_c\rangle$ via a $\pi$ pulse and monitoring the subsequent free evolution [Fig.~\ref{fig:Fig2}(f)]. 
The observed dynamics quantitatively match our theoretical simulations, confirming the star-coupling architecture required for the subsequent tomographic protocol.
However, a direct readout of $\langle\hat{\sigma}_z\rangle$ only yields on-site populations that mix contributions from different energy eigenstates, making it insufficient to isolate the zero-energy eigenstate, let alone extract its encoded winding phase $\phi_q$. 
To overcome this, we complement the population measurement with simultaneous single-qubit quantum state tomography and temporal spectroscopy. Crucially, the spectroscopic analysis provides the necessary energy resolution to filter out the zero-energy mode from the complex dynamics, while the tomography captures the transverse expectation values $\langle\hat{\sigma}_x\rangle$ and $\langle\hat{\sigma}_y\rangle$ across the lattice to reconstruct the winding phase \citep{xiang2023simulating,Roushan2017Spectroscopic} in what follows (Methods).

\subsection{Band-topology tomography of the standard SSH model}

\begin{figure*}[tb]
\includegraphics{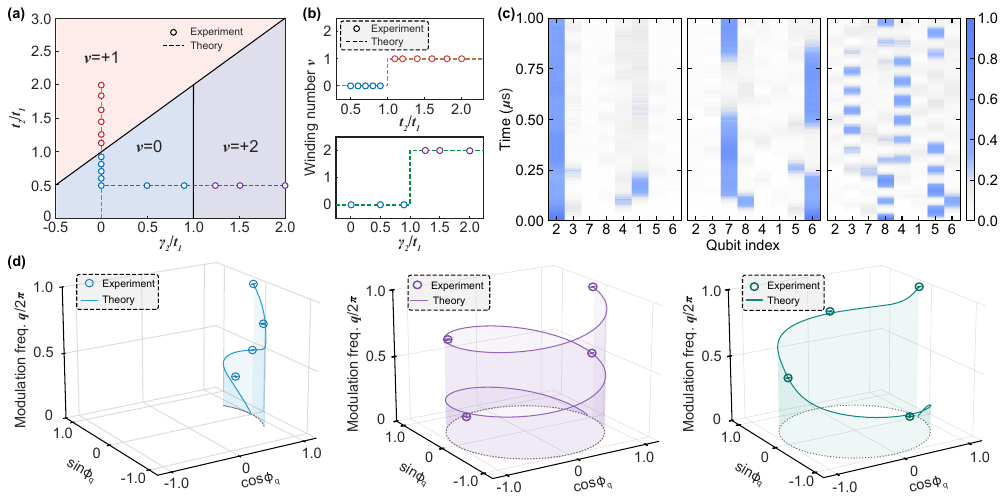}
\caption{\textbf{Measurement of the topological phase diagram.} (a) Experimentally reconstructed phase diagram as a function of $t_2/t_1$ and $\gamma_2/t_1$, with $\gamma_1$ set to zero. Colour encodes the measured winding number $\nu$ (circles, experiment; solid lines, theoretical phase boundaries). The $\nu=-1$ phase, which can be realized by re-defining the unit cell in the present model, is not shown in the phase diagram. (b) Winding number $\nu$ extracted along the brown and green trajectories indicated in (a) (circles, experiment; dashed lines, theory). (c) Verification of the topological edge states in the $\nu =+2$ phase under open boundary conditions. Site-resolved population dynamics (top, experiment; bottom, theory) following Fock-state initialization at the left edge, in the bulk and at the right edge (left to right). (d) Representative winding phase $\phi_q$ versus the modulation frequency $q$ within distinct phases ($\nu=0, +2, -1$, left to right), extracted from the dynamical evolution results (open circles with error bars, experiment; solid line, theory). Parameters are respectively chosen as: $t_2/t_1=0.5, \gamma_{1,2}=0$; $t_2/t_1=0.5, \gamma_1 = 0, \gamma_2/t_1 = 2$; and $t_2/t_1=2, \gamma_1/t_1 = 4, \gamma_2 = 0$. 
}
\label{fig:Fig4}
\end{figure*}

To extract the topological winding phase $\phi_q$ encoded in the zero-energy wavefunction, we apply the three-step protocol\citep{xiang2023simulating,Roushan2017Spectroscopic} illustrated in Fig.~\ref{fig:Fig3}(a). (i) A $Y_{\pi/2}$-pulse on $Q_0$ prepares the system in $|\Psi(0)\rangle = (|1_c\rangle + |\text{vac}\rangle)/\sqrt{2}$. (ii) All pre-calibrated couplings $t_1$, $t_2$, $\gamma_2$, and $\eta_B(x)$ are switched on simultaneously, allowing the system to evolve coherently under the engineered Hamiltonian $H$ [Fig.~\ref{fig:Fig2}(e)] for a duration $t$. (iii) Simultaneous single-qubit quantum state tomography on each $A$-sublattice qubit $Q_j$ yields the expectation values $\langle\hat{\sigma}_x^{(j)}(t)\rangle$ and $\langle\hat{\sigma}_y^{(j)}(t)\rangle$. Figure~\ref{fig:Fig3}(b) illustrates the raw space–time dataset obtained at the standard-SSH point ($t_2/t_1 = 2$, $\gamma_{1,2} = 0$) for several target momenta $q$. To isolate the zero-energy eigenstates, we define a complex parameter $\xi(j, t) = \langle\hat{\sigma}_x^{(j)}(t)\rangle + \mathrm{i}\langle\hat{\sigma}_y^{(j)}(t)\rangle$. A temporal Fourier transform then resolves the spatial distribution of distinct eigenenergy components, yielding $\rho_{q,n}(j) = \frac{1}{T}\int_0^T dt\,\xi(j, t) e^{\mathrm{i}E_{q,n}t} \approx \langle \Psi_{q,n} \mid 1_c \rangle \Psi_{q,n}(j)$ [Fig.~\ref{fig:Fig3}(c)]. Detailed derivations can be found in Methods.

According to Eq.~(\ref{eqn:zero_energy_Eigenstate}), we focus on the zero-energy component ($E_{q,n} = 0$), whose spatial profile follows the envelope $\rho_{q,n=0}(x) \propto C_1 \text{Re}[e^{\mathrm{i}(qx - \phi_q)}] + C_2 e^{-\mathrm{i}\phi_0}$. It is worth noting the gauge uncertainty in the measured wavefunctions. To eliminate this ambiguity and establish a well-defined reference frame, we implement a gauge-fixing procedure defined by $\phi_q = -\arg\sum_j |\rho_{q,n=0}(j)| e^{-\mathrm{i}qj}$ to obtain the winding phase. Repeating the protocol across the entire Brillouin zone establishes the mapping between $\phi_q$ and $q$, see Fig.~\ref{fig:Fig3}(e). As $q$ is swept across the Brillouin zone, the experimental data (blue circles) closely follow the analytical trajectory of Eq.~(\ref{eqn:winding_number_formula}) (solid line), together tracking a distinct single-turn helix. It can be seen that the experimental result is well consistent with the analytical prediction. Its projection onto the $(\cos\phi_q, \sin\phi_q)$ plane closes into a unit circle that explicitly encircles the origin exactly once. This geometric configuration corresponds directly to a quantized winding number $\nu = \frac{1}{2\pi}\oint_{\text{BZ}} d\phi_q = +1$, validating the Hamiltonian implementation, the momentum-selective coupling, and the nested FFT underpinning the full phase-diagram scan of the following section.

Moreover, we evaluate the resilience of our scheme to experimental imperfections. We introduce the coupling disorder to the extended SSH model, i.e., ${W}\sum\nolimits_x {\left( {\delta {t_{1,x}}\hat c_{B,x}^\dag {{\hat c}_{A,x}} + \delta {t_{2,x}}\hat c_{A,x+1}^\dag {{\hat c}_{B,x}} + \mathrm{H.c.}} \right)}$, where $W$ sets the absolute disorder strength and the dimensionless random variables $\delta {t_{i,x} (i=1,2)}$ are drawn uniformly from $[-0.5, 0.5]$. For each disorder strength, we extract the winding number $\nu$ across $\sim 15$ random realizations, reporting its mean and standard deviation. As depicted in Fig.~\ref{fig:Fig3}(f), the measured $\nu$ (blue open circles) remains pinned to the quantized value $\nu = +1$ throughout the weak-disorder regime up to $W/t_1 \sim 1.375$. This observation is in excellent agreement with numerical simulations (red solid line with shaded $\pm\sigma$ band), where the strongly suppressed variance reflects a robust topological plateau. Beyond this critical threshold, both experiment and simulation exhibit strong fluctuations, marking the onset of a disorder-induced topological breakdown. This robustness is fundamentally rooted in our extraction principle: the nested Fourier analysis decodes the globally protected winding phase. Consequently, moderate fluctuations in the extracted phase at individual momenta do not alter the quantized winding number.

\subsection{Band-topology tomography of an extended SSH model with higher winding phase}

Having validated the band-topology tomography protocol at a single parameter point, we exploit the connectivity and tunability of the quantum processor to systematically map the topological phase diagram of an extended SSH model. 
Owing to the inter-cell long-range couplings $\gamma_{1,2}$, the extended SSH model hosts more topological phases ($\nu = -1, 2$). Here, we sweep the coupling strengths $t_2/t_1$ and $\gamma_2/t_1$ experimentally and examine the phase diagram in Fig.~\ref{fig:Fig4}(a).

We probe the phase diagram along two parameter cuts (brown and green trajectories in Fig.~\ref{fig:Fig4}(a)).
Along the vertical cut at $\gamma_2/t_1 = 0$, varying $t_2/t_1$ drives a conventional transition from the trivial phase $\nu = 0$ to $\nu = +1$ across the standard-SSH boundary $t_2 = t_1$ (Fig.~\ref{fig:Fig4}(b), top). Along the horizontal cut at $t_2/t_1 = 0.5$, increasing the inter-cell long-range coupling $\gamma_2/t_1$ takes the system from $\nu = 0$ to the higher-winding phase $\nu = +2$ (Fig.~\ref{fig:Fig4}(b), bottom). On both cuts, the extracted $\nu$ exhibits sharp integer steps and quantized plateaus, which is consistent with analytical predictions. We further confirm the $\nu = +2$ non-trivial phase by detecting the topological edge states under open boundary condition. Excitations initialized at the edge and in the bulk exhibit qualitatively distinct time evolution in Fig.~\ref{fig:Fig4}(c).

Our band-topology tomography protocol further allows us to resolve the detailed structure of the winding phase $\phi_q$ across distinct topological phases (Fig.~\ref{fig:Fig4}(d)). In the trivial phase ($\nu = 0$), $\phi_q$ exhibits non-winding oscillations. 
Crucially, the long-range coupling architecture enables the verification of the higher-winding phase characterized by $\nu = +2$. Here, the extracted $\phi_q$ displays a continuous $4\pi$ phase progression across the Brillouin zone, successfully resolving the geometric signature of a double-turn helical winding. For completeness, we additionally probe the $\nu = - 1$ phase, which can be realized by re-defining the unit cell in our present experimental scheme and is therefore not included in the phase diagram of Fig.~\ref{fig:Fig4}(a). Similar to the $\nu = +1$ case in Fig.~\ref{fig:Fig3}(e), $\phi_q$ acquires a $2\pi$ phase winding across the Brillouin zone, but in the opposite direction. 
Remarkably, across all observed phases, the experimental data obtained from the compact system ($L = 4$) are not only self-consistent with finite-size exact diagonalization but also match the theoretical prediction derived in the long-chain limit ($L = 100$).
This agreement validates the momentum-selective extraction of the topological invariant on the compact lattice and underscores the efficiency of our protocol for
resource-limited near-term quantum processors.

\section{Discussions}

In summary, we have theoretically proposed and experimentally demonstrated a momentum-selective tomography that directly extracts topological invariants (winding numbers) on a superconducting quantum processor. 
Crucially, these results establish momentum-selective excitation as a robust tool for band-topology tomography based upon state-by-state measurement of momentum-resolved geometric information.
This approach offers distinct advantages over existing techniques: (i) unlike established approaches based on Bloch oscillations, quantum walks, and charge pumping, it explicitly resolves the momentum-dependent geometric information (the winding phase $\phi_q$) of each Bloch state; and (ii) in contrast to full tomography of Bloch states via quantum quench, it bypasses the need for filling the entire band.
The robustness of our protocol has been verified in experiments: the extracted winding number remains quantized against moderate disorder. 
This resilience is a natural consequence of our extraction protocol, wherein local errors merely shift the individual winding phases $\phi_k$ and leave the topological invariant unchanged unless the perturbation is strong enough to close the bulk gap.

The architecture can also be extended to larger lattices and higher winding phases (see Fig. S5 in Supplementary Material). While demonstrated here within a 1D chiral-symmetric framework, this momentum-selective tomography protocol can be generalized to other symmetry classes and higher-dimensional systems, such as the Chern insulator. The star-like geometric framework can further be extended to multi-particle systems. More broadly, the ability to selectively address momentum states and read out their topological invariant establishes a versatile foundation for exploring topological quantum information processing, offering a potential pathway toward robust quantum computing architectures.

\section{Methods}

\subsection{Sample design and device fabrication}

Our quantum processor employs a flip-chip \citep{Rosenberg2017FlipChip,Foxen2018Interconnects,Kosen2022FlipChip} architecture comprising two stacked sapphire substrates. The top layer (qubit chip) hosts nine frequency-tunable transmon qubits ($Q_0-Q_8$) together with 24 tunable floating couplers. The bottom layer (routing chip) accommodates the corresponding readout resonators and microwave control lines. The two chips are electrically and mechanically bonded through precisely aligned indium bumps. Each transmon qubit is equipped with a combined XYZ line \citep{Manenti2021CombinedXYZ}, and each of the 24 tunable couplers is addressed by its own Z control line to precisely tune the effective coupling strength between specific qubit pairs. Quantum state readout is carried out through two independent microwave feedlines, onto which the readout resonators of individual qubits are frequency-multiplexed.

Fabrication begins with the deposition of a $200~nm$-thick $\alpha$-phase tantalum film \citep{Place2021TantalumTransmon,Wang2022TantalumLifetime} onto a sapphire substrate via high-temperature magnetron sputtering. All base circuit features, including readout resonators, control lines, and shunt capacitors are defined using laser direct-write lithography. The Manhattan-style Al/AlO$_{x}$/Al Josephson junctions on qubit chip are patterned by electron-beam lithography on a PMMA/MAA bilayer photoresist, followed by double-angle evaporation and in situ oxidation. The fully capped Ta airbridges \citep{Chen2014Airbridges,Dunsworth2018MultiLayerWiring,Girgis2006MetallicAirBridges,Stavenga2023GrayscaleAirbridges,Alegria2025TaAirbridge} for the control lines on routing chip are fabricated using a previously detailed lift-off method~\cite{Bu2025TantalumAirbridges}. Specifically, a stiffener-reinforced \citep{Bolgar2025StiffenerAirbridges,Chen2026InPrep} Ta airbridges serving for the 2.5D non-local coupling are patterned  on qubit chip via Gray-scale lithography \citep{Stavenga2023GrayscaleAirbridges,Sun2022GradientExposure}. This advanced process shapes the photoresist scaffold into a periodic corrugated structure along the transverse axis, achieving a peak-to-valley height difference exceeding $2~\mathrm{\mu m}$  (Fig.~\ref{fig:Fig2} (c)). The resulting mechanical rigidity yields $200~\mathrm{\mu m}$-long airbridges spanning non-adjacent qubits, with airbridge height below $5~\mu m$. Finally, $10~\mathrm{\mu m}$-tall indium bumps are deposited on both chips to define the inter-chip spacing post-hybridization. The fully assembled flip-chip device is then wire-bonded to an aluminum sample holder and anchored in a dilution refrigerator for cryogenic characterization.

\subsection{Measurement setup}

Inside the refrigerator, the sample holder is further enclosed in nested aluminium and Cryoperm shields to screen residual magnetic noise \citep{Malevannaya2025Shielding}, and thermally anchored to the mixing-chamber plate at a base temperature of $\sim\!15~\mathrm{mK}$. Each transmon is controlled by separate XY and Z lines that are combined by a low-temperature diplexer immediately before entering the chip, while each of the 24 tunable couplers is driven by its own Z line to set the effective coupling strength $g$. For readout, the resonators are frequency-multiplexed onto two microwave feedlines. The reflected readout signal is amplified by a three-stage chain—a Josephson parametric amplifier at the mixing chamber, a high-electron-mobility transistor at 4 K, and a room-temperature amplifier—before being down-converted and digitized for single-shot state discrimination \citep{Krantz2019Guide}. The complete wiring diagram and component specifications are provided in Supplementary Fig.~S1.

\subsection{Detection of the zero-energy state}

To experimentally extract the winding phase $\phi_q$, we employ a temporal spectroscopy protocol optimized for superconducting quantum processors. In standard population measurements, the longitudinal observable $\langle\hat{\sigma}_z\rangle$ merely extracts the on-site probability density, providing neither the phase information nor the energy resolution needed to isolate specific eigenstates. 
To recover both, we track $\langle  \hat\sigma_x \rangle$ and $\langle \hat\sigma_y \rangle$ during the dynamical evolution and apply a temporal Fourier transform, which simultaneously reconstructs the complex wavefunction and resolves individual energy components.

A $\pi/2$ pulse on the auxiliary probe site $Q_0$ drives the system into a coherent superposition of the vacuum state and a single-site excitation:
\begin{equation}
|\Psi(0)\rangle = \frac{1}{\sqrt{2}} \left( |1_{c}\rangle + |\text{vac}\rangle \right),
\end{equation}
where $|\text{vac}\rangle \equiv |0,0,\cdots,0\rangle$ denotes the collective vacuum state of the qubit lattice, and $|1_{c}\rangle \equiv \hat{c}_0^\dag |\text{vac}\rangle$ represents the single-excitation state residing at the central probe site. Under the application of fast flux bias pulses, the system is brought into mutual resonance, initiating free coherent evolution governed by the modulated Hamiltonian $\hat{H}$. Because $\hat{H}$ commutes with the total excitation number operator $\hat{N} = \hat{c}_0^\dag \hat{c}_0 + \sum_{x,\sigma} \hat{c}_{\sigma,x}^\dag \hat{c}_{\sigma,x}$, the dynamics remain strictly constrained within the single-excitation subspace. 
Consequently, the time-evolved state at an arbitrary duration $t$ can be expanded in the eigenbasis of the coupled system as:
\begin{equation}
|\Psi(t)\rangle = \frac{1}{\sqrt{2}} \left( \sum_{n} \psi_{n}^{(c)} e^{-\mathrm{i}E_{n}t} |\Psi_{n}\rangle + |\text{vac}\rangle \right),
\end{equation}
where $|\Psi_{n}\rangle$ is the $n$-th eigenstate of the total Hamiltonian with eigenenergy $E_n$, and $\psi_{n}^{(c)} = \langle\Psi_{n} | 1_{c}\rangle$ denotes the corresponding expansion coefficient mapping the projection of the central excitation.

To resolve the complex wavefunctions of these distinct energy bands, we perform simultaneous single-qubit quantum state tomography on each lattice site during the dynamical evolution\citep{xiang2023simulating,Roushan2017Spectroscopic}. Specifically, we monitor the time-resolved expectation values of the Pauli operators $\hat{\sigma}_x^{(j)}$ and $\hat{\sigma}_y^{(j)}$ at the $j$-th qubit. In the single-excitation subspace, the transverse Pauli operators reduce to $\hat{\sigma}_x^{(j)} = |1_j\rangle\langle \text{vac}| + \text{H.c.}$ and $\hat{\sigma}_y^{(j)} = \mathrm{i}|1_j\rangle\langle \text{vac}| + \text{H.c.}$, allowing their expectation values to be analytically expressed as:
\begin{eqnarray}
    \langle\hat{\sigma}_x^{(j)}\rangle(t) = \text{Re} \left( \sum_{n} \psi_{n}^{(c)} \Psi_{n}(j) e^{-\mathrm{i}E_{n}t} \right), \nonumber \\
    \langle\hat{\sigma}_y^{(j)}\rangle(t) = \text{Im} \left( \sum_{n} \psi_{n}^{(c)} \Psi_{n}(j) e^{-\mathrm{i}E_{n}t} \right),
\end{eqnarray}
where $\Psi_{n}(j) = \langle 1_j | \Psi_{n}\rangle$ defines the real-space wavefunction amplitude of the $n$-th eigenstate at site $j$. By constructing a complex dynamical observable $\xi(j, t) \equiv \langle\hat{\sigma}_x^{(j)}\rangle(t) + \mathrm{i}\langle\hat{\sigma}_y^{(j)}\rangle(t)$, we reconstruct the full time-dependent complex wavefunction of the system without requiring multi-qubit entanglement measurements. The overlapping contributions from different energy bands can be decoupled through a nested temporal Fourier transform:
\begin{equation}
\rho_{n}(j) = \frac{1}{T}\int_{0}^{T} dt\; \xi(j, t) e^{\mathrm{i}E_{n}t} \approx \psi_{n}^*(c) \Psi_{n}(j),
\end{equation}
where $T$ represents the total coherent integration window. 

Under the constraint of chiral symmetry, the system hosts a unique zero-energy mode ($E_{0} = 0$) in the mid-gap spectrum. By selectively filtering the zero-frequency component ($E_{n}=0$), we isolate the stationary spatial profile of this mid-gap state according to Eq.~\eqref{eqn:zero_energy_Eigenstate}:
\begin{equation}
\rho_{n=0}(j) \propto \Psi_{E=0}(j) = C_{1} \text{Re} \left( e^{\mathrm{i}(qj - \phi_q)} \right) + C_2 e^{-\mathrm{i}\phi_0},
\end{equation}
which is supported exclusively on the $A$-sublattice qubits of the ring. In practice, since the zero-energy mode is protected by a finite spectral gap, we integrate the Fourier components within a narrow window around $E=0$ to reconstruct the state, thereby maximizing the signal-to-noise ratio and mitigating finite-time truncation errors. 
Finally, we perform a spatial Fourier transform on the isolated zero-energy component to extract the winding phase. 
Due to the gauge freedom, the measured wavefunction may have a random global phase. To fix this gauge freedom, we take the magnitude $|\rho_{n=0}(j)| $ before the spatial Fourier transformation. This removes the arbitrary global phase at the cost of only a global sign ambiguity, i.e., a uniform $\pi$ shift of all extracted $\phi_q$.  Such a rigid offset leaves the winding number unchanged. 
The target winding phase then reads $\phi_q = -\arg \sum_j |\rho_{n=0}(j)| e^{-\mathrm{i}qj}$.
Importantly, because this nested Fourier analysis intrinsically decodes the globally protected relative phase rather than relying on local wave amplitudes, moderate fluctuations in the measured phase $\phi_q$ at individual momenta do not change the winding number. This is exactly what makes it a topological invariant: it remains unchanged under small perturbations.

\makeatletter
\renewcommand\@biblabel[1]{#1.}
\makeatother
\bibliography{Citation}

\begin{thebibliography}{85}%
\makeatletter
\providecommand \@ifxundefined [1]{%
 \@ifx{#1\undefined}
}%
\providecommand \@ifnum [1]{%
 \ifnum #1\expandafter \@firstoftwo
 \else \expandafter \@secondoftwo
 \fi
}%
\providecommand \@ifx [1]{%
 \ifx #1\expandafter \@firstoftwo
 \else \expandafter \@secondoftwo
 \fi
}%
\providecommand \natexlab [1]{#1}%
\providecommand \enquote  [1]{``#1''}%
\providecommand \bibnamefont  [1]{#1}%
\providecommand \bibfnamefont [1]{#1}%
\providecommand \citenamefont [1]{#1}%
\providecommand \href@noop [0]{\@secondoftwo}%
\providecommand \href [0]{\begingroup \@sanitize@url \@href}%
\providecommand \@href[1]{\@@startlink{#1}\@@href}%
\providecommand \@@href[1]{\endgroup#1\@@endlink}%
\providecommand \@sanitize@url [0]{\catcode `\\12\catcode `\$12\catcode
  `\&12\catcode `\#12\catcode `\^12\catcode `\_12\catcode `\%12\relax}%
\providecommand \@@startlink[1]{}%
\providecommand \@@endlink[0]{}%
\providecommand \url  [0]{\begingroup\@sanitize@url \@url }%
\providecommand \@url [1]{\endgroup\@href {#1}{\urlprefix }}%
\providecommand \urlprefix  [0]{URL }%
\providecommand \Eprint [0]{\href }%
\providecommand \doibase [0]{https://doi.org/}%
\providecommand \selectlanguage [0]{\@gobble}%
\providecommand \bibinfo  [0]{\@secondoftwo}%
\providecommand \bibfield  [0]{\@secondoftwo}%
\providecommand \translation [1]{[#1]}%
\providecommand \BibitemOpen [0]{}%
\providecommand \bibitemStop [0]{}%
\providecommand \bibitemNoStop [0]{.\EOS\space}%
\providecommand \EOS [0]{\spacefactor3000\relax}%
\providecommand \BibitemShut  [1]{\csname bibitem#1\endcsname}%
\let\auto@bib@innerbib\@empty
\bibitem [{\citenamefont {Klitzing}\ \emph {et~al.}(1980)\citenamefont
  {Klitzing}, \citenamefont {Dorda},\ and\ \citenamefont
  {Pepper}}]{PhysRevLett.45.494}%
  \BibitemOpen
  \bibfield  {author} {\bibinfo {author} {\bibfnamefont {K.~v.}\ \bibnamefont
  {Klitzing}}, \bibinfo {author} {\bibfnamefont {G.}~\bibnamefont {Dorda}},\
  and\ \bibinfo {author} {\bibfnamefont {M.}~\bibnamefont {Pepper}},\
  }\bibfield  {title} {\bibinfo {title} {New method for high-accuracy
  determination of the fine-structure constant based on quantized {Hall}
  resistance},\ }\href {https://doi.org/10.1103/PhysRevLett.45.494} {\bibfield
  {journal} {\bibinfo  {journal} {Phys. Rev. Lett.}\ }\textbf {\bibinfo
  {volume} {45}},\ \bibinfo {pages} {494} (\bibinfo {year} {1980})}\BibitemShut
  {NoStop}%
\bibitem [{\citenamefont {Thouless}\ \emph {et~al.}(1982)\citenamefont
  {Thouless}, \citenamefont {Kohmoto}, \citenamefont {Nightingale},\ and\
  \citenamefont {den Nijs}}]{PhysRevLett.49.405}%
  \BibitemOpen
  \bibfield  {author} {\bibinfo {author} {\bibfnamefont {D.~J.}\ \bibnamefont
  {Thouless}}, \bibinfo {author} {\bibfnamefont {M.}~\bibnamefont {Kohmoto}},
  \bibinfo {author} {\bibfnamefont {M.~P.}\ \bibnamefont {Nightingale}},\ and\
  \bibinfo {author} {\bibfnamefont {M.}~\bibnamefont {den Nijs}},\ }\bibfield
  {title} {\bibinfo {title} {Quantized {Hall} conductance in a two-dimensional
  periodic potential},\ }\href {https://doi.org/10.1103/PhysRevLett.49.405}
  {\bibfield  {journal} {\bibinfo  {journal} {Phys. Rev. Lett.}\ }\textbf
  {\bibinfo {volume} {49}},\ \bibinfo {pages} {405} (\bibinfo {year}
  {1982})}\BibitemShut {NoStop}%
\bibitem [{\citenamefont {Haldane}(1988)}]{PhysRevLett.61.2015}%
  \BibitemOpen
  \bibfield  {author} {\bibinfo {author} {\bibfnamefont {F.~D.~M.}\
  \bibnamefont {Haldane}},\ }\bibfield  {title} {\bibinfo {title} {Model for a
  quantum {Hall} effect without {Landau} levels: Condensed-matter realization
  of the "parity anomaly"},\ }\href
  {https://doi.org/10.1103/PhysRevLett.61.2015} {\bibfield  {journal} {\bibinfo
   {journal} {Phys. Rev. Lett.}\ }\textbf {\bibinfo {volume} {61}},\ \bibinfo
  {pages} {2015} (\bibinfo {year} {1988})}\BibitemShut {NoStop}%
\bibitem [{\citenamefont {Kane}\ and\ \citenamefont
  {Mele}(2005)}]{PhysRevLett.95.146802}%
  \BibitemOpen
  \bibfield  {author} {\bibinfo {author} {\bibfnamefont {C.~L.}\ \bibnamefont
  {Kane}}\ and\ \bibinfo {author} {\bibfnamefont {E.~J.}\ \bibnamefont
  {Mele}},\ }\bibfield  {title} {\bibinfo {title} {${Z}_{2}$ topological order
  and the quantum spin {Hall} effect},\ }\href
  {https://doi.org/10.1103/PhysRevLett.95.146802} {\bibfield  {journal}
  {\bibinfo  {journal} {Phys. Rev. Lett.}\ }\textbf {\bibinfo {volume} {95}},\
  \bibinfo {pages} {146802} (\bibinfo {year} {2005})}\BibitemShut {NoStop}%
\bibitem [{\citenamefont {Chiu}\ \emph {et~al.}(2016)\citenamefont {Chiu},
  \citenamefont {Teo}, \citenamefont {Schnyder},\ and\ \citenamefont
  {Ryu}}]{RevModPhys.88.035005}%
  \BibitemOpen
  \bibfield  {author} {\bibinfo {author} {\bibfnamefont {C.-K.}\ \bibnamefont
  {Chiu}}, \bibinfo {author} {\bibfnamefont {J.~C.~Y.}\ \bibnamefont {Teo}},
  \bibinfo {author} {\bibfnamefont {A.~P.}\ \bibnamefont {Schnyder}},\ and\
  \bibinfo {author} {\bibfnamefont {S.}~\bibnamefont {Ryu}},\ }\bibfield
  {title} {\bibinfo {title} {{Classification of topological quantum matter with
  symmetries}},\ }\href {https://doi.org/10.1103/RevModPhys.88.035005}
  {\bibfield  {journal} {\bibinfo  {journal} {Rev. Mod. Phys.}\ }\textbf
  {\bibinfo {volume} {88}},\ \bibinfo {pages} {035005} (\bibinfo {year}
  {2016})}\BibitemShut {NoStop}%
\bibitem [{\citenamefont {Kitaev}(2003)}]{KITAEV20032}%
  \BibitemOpen
  \bibfield  {author} {\bibinfo {author} {\bibfnamefont {A.}~\bibnamefont
  {Kitaev}},\ }\bibfield  {title} {\bibinfo {title} {Fault-tolerant quantum
  computation by anyons},\ }\href
  {https://doi.org/https://doi.org/10.1016/S0003-4916(02)00018-0} {\bibfield
  {journal} {\bibinfo  {journal} {Annals of Physics}\ }\textbf {\bibinfo
  {volume} {303}},\ \bibinfo {pages} {2} (\bibinfo {year} {2003})}\BibitemShut
  {NoStop}%
\bibitem [{\citenamefont {Nayak}\ \emph {et~al.}(2008)\citenamefont {Nayak},
  \citenamefont {Simon}, \citenamefont {Stern}, \citenamefont {Freedman},\ and\
  \citenamefont {Das~Sarma}}]{RevModPhys.80.1083}%
  \BibitemOpen
  \bibfield  {author} {\bibinfo {author} {\bibfnamefont {C.}~\bibnamefont
  {Nayak}}, \bibinfo {author} {\bibfnamefont {S.~H.}\ \bibnamefont {Simon}},
  \bibinfo {author} {\bibfnamefont {A.}~\bibnamefont {Stern}}, \bibinfo
  {author} {\bibfnamefont {M.}~\bibnamefont {Freedman}},\ and\ \bibinfo
  {author} {\bibfnamefont {S.}~\bibnamefont {Das~Sarma}},\ }\bibfield  {title}
  {\bibinfo {title} {Non-{Abelian} anyons and topological quantum
  computation},\ }\href {https://doi.org/10.1103/RevModPhys.80.1083} {\bibfield
   {journal} {\bibinfo  {journal} {Rev. Mod. Phys.}\ }\textbf {\bibinfo
  {volume} {80}},\ \bibinfo {pages} {1083} (\bibinfo {year}
  {2008})}\BibitemShut {NoStop}%
\bibitem [{\citenamefont {Goldman}\ \emph {et~al.}(2016)\citenamefont
  {Goldman}, \citenamefont {Budich},\ and\ \citenamefont
  {Zoller}}]{Goldman2016}%
  \BibitemOpen
  \bibfield  {author} {\bibinfo {author} {\bibfnamefont {N.}~\bibnamefont
  {Goldman}}, \bibinfo {author} {\bibfnamefont {J.~C.}\ \bibnamefont
  {Budich}},\ and\ \bibinfo {author} {\bibfnamefont {P.}~\bibnamefont
  {Zoller}},\ }\bibfield  {title} {\bibinfo {title} {Topological quantum matter
  with ultracold gases in optical lattices},\ }\href
  {https://doi.org/10.1038/nphys3803} {\bibfield  {journal} {\bibinfo
  {journal} {Nature Physics}\ }\textbf {\bibinfo {volume} {12}},\ \bibinfo
  {pages} {639} (\bibinfo {year} {2016})}\BibitemShut {NoStop}%
\bibitem [{\citenamefont {König}\ \emph {et~al.}(2007)\citenamefont {König},
  \citenamefont {Wiedmann}, \citenamefont {Brüne}, \citenamefont {Roth},
  \citenamefont {Buhmann}, \citenamefont {Molenkamp}, \citenamefont {Qi},\ and\
  \citenamefont {Zhang}}]{science.1148047}%
  \BibitemOpen
  \bibfield  {author} {\bibinfo {author} {\bibfnamefont {M.}~\bibnamefont
  {König}}, \bibinfo {author} {\bibfnamefont {S.}~\bibnamefont {Wiedmann}},
  \bibinfo {author} {\bibfnamefont {C.}~\bibnamefont {Brüne}}, \bibinfo
  {author} {\bibfnamefont {A.}~\bibnamefont {Roth}}, \bibinfo {author}
  {\bibfnamefont {H.}~\bibnamefont {Buhmann}}, \bibinfo {author} {\bibfnamefont
  {L.~W.}\ \bibnamefont {Molenkamp}}, \bibinfo {author} {\bibfnamefont {X.-L.}\
  \bibnamefont {Qi}},\ and\ \bibinfo {author} {\bibfnamefont {S.-C.}\
  \bibnamefont {Zhang}},\ }\bibfield  {title} {\bibinfo {title} {Quantum spin
  {Hall} insulator state in {HgTe} quantum wells},\ }\href
  {https://doi.org/10.1126/science.1148047} {\bibfield  {journal} {\bibinfo
  {journal} {Science}\ }\textbf {\bibinfo {volume} {318}},\ \bibinfo {pages}
  {766} (\bibinfo {year} {2007})}\BibitemShut {NoStop}%
\bibitem [{\citenamefont {Hsieh}\ \emph {et~al.}(2008)\citenamefont {Hsieh},
  \citenamefont {Qian}, \citenamefont {Wray}, \citenamefont {Xia},
  \citenamefont {Hor}, \citenamefont {Cava},\ and\ \citenamefont
  {Hasan}}]{Hsieh2008}%
  \BibitemOpen
  \bibfield  {author} {\bibinfo {author} {\bibfnamefont {D.}~\bibnamefont
  {Hsieh}}, \bibinfo {author} {\bibfnamefont {D.}~\bibnamefont {Qian}},
  \bibinfo {author} {\bibfnamefont {L.}~\bibnamefont {Wray}}, \bibinfo {author}
  {\bibfnamefont {Y.}~\bibnamefont {Xia}}, \bibinfo {author} {\bibfnamefont
  {Y.~S.}\ \bibnamefont {Hor}}, \bibinfo {author} {\bibfnamefont {R.~J.}\
  \bibnamefont {Cava}},\ and\ \bibinfo {author} {\bibfnamefont {M.~Z.}\
  \bibnamefont {Hasan}},\ }\bibfield  {title} {\bibinfo {title} {A topological
  {Dirac} insulator in a quantum spin {Hall} phase},\ }\href
  {https://doi.org/10.1038/nature06843} {\bibfield  {journal} {\bibinfo
  {journal} {Nature}\ }\textbf {\bibinfo {volume} {452}},\ \bibinfo {pages}
  {970} (\bibinfo {year} {2008})}\BibitemShut {NoStop}%
\bibitem [{\citenamefont {Rechtsman}\ \emph {et~al.}(2013)\citenamefont
  {Rechtsman}, \citenamefont {Zeuner}, \citenamefont {Plotnik}, \citenamefont
  {Lumer}, \citenamefont {Podolsky}, \citenamefont {Dreisow}, \citenamefont
  {Nolte}, \citenamefont {Segev},\ and\ \citenamefont
  {Szameit}}]{rechtsman2013photonic}%
  \BibitemOpen
  \bibfield  {author} {\bibinfo {author} {\bibfnamefont {M.~C.}\ \bibnamefont
  {Rechtsman}}, \bibinfo {author} {\bibfnamefont {J.~M.}\ \bibnamefont
  {Zeuner}}, \bibinfo {author} {\bibfnamefont {Y.}~\bibnamefont {Plotnik}},
  \bibinfo {author} {\bibfnamefont {Y.}~\bibnamefont {Lumer}}, \bibinfo
  {author} {\bibfnamefont {D.}~\bibnamefont {Podolsky}}, \bibinfo {author}
  {\bibfnamefont {F.}~\bibnamefont {Dreisow}}, \bibinfo {author} {\bibfnamefont
  {S.}~\bibnamefont {Nolte}}, \bibinfo {author} {\bibfnamefont
  {M.}~\bibnamefont {Segev}},\ and\ \bibinfo {author} {\bibfnamefont
  {A.}~\bibnamefont {Szameit}},\ }\bibfield  {title} {\bibinfo {title}
  {Photonic {Floquet} topological insulators},\ }\href
  {https://doi.org/10.1038/nature12066} {\bibfield  {journal} {\bibinfo
  {journal} {Nature}\ }\textbf {\bibinfo {volume} {496}},\ \bibinfo {pages}
  {196} (\bibinfo {year} {2013})}\BibitemShut {NoStop}%
\bibitem [{\citenamefont {Yang}\ \emph {et~al.}(2018)\citenamefont {Yang},
  \citenamefont {Guo}, \citenamefont {Tremain}, \citenamefont {Liu},
  \citenamefont {Barr}, \citenamefont {Yan}, \citenamefont {Gao}, \citenamefont
  {Liu}, \citenamefont {Xiang}, \citenamefont {Chen}, \citenamefont {Fang},
  \citenamefont {Hibbins}, \citenamefont {Lu},\ and\ \citenamefont
  {Zhang}}]{Yang2018ideal}%
  \BibitemOpen
  \bibfield  {author} {\bibinfo {author} {\bibfnamefont {B.}~\bibnamefont
  {Yang}}, \bibinfo {author} {\bibfnamefont {Q.}~\bibnamefont {Guo}}, \bibinfo
  {author} {\bibfnamefont {B.}~\bibnamefont {Tremain}}, \bibinfo {author}
  {\bibfnamefont {R.}~\bibnamefont {Liu}}, \bibinfo {author} {\bibfnamefont
  {L.~E.}\ \bibnamefont {Barr}}, \bibinfo {author} {\bibfnamefont
  {Q.}~\bibnamefont {Yan}}, \bibinfo {author} {\bibfnamefont {W.}~\bibnamefont
  {Gao}}, \bibinfo {author} {\bibfnamefont {H.}~\bibnamefont {Liu}}, \bibinfo
  {author} {\bibfnamefont {Y.}~\bibnamefont {Xiang}}, \bibinfo {author}
  {\bibfnamefont {J.}~\bibnamefont {Chen}}, \bibinfo {author} {\bibfnamefont
  {C.}~\bibnamefont {Fang}}, \bibinfo {author} {\bibfnamefont {A.}~\bibnamefont
  {Hibbins}}, \bibinfo {author} {\bibfnamefont {L.}~\bibnamefont {Lu}},\ and\
  \bibinfo {author} {\bibfnamefont {S.}~\bibnamefont {Zhang}},\ }\bibfield
  {title} {\bibinfo {title} {Ideal {Weyl} points and helicoid surface states in
  artificial photonic crystal structures},\ }\href
  {https://doi.org/10.1126/science.aaq1221} {\bibfield  {journal} {\bibinfo
  {journal} {Science}\ }\textbf {\bibinfo {volume} {359}},\ \bibinfo {pages}
  {1013} (\bibinfo {year} {2018})}\BibitemShut {NoStop}%
\bibitem [{\citenamefont {Zhirihin}\ and\ \citenamefont
  {Kivshar}(2021)}]{Zhirihin2021}%
  \BibitemOpen
  \bibfield  {author} {\bibinfo {author} {\bibfnamefont {D.~V.}\ \bibnamefont
  {Zhirihin}}\ and\ \bibinfo {author} {\bibfnamefont {Y.~S.}\ \bibnamefont
  {Kivshar}},\ }\bibfield  {title} {\bibinfo {title} {Topological photonics on
  a small scale},\ }\href
  {https://doi.org/https://doi.org/10.1002/smsc.202170032} {\bibfield
  {journal} {\bibinfo  {journal} {Small Science}\ }\textbf {\bibinfo {volume}
  {1}},\ \bibinfo {pages} {2170032} (\bibinfo {year} {2021})}\BibitemShut
  {NoStop}%
\bibitem [{\citenamefont {Braun}\ \emph {et~al.}(2024)\citenamefont {Braun},
  \citenamefont {Saint-Jalm}, \citenamefont {Hesse}, \citenamefont {Arceri},
  \citenamefont {Bloch},\ and\ \citenamefont {Aidelsburger}}]{braun2024real}%
  \BibitemOpen
  \bibfield  {author} {\bibinfo {author} {\bibfnamefont {C.}~\bibnamefont
  {Braun}}, \bibinfo {author} {\bibfnamefont {R.}~\bibnamefont {Saint-Jalm}},
  \bibinfo {author} {\bibfnamefont {A.}~\bibnamefont {Hesse}}, \bibinfo
  {author} {\bibfnamefont {J.}~\bibnamefont {Arceri}}, \bibinfo {author}
  {\bibfnamefont {I.}~\bibnamefont {Bloch}},\ and\ \bibinfo {author}
  {\bibfnamefont {M.}~\bibnamefont {Aidelsburger}},\ }\bibfield  {title}
  {\bibinfo {title} {Real-space detection and manipulation of topological edge
  modes with ultracold atoms},\ }\href
  {https://doi.org/10.1038/s41567-024-02506-z} {\bibfield  {journal} {\bibinfo
  {journal} {Nature Physics}\ }\textbf {\bibinfo {volume} {20}},\ \bibinfo
  {pages} {1306} (\bibinfo {year} {2024})}\BibitemShut {NoStop}%
\bibitem [{\citenamefont {Atala}\ \emph {et~al.}(2013)\citenamefont {Atala},
  \citenamefont {Aidelsburger}, \citenamefont {Barreiro}, \citenamefont
  {Abanin}, \citenamefont {Kitagawa}, \citenamefont {Demler},\ and\
  \citenamefont {Bloch}}]{atala2013direct}%
  \BibitemOpen
  \bibfield  {author} {\bibinfo {author} {\bibfnamefont {M.}~\bibnamefont
  {Atala}}, \bibinfo {author} {\bibfnamefont {M.}~\bibnamefont {Aidelsburger}},
  \bibinfo {author} {\bibfnamefont {J.~T.}\ \bibnamefont {Barreiro}}, \bibinfo
  {author} {\bibfnamefont {D.}~\bibnamefont {Abanin}}, \bibinfo {author}
  {\bibfnamefont {T.}~\bibnamefont {Kitagawa}}, \bibinfo {author}
  {\bibfnamefont {E.}~\bibnamefont {Demler}},\ and\ \bibinfo {author}
  {\bibfnamefont {I.}~\bibnamefont {Bloch}},\ }\bibfield  {title} {\bibinfo
  {title} {Direct measurement of the {Zak} phase in topological {Bloch}
  bands},\ }\href {https://doi.org/10.1038/nphys2790} {\bibfield  {journal}
  {\bibinfo  {journal} {Nature Physics}\ }\textbf {\bibinfo {volume} {9}},\
  \bibinfo {pages} {795} (\bibinfo {year} {2013})}\BibitemShut {NoStop}%
\bibitem [{\citenamefont {Duca}\ \emph {et~al.}(2015)\citenamefont {Duca},
  \citenamefont {Li}, \citenamefont {Reitter}, \citenamefont {Bloch},
  \citenamefont {Schleier-Smith},\ and\ \citenamefont
  {Schneider}}]{science.1259052}%
  \BibitemOpen
  \bibfield  {author} {\bibinfo {author} {\bibfnamefont {L.}~\bibnamefont
  {Duca}}, \bibinfo {author} {\bibfnamefont {T.}~\bibnamefont {Li}}, \bibinfo
  {author} {\bibfnamefont {M.}~\bibnamefont {Reitter}}, \bibinfo {author}
  {\bibfnamefont {I.}~\bibnamefont {Bloch}}, \bibinfo {author} {\bibfnamefont
  {M.}~\bibnamefont {Schleier-Smith}},\ and\ \bibinfo {author} {\bibfnamefont
  {U.}~\bibnamefont {Schneider}},\ }\bibfield  {title} {\bibinfo {title} {An
  {Aharonov-Bohm} interferometer for determining {Bloch} band topology},\
  }\href {https://doi.org/10.1126/science.1259052} {\bibfield  {journal}
  {\bibinfo  {journal} {Science}\ }\textbf {\bibinfo {volume} {347}},\ \bibinfo
  {pages} {288} (\bibinfo {year} {2015})}\BibitemShut {NoStop}%
\bibitem [{\citenamefont {Jotzu}\ \emph {et~al.}(2014)\citenamefont {Jotzu},
  \citenamefont {Messer}, \citenamefont {Desbuquois}, \citenamefont {Lebrat},
  \citenamefont {Uehlinger}, \citenamefont {Greif},\ and\ \citenamefont
  {Esslinger}}]{jotzu2014experimental}%
  \BibitemOpen
  \bibfield  {author} {\bibinfo {author} {\bibfnamefont {G.}~\bibnamefont
  {Jotzu}}, \bibinfo {author} {\bibfnamefont {M.}~\bibnamefont {Messer}},
  \bibinfo {author} {\bibfnamefont {R.}~\bibnamefont {Desbuquois}}, \bibinfo
  {author} {\bibfnamefont {M.}~\bibnamefont {Lebrat}}, \bibinfo {author}
  {\bibfnamefont {T.}~\bibnamefont {Uehlinger}}, \bibinfo {author}
  {\bibfnamefont {D.}~\bibnamefont {Greif}},\ and\ \bibinfo {author}
  {\bibfnamefont {T.}~\bibnamefont {Esslinger}},\ }\bibfield  {title} {\bibinfo
  {title} {Experimental realization of the topological {Haldane} model with
  ultracold fermions},\ }\href {https://doi.org/10.1038/nature13915} {\bibfield
   {journal} {\bibinfo  {journal} {Nature}\ }\textbf {\bibinfo {volume}
  {515}},\ \bibinfo {pages} {237} (\bibinfo {year} {2014})}\BibitemShut
  {NoStop}%
\bibitem [{\citenamefont {Aidelsburger}\ \emph {et~al.}(2015)\citenamefont
  {Aidelsburger}, \citenamefont {Lohse}, \citenamefont {Schweizer},
  \citenamefont {Atala}, \citenamefont {Barreiro}, \citenamefont
  {Nascimb{\`e}ne}, \citenamefont {Cooper}, \citenamefont {Bloch},\ and\
  \citenamefont {Goldman}}]{aidelsburger2015measuring}%
  \BibitemOpen
  \bibfield  {author} {\bibinfo {author} {\bibfnamefont {M.}~\bibnamefont
  {Aidelsburger}}, \bibinfo {author} {\bibfnamefont {M.}~\bibnamefont {Lohse}},
  \bibinfo {author} {\bibfnamefont {C.}~\bibnamefont {Schweizer}}, \bibinfo
  {author} {\bibfnamefont {M.}~\bibnamefont {Atala}}, \bibinfo {author}
  {\bibfnamefont {J.~T.}\ \bibnamefont {Barreiro}}, \bibinfo {author}
  {\bibfnamefont {S.}~\bibnamefont {Nascimb{\`e}ne}}, \bibinfo {author}
  {\bibfnamefont {N.}~\bibnamefont {Cooper}}, \bibinfo {author} {\bibfnamefont
  {I.}~\bibnamefont {Bloch}},\ and\ \bibinfo {author} {\bibfnamefont
  {N.}~\bibnamefont {Goldman}},\ }\bibfield  {title} {\bibinfo {title}
  {Measuring the {Chern} number of {Hofstadter} bands with ultracold bosonic
  atoms},\ }\href {https://doi.org/10.1038/nphys3171} {\bibfield  {journal}
  {\bibinfo  {journal} {Nature Physics}\ }\textbf {\bibinfo {volume} {11}},\
  \bibinfo {pages} {162} (\bibinfo {year} {2015})}\BibitemShut {NoStop}%
\bibitem [{\citenamefont {Li}\ \emph {et~al.}(2016)\citenamefont {Li},
  \citenamefont {Duca}, \citenamefont {Reitter}, \citenamefont {Grusdt},
  \citenamefont {Demler}, \citenamefont {Endres}, \citenamefont
  {Schleier-Smith}, \citenamefont {Bloch},\ and\ \citenamefont
  {Schneider}}]{science.aad5812}%
  \BibitemOpen
  \bibfield  {author} {\bibinfo {author} {\bibfnamefont {T.}~\bibnamefont
  {Li}}, \bibinfo {author} {\bibfnamefont {L.}~\bibnamefont {Duca}}, \bibinfo
  {author} {\bibfnamefont {M.}~\bibnamefont {Reitter}}, \bibinfo {author}
  {\bibfnamefont {F.}~\bibnamefont {Grusdt}}, \bibinfo {author} {\bibfnamefont
  {E.}~\bibnamefont {Demler}}, \bibinfo {author} {\bibfnamefont
  {M.}~\bibnamefont {Endres}}, \bibinfo {author} {\bibfnamefont
  {M.}~\bibnamefont {Schleier-Smith}}, \bibinfo {author} {\bibfnamefont
  {I.}~\bibnamefont {Bloch}},\ and\ \bibinfo {author} {\bibfnamefont
  {U.}~\bibnamefont {Schneider}},\ }\bibfield  {title} {\bibinfo {title}
  {{Bloch} state tomography using {Wilson} lines},\ }\href
  {https://doi.org/10.1126/science.aad5812} {\bibfield  {journal} {\bibinfo
  {journal} {Science}\ }\textbf {\bibinfo {volume} {352}},\ \bibinfo {pages}
  {1094} (\bibinfo {year} {2016})}\BibitemShut {NoStop}%
\bibitem [{\citenamefont {Cardano}\ \emph {et~al.}(2017)\citenamefont
  {Cardano}, \citenamefont {D’Errico}, \citenamefont {Dauphin}, \citenamefont
  {Maffei}, \citenamefont {Piccirillo}, \citenamefont {de~Lisio}, \citenamefont
  {De~Filippis}, \citenamefont {Cataudella}, \citenamefont {Santamato},
  \citenamefont {Marrucci}, \citenamefont {Lewenstein},\ and\ \citenamefont
  {Massignan}}]{cardano2017detection}%
  \BibitemOpen
  \bibfield  {author} {\bibinfo {author} {\bibfnamefont {F.}~\bibnamefont
  {Cardano}}, \bibinfo {author} {\bibfnamefont {A.}~\bibnamefont {D’Errico}},
  \bibinfo {author} {\bibfnamefont {A.}~\bibnamefont {Dauphin}}, \bibinfo
  {author} {\bibfnamefont {M.}~\bibnamefont {Maffei}}, \bibinfo {author}
  {\bibfnamefont {B.}~\bibnamefont {Piccirillo}}, \bibinfo {author}
  {\bibfnamefont {C.}~\bibnamefont {de~Lisio}}, \bibinfo {author}
  {\bibfnamefont {G.}~\bibnamefont {De~Filippis}}, \bibinfo {author}
  {\bibfnamefont {V.}~\bibnamefont {Cataudella}}, \bibinfo {author}
  {\bibfnamefont {E.}~\bibnamefont {Santamato}}, \bibinfo {author}
  {\bibfnamefont {L.}~\bibnamefont {Marrucci}}, \bibinfo {author}
  {\bibfnamefont {M.}~\bibnamefont {Lewenstein}},\ and\ \bibinfo {author}
  {\bibfnamefont {P.}~\bibnamefont {Massignan}},\ }\bibfield  {title} {\bibinfo
  {title} {Detection of {Zak} phases and topological invariants in a chiral
  quantum walk of twisted photons},\ }\href
  {https://doi.org/10.1038/ncomms15516} {\bibfield  {journal} {\bibinfo
  {journal} {Nature communications}\ }\textbf {\bibinfo {volume} {8}},\
  \bibinfo {pages} {15516} (\bibinfo {year} {2017})}\BibitemShut {NoStop}%
\bibitem [{\citenamefont {Jiao}\ \emph {et~al.}(2021)\citenamefont {Jiao},
  \citenamefont {Longhi}, \citenamefont {Wang}, \citenamefont {Gao},
  \citenamefont {Zhou}, \citenamefont {Wang}, \citenamefont {Fu}, \citenamefont
  {Wang}, \citenamefont {Ren}, \citenamefont {Qiao},\ and\ \citenamefont
  {Jin}}]{PhysRevLett.127.147401}%
  \BibitemOpen
  \bibfield  {author} {\bibinfo {author} {\bibfnamefont {Z.-Q.}\ \bibnamefont
  {Jiao}}, \bibinfo {author} {\bibfnamefont {S.}~\bibnamefont {Longhi}},
  \bibinfo {author} {\bibfnamefont {X.-W.}\ \bibnamefont {Wang}}, \bibinfo
  {author} {\bibfnamefont {J.}~\bibnamefont {Gao}}, \bibinfo {author}
  {\bibfnamefont {W.-H.}\ \bibnamefont {Zhou}}, \bibinfo {author}
  {\bibfnamefont {Y.}~\bibnamefont {Wang}}, \bibinfo {author} {\bibfnamefont
  {Y.-X.}\ \bibnamefont {Fu}}, \bibinfo {author} {\bibfnamefont
  {L.}~\bibnamefont {Wang}}, \bibinfo {author} {\bibfnamefont {R.-J.}\
  \bibnamefont {Ren}}, \bibinfo {author} {\bibfnamefont {L.-F.}\ \bibnamefont
  {Qiao}},\ and\ \bibinfo {author} {\bibfnamefont {X.-M.}\ \bibnamefont
  {Jin}},\ }\bibfield  {title} {\bibinfo {title} {Experimentally detecting
  quantized {Zak} phases without chiral symmetry in photonic lattices},\ }\href
  {https://doi.org/10.1103/PhysRevLett.127.147401} {\bibfield  {journal}
  {\bibinfo  {journal} {Phys. Rev. Lett.}\ }\textbf {\bibinfo {volume} {127}},\
  \bibinfo {pages} {147401} (\bibinfo {year} {2021})}\BibitemShut {NoStop}%
\bibitem [{\citenamefont {Flurin}\ \emph {et~al.}(2017)\citenamefont {Flurin},
  \citenamefont {Ramasesh}, \citenamefont {Hacohen-Gourgy}, \citenamefont
  {Martin}, \citenamefont {Yao},\ and\ \citenamefont
  {Siddiqi}}]{PhysRevX.7.031023}%
  \BibitemOpen
  \bibfield  {author} {\bibinfo {author} {\bibfnamefont {E.}~\bibnamefont
  {Flurin}}, \bibinfo {author} {\bibfnamefont {V.~V.}\ \bibnamefont
  {Ramasesh}}, \bibinfo {author} {\bibfnamefont {S.}~\bibnamefont
  {Hacohen-Gourgy}}, \bibinfo {author} {\bibfnamefont {L.~S.}\ \bibnamefont
  {Martin}}, \bibinfo {author} {\bibfnamefont {N.~Y.}\ \bibnamefont {Yao}},\
  and\ \bibinfo {author} {\bibfnamefont {I.}~\bibnamefont {Siddiqi}},\
  }\bibfield  {title} {\bibinfo {title} {Observing topological invariants using
  quantum walks in superconducting circuits},\ }\href
  {https://doi.org/10.1103/PhysRevX.7.031023} {\bibfield  {journal} {\bibinfo
  {journal} {Phys. Rev. X}\ }\textbf {\bibinfo {volume} {7}},\ \bibinfo {pages}
  {031023} (\bibinfo {year} {2017})}\BibitemShut {NoStop}%
\bibitem [{\citenamefont {Lohse}\ \emph {et~al.}(2016)\citenamefont {Lohse},
  \citenamefont {Schweizer}, \citenamefont {Zilberberg}, \citenamefont
  {Aidelsburger},\ and\ \citenamefont {Bloch}}]{lohse2016thouless}%
  \BibitemOpen
  \bibfield  {author} {\bibinfo {author} {\bibfnamefont {M.}~\bibnamefont
  {Lohse}}, \bibinfo {author} {\bibfnamefont {C.}~\bibnamefont {Schweizer}},
  \bibinfo {author} {\bibfnamefont {O.}~\bibnamefont {Zilberberg}}, \bibinfo
  {author} {\bibfnamefont {M.}~\bibnamefont {Aidelsburger}},\ and\ \bibinfo
  {author} {\bibfnamefont {I.}~\bibnamefont {Bloch}},\ }\bibfield  {title}
  {\bibinfo {title} {A {Thouless} quantum pump with ultracold bosonic atoms in
  an optical superlattice},\ }\href {https://doi.org/10.1038/NPHYS3584}
  {\bibfield  {journal} {\bibinfo  {journal} {Nature Physics}\ }\textbf
  {\bibinfo {volume} {12}},\ \bibinfo {pages} {350} (\bibinfo {year}
  {2016})}\BibitemShut {NoStop}%
\bibitem [{\citenamefont {Nakajima}\ \emph {et~al.}(2016)\citenamefont
  {Nakajima}, \citenamefont {Tomita}, \citenamefont {Taie}, \citenamefont
  {Ichinose}, \citenamefont {Ozawa}, \citenamefont {Wang}, \citenamefont
  {Troyer},\ and\ \citenamefont {Takahashi}}]{nakajima2016topological}%
  \BibitemOpen
  \bibfield  {author} {\bibinfo {author} {\bibfnamefont {S.}~\bibnamefont
  {Nakajima}}, \bibinfo {author} {\bibfnamefont {T.}~\bibnamefont {Tomita}},
  \bibinfo {author} {\bibfnamefont {S.}~\bibnamefont {Taie}}, \bibinfo {author}
  {\bibfnamefont {T.}~\bibnamefont {Ichinose}}, \bibinfo {author}
  {\bibfnamefont {H.}~\bibnamefont {Ozawa}}, \bibinfo {author} {\bibfnamefont
  {L.}~\bibnamefont {Wang}}, \bibinfo {author} {\bibfnamefont {M.}~\bibnamefont
  {Troyer}},\ and\ \bibinfo {author} {\bibfnamefont {Y.}~\bibnamefont
  {Takahashi}},\ }\bibfield  {title} {\bibinfo {title} {Topological {Thouless}
  pumping of ultracold fermions},\ }\href {https://doi.org/10.1038/NPHYS3622}
  {\bibfield  {journal} {\bibinfo  {journal} {Nature Physics}\ }\textbf
  {\bibinfo {volume} {12}},\ \bibinfo {pages} {296} (\bibinfo {year}
  {2016})}\BibitemShut {NoStop}%
\bibitem [{\citenamefont {Xiang}\ \emph {et~al.}(2023)\citenamefont {Xiang},
  \citenamefont {Huang}, \citenamefont {Zhang}, \citenamefont {Liu},
  \citenamefont {Shi}, \citenamefont {Deng}, \citenamefont {Liu}, \citenamefont
  {Li}, \citenamefont {Liang}, \citenamefont {Mei}, \citenamefont {Yu},
  \citenamefont {Xue}, \citenamefont {Tian}, \citenamefont {Song},
  \citenamefont {Liu}, \citenamefont {Xu}, \citenamefont {Zheng}, \citenamefont
  {Nori},\ and\ \citenamefont {Fan}}]{xiang2023simulating}%
  \BibitemOpen
  \bibfield  {author} {\bibinfo {author} {\bibfnamefont {Z.-C.}\ \bibnamefont
  {Xiang}}, \bibinfo {author} {\bibfnamefont {K.}~\bibnamefont {Huang}},
  \bibinfo {author} {\bibfnamefont {Y.-R.}\ \bibnamefont {Zhang}}, \bibinfo
  {author} {\bibfnamefont {T.}~\bibnamefont {Liu}}, \bibinfo {author}
  {\bibfnamefont {Y.-H.}\ \bibnamefont {Shi}}, \bibinfo {author} {\bibfnamefont
  {C.-L.}\ \bibnamefont {Deng}}, \bibinfo {author} {\bibfnamefont
  {T.}~\bibnamefont {Liu}}, \bibinfo {author} {\bibfnamefont {H.}~\bibnamefont
  {Li}}, \bibinfo {author} {\bibfnamefont {G.-H.}\ \bibnamefont {Liang}},
  \bibinfo {author} {\bibfnamefont {Z.-Y.}\ \bibnamefont {Mei}}, \bibinfo
  {author} {\bibfnamefont {H.}~\bibnamefont {Yu}}, \bibinfo {author}
  {\bibfnamefont {G.}~\bibnamefont {Xue}}, \bibinfo {author} {\bibfnamefont
  {Y.}~\bibnamefont {Tian}}, \bibinfo {author} {\bibfnamefont {X.}~\bibnamefont
  {Song}}, \bibinfo {author} {\bibfnamefont {Z.-B.}\ \bibnamefont {Liu}},
  \bibinfo {author} {\bibfnamefont {K.}~\bibnamefont {Xu}}, \bibinfo {author}
  {\bibfnamefont {D.}~\bibnamefont {Zheng}}, \bibinfo {author} {\bibfnamefont
  {F.}~\bibnamefont {Nori}},\ and\ \bibinfo {author} {\bibfnamefont
  {H.}~\bibnamefont {Fan}},\ }\bibfield  {title} {\bibinfo {title} {Simulating
  {Chern} insulators on a superconducting quantum processor},\ }\href
  {https://doi.org/10.1038/s41467-023-41230-9} {\bibfield  {journal} {\bibinfo
  {journal} {Nature Communications}\ }\textbf {\bibinfo {volume} {14}},\
  \bibinfo {pages} {5433} (\bibinfo {year} {2023})}\BibitemShut {NoStop}%
\bibitem [{\citenamefont {Meier}\ \emph {et~al.}(2018)\citenamefont {Meier},
  \citenamefont {An}, \citenamefont {Dauphin}, \citenamefont {Maffei},
  \citenamefont {Massignan}, \citenamefont {Hughes},\ and\ \citenamefont
  {Gadway}}]{Science362}%
  \BibitemOpen
  \bibfield  {author} {\bibinfo {author} {\bibfnamefont {E.~J.}\ \bibnamefont
  {Meier}}, \bibinfo {author} {\bibfnamefont {F.~A.}\ \bibnamefont {An}},
  \bibinfo {author} {\bibfnamefont {A.}~\bibnamefont {Dauphin}}, \bibinfo
  {author} {\bibfnamefont {M.}~\bibnamefont {Maffei}}, \bibinfo {author}
  {\bibfnamefont {P.}~\bibnamefont {Massignan}}, \bibinfo {author}
  {\bibfnamefont {T.~L.}\ \bibnamefont {Hughes}},\ and\ \bibinfo {author}
  {\bibfnamefont {B.}~\bibnamefont {Gadway}},\ }\bibfield  {title} {\bibinfo
  {title} {{Observation of the topological {Anderson} insulator in disordered
  atomic wires}},\ }\href {https://doi.org/10.1126/science.aat3406} {\bibfield
  {journal} {\bibinfo  {journal} {Science}\ }\textbf {\bibinfo {volume}
  {362}},\ \bibinfo {pages} {929} (\bibinfo {year} {2018})}\BibitemShut
  {NoStop}%
\bibitem [{\citenamefont {Hauke}\ \emph {et~al.}(2014)\citenamefont {Hauke},
  \citenamefont {Lewenstein},\ and\ \citenamefont
  {Eckardt}}]{PhysRevLett.113.045303}%
  \BibitemOpen
  \bibfield  {author} {\bibinfo {author} {\bibfnamefont {P.}~\bibnamefont
  {Hauke}}, \bibinfo {author} {\bibfnamefont {M.}~\bibnamefont {Lewenstein}},\
  and\ \bibinfo {author} {\bibfnamefont {A.}~\bibnamefont {Eckardt}},\
  }\bibfield  {title} {\bibinfo {title} {Tomography of band insulators from
  quench dynamics},\ }\href {https://doi.org/10.1103/PhysRevLett.113.045303}
  {\bibfield  {journal} {\bibinfo  {journal} {Phys. Rev. Lett.}\ }\textbf
  {\bibinfo {volume} {113}},\ \bibinfo {pages} {045303} (\bibinfo {year}
  {2014})}\BibitemShut {NoStop}%
\bibitem [{\citenamefont {Fläschner}\ \emph {et~al.}(2016)\citenamefont
  {Fläschner}, \citenamefont {Rem}, \citenamefont {Tarnowski}, \citenamefont
  {Vogel}, \citenamefont {Lühmann}, \citenamefont {Sengstock},\ and\
  \citenamefont {Weitenberg}}]{science.aad4568}%
  \BibitemOpen
  \bibfield  {author} {\bibinfo {author} {\bibfnamefont {N.}~\bibnamefont
  {Fläschner}}, \bibinfo {author} {\bibfnamefont {B.~S.}\ \bibnamefont {Rem}},
  \bibinfo {author} {\bibfnamefont {M.}~\bibnamefont {Tarnowski}}, \bibinfo
  {author} {\bibfnamefont {D.}~\bibnamefont {Vogel}}, \bibinfo {author}
  {\bibfnamefont {D.-S.}\ \bibnamefont {Lühmann}}, \bibinfo {author}
  {\bibfnamefont {K.}~\bibnamefont {Sengstock}},\ and\ \bibinfo {author}
  {\bibfnamefont {C.}~\bibnamefont {Weitenberg}},\ }\bibfield  {title}
  {\bibinfo {title} {Experimental reconstruction of the {Berry} curvature in a
  {Floquet Bloch} band},\ }\href {https://doi.org/10.1126/science.aad4568}
  {\bibfield  {journal} {\bibinfo  {journal} {Science}\ }\textbf {\bibinfo
  {volume} {352}},\ \bibinfo {pages} {1091} (\bibinfo {year}
  {2016})}\BibitemShut {NoStop}%
\bibitem [{\citenamefont {Wang}\ \emph {et~al.}(2017)\citenamefont {Wang},
  \citenamefont {Zhang}, \citenamefont {Chen}, \citenamefont {Yu},\ and\
  \citenamefont {Zhai}}]{PhysRevLett.118.185701}%
  \BibitemOpen
  \bibfield  {author} {\bibinfo {author} {\bibfnamefont {C.}~\bibnamefont
  {Wang}}, \bibinfo {author} {\bibfnamefont {P.}~\bibnamefont {Zhang}},
  \bibinfo {author} {\bibfnamefont {X.}~\bibnamefont {Chen}}, \bibinfo {author}
  {\bibfnamefont {J.}~\bibnamefont {Yu}},\ and\ \bibinfo {author}
  {\bibfnamefont {H.}~\bibnamefont {Zhai}},\ }\bibfield  {title} {\bibinfo
  {title} {Scheme to measure the topological number of a {Chern} insulator from
  quench dynamics},\ }\href {https://doi.org/10.1103/PhysRevLett.118.185701}
  {\bibfield  {journal} {\bibinfo  {journal} {Phys. Rev. Lett.}\ }\textbf
  {\bibinfo {volume} {118}},\ \bibinfo {pages} {185701} (\bibinfo {year}
  {2017})}\BibitemShut {NoStop}%
\bibitem [{\citenamefont {Fl{\"a}schner}\ \emph {et~al.}(2018)\citenamefont
  {Fl{\"a}schner}, \citenamefont {Vogel}, \citenamefont {Tarnowski},
  \citenamefont {Rem}, \citenamefont {L{\"u}hmann}, \citenamefont {Heyl},
  \citenamefont {Budich}, \citenamefont {Mathey}, \citenamefont {Sengstock},\
  and\ \citenamefont {Weitenberg}}]{flaschner2018observation}%
  \BibitemOpen
  \bibfield  {author} {\bibinfo {author} {\bibfnamefont {N.}~\bibnamefont
  {Fl{\"a}schner}}, \bibinfo {author} {\bibfnamefont {D.}~\bibnamefont
  {Vogel}}, \bibinfo {author} {\bibfnamefont {M.}~\bibnamefont {Tarnowski}},
  \bibinfo {author} {\bibfnamefont {B.~S.}\ \bibnamefont {Rem}}, \bibinfo
  {author} {\bibfnamefont {D.-S.}\ \bibnamefont {L{\"u}hmann}}, \bibinfo
  {author} {\bibfnamefont {M.}~\bibnamefont {Heyl}}, \bibinfo {author}
  {\bibfnamefont {J.~C.}\ \bibnamefont {Budich}}, \bibinfo {author}
  {\bibfnamefont {L.}~\bibnamefont {Mathey}}, \bibinfo {author} {\bibfnamefont
  {K.}~\bibnamefont {Sengstock}},\ and\ \bibinfo {author} {\bibfnamefont
  {C.}~\bibnamefont {Weitenberg}},\ }\bibfield  {title} {\bibinfo {title}
  {Observation of dynamical vortices after quenches in a system with
  topology},\ }\href {https://doi.org/10.1038/s41567-017-0013-8} {\bibfield
  {journal} {\bibinfo  {journal} {Nature Physics}\ }\textbf {\bibinfo {volume}
  {14}},\ \bibinfo {pages} {265} (\bibinfo {year} {2018})}\BibitemShut
  {NoStop}%
\bibitem [{\citenamefont {Tarnowski}\ \emph {et~al.}(2019)\citenamefont
  {Tarnowski}, \citenamefont {{\"U}nal}, \citenamefont {Fl{\"a}schner},
  \citenamefont {Rem}, \citenamefont {Eckardt}, \citenamefont {Sengstock},\
  and\ \citenamefont {Weitenberg}}]{tarnowski2019measuring}%
  \BibitemOpen
  \bibfield  {author} {\bibinfo {author} {\bibfnamefont {M.}~\bibnamefont
  {Tarnowski}}, \bibinfo {author} {\bibfnamefont {F.~N.}\ \bibnamefont
  {{\"U}nal}}, \bibinfo {author} {\bibfnamefont {N.}~\bibnamefont
  {Fl{\"a}schner}}, \bibinfo {author} {\bibfnamefont {B.~S.}\ \bibnamefont
  {Rem}}, \bibinfo {author} {\bibfnamefont {A.}~\bibnamefont {Eckardt}},
  \bibinfo {author} {\bibfnamefont {K.}~\bibnamefont {Sengstock}},\ and\
  \bibinfo {author} {\bibfnamefont {C.}~\bibnamefont {Weitenberg}},\ }\bibfield
   {title} {\bibinfo {title} {Measuring topology from dynamics by obtaining the
  {Chern} number from a linking number},\ }\href
  {https://doi.org/10.1038/s41467-019-09668-y} {\bibfield  {journal} {\bibinfo
  {journal} {Nature communications}\ }\textbf {\bibinfo {volume} {10}},\
  \bibinfo {pages} {1728} (\bibinfo {year} {2019})}\BibitemShut {NoStop}%
\bibitem [{\citenamefont {Sun}\ \emph {et~al.}(2018)\citenamefont {Sun},
  \citenamefont {Yi}, \citenamefont {Wang}, \citenamefont {Zhang},
  \citenamefont {Sanders}, \citenamefont {Xu}, \citenamefont {Wang},
  \citenamefont {Schmiedmayer}, \citenamefont {Deng}, \citenamefont {Liu},
  \citenamefont {Chen},\ and\ \citenamefont {Pan}}]{PhysRevLett.121.250403}%
  \BibitemOpen
  \bibfield  {author} {\bibinfo {author} {\bibfnamefont {W.}~\bibnamefont
  {Sun}}, \bibinfo {author} {\bibfnamefont {C.-R.}\ \bibnamefont {Yi}},
  \bibinfo {author} {\bibfnamefont {B.-Z.}\ \bibnamefont {Wang}}, \bibinfo
  {author} {\bibfnamefont {W.-W.}\ \bibnamefont {Zhang}}, \bibinfo {author}
  {\bibfnamefont {B.~C.}\ \bibnamefont {Sanders}}, \bibinfo {author}
  {\bibfnamefont {X.-T.}\ \bibnamefont {Xu}}, \bibinfo {author} {\bibfnamefont
  {Z.-Y.}\ \bibnamefont {Wang}}, \bibinfo {author} {\bibfnamefont
  {J.}~\bibnamefont {Schmiedmayer}}, \bibinfo {author} {\bibfnamefont
  {Y.}~\bibnamefont {Deng}}, \bibinfo {author} {\bibfnamefont {X.-J.}\
  \bibnamefont {Liu}}, \bibinfo {author} {\bibfnamefont {S.}~\bibnamefont
  {Chen}},\ and\ \bibinfo {author} {\bibfnamefont {J.-W.}\ \bibnamefont
  {Pan}},\ }\bibfield  {title} {\bibinfo {title} {Uncover topology by quantum
  quench dynamics},\ }\href {https://doi.org/10.1103/PhysRevLett.121.250403}
  {\bibfield  {journal} {\bibinfo  {journal} {Phys. Rev. Lett.}\ }\textbf
  {\bibinfo {volume} {121}},\ \bibinfo {pages} {250403} (\bibinfo {year}
  {2018})}\BibitemShut {NoStop}%
\bibitem [{\citenamefont {Cai}\ \emph {et~al.}(2019)\citenamefont {Cai},
  \citenamefont {Han}, \citenamefont {Mei}, \citenamefont {Xu}, \citenamefont
  {Ma}, \citenamefont {Li}, \citenamefont {Wang}, \citenamefont {Song},
  \citenamefont {Xue}, \citenamefont {Yin}, \citenamefont {Jia},\ and\
  \citenamefont {Sun}}]{PhysRevLett.123.080501}%
  \BibitemOpen
  \bibfield  {author} {\bibinfo {author} {\bibfnamefont {W.}~\bibnamefont
  {Cai}}, \bibinfo {author} {\bibfnamefont {J.}~\bibnamefont {Han}}, \bibinfo
  {author} {\bibfnamefont {F.}~\bibnamefont {Mei}}, \bibinfo {author}
  {\bibfnamefont {Y.}~\bibnamefont {Xu}}, \bibinfo {author} {\bibfnamefont
  {Y.}~\bibnamefont {Ma}}, \bibinfo {author} {\bibfnamefont {X.}~\bibnamefont
  {Li}}, \bibinfo {author} {\bibfnamefont {H.}~\bibnamefont {Wang}}, \bibinfo
  {author} {\bibfnamefont {Y.~P.}\ \bibnamefont {Song}}, \bibinfo {author}
  {\bibfnamefont {Z.-Y.}\ \bibnamefont {Xue}}, \bibinfo {author} {\bibfnamefont
  {Z.-q.}\ \bibnamefont {Yin}}, \bibinfo {author} {\bibfnamefont
  {S.}~\bibnamefont {Jia}},\ and\ \bibinfo {author} {\bibfnamefont
  {L.}~\bibnamefont {Sun}},\ }\bibfield  {title} {\bibinfo {title} {Observation
  of topological magnon insulator states in a superconducting circuit},\ }\href
  {https://doi.org/10.1103/PhysRevLett.123.080501} {\bibfield  {journal}
  {\bibinfo  {journal} {Phys. Rev. Lett.}\ }\textbf {\bibinfo {volume} {123}},\
  \bibinfo {pages} {080501} (\bibinfo {year} {2019})}\BibitemShut {NoStop}%
\bibitem [{\citenamefont {Zhang}\ \emph {et~al.}(2020)\citenamefont {Zhang},
  \citenamefont {Zhang},\ and\ \citenamefont {Liu}}]{PhysRevLett.125.183001}%
  \BibitemOpen
  \bibfield  {author} {\bibinfo {author} {\bibfnamefont {L.}~\bibnamefont
  {Zhang}}, \bibinfo {author} {\bibfnamefont {L.}~\bibnamefont {Zhang}},\ and\
  \bibinfo {author} {\bibfnamefont {X.-J.}\ \bibnamefont {Liu}},\ }\bibfield
  {title} {\bibinfo {title} {Unified theory to characterize {Floquet}
  topological phases by quench dynamics},\ }\href
  {https://doi.org/10.1103/PhysRevLett.125.183001} {\bibfield  {journal}
  {\bibinfo  {journal} {Phys. Rev. Lett.}\ }\textbf {\bibinfo {volume} {125}},\
  \bibinfo {pages} {183001} (\bibinfo {year} {2020})}\BibitemShut {NoStop}%
\bibitem [{\citenamefont {Qian}\ \emph {et~al.}(2025)\citenamefont {Qian},
  \citenamefont {Gong}, \citenamefont {Zhang}, \citenamefont {Guo},
  \citenamefont {Zha}, \citenamefont {Chen}, \citenamefont {Ye}, \citenamefont
  {Wu}, \citenamefont {Cao}, \citenamefont {Ying}, \citenamefont {Zhu},
  \citenamefont {Huang}, \citenamefont {Zhao}, \citenamefont {Li},
  \citenamefont {Yu}, \citenamefont {Fan}, \citenamefont {Wu}, \citenamefont
  {Su}, \citenamefont {Deng}, \citenamefont {Rong}, \citenamefont {Li},
  \citenamefont {Zhang}, \citenamefont {Chung}, \citenamefont {Liang},
  \citenamefont {Lin}, \citenamefont {Xu}, \citenamefont {Guo}, \citenamefont
  {Li}, \citenamefont {Yan}, \citenamefont {Su}, \citenamefont {Wu},
  \citenamefont {Huo}, \citenamefont {Peng}, \citenamefont {Lu}, \citenamefont
  {Mei}, \citenamefont {Jia}, \citenamefont {Zhu},\ and\ \citenamefont
  {Pan}}]{science.adp6802}%
  \BibitemOpen
  \bibfield  {author} {\bibinfo {author} {\bibfnamefont {H.}~\bibnamefont
  {Qian}}, \bibinfo {author} {\bibfnamefont {M.}~\bibnamefont {Gong}}, \bibinfo
  {author} {\bibfnamefont {J.}~\bibnamefont {Zhang}}, \bibinfo {author}
  {\bibfnamefont {S.}~\bibnamefont {Guo}}, \bibinfo {author} {\bibfnamefont
  {C.}~\bibnamefont {Zha}}, \bibinfo {author} {\bibfnamefont {F.}~\bibnamefont
  {Chen}}, \bibinfo {author} {\bibfnamefont {Y.}~\bibnamefont {Ye}}, \bibinfo
  {author} {\bibfnamefont {Y.}~\bibnamefont {Wu}}, \bibinfo {author}
  {\bibfnamefont {S.}~\bibnamefont {Cao}}, \bibinfo {author} {\bibfnamefont
  {C.}~\bibnamefont {Ying}}, \bibinfo {author} {\bibfnamefont {Q.}~\bibnamefont
  {Zhu}}, \bibinfo {author} {\bibfnamefont {H.-L.}\ \bibnamefont {Huang}},
  \bibinfo {author} {\bibfnamefont {Y.}~\bibnamefont {Zhao}}, \bibinfo {author}
  {\bibfnamefont {S.}~\bibnamefont {Li}}, \bibinfo {author} {\bibfnamefont
  {J.}~\bibnamefont {Yu}}, \bibinfo {author} {\bibfnamefont {D.}~\bibnamefont
  {Fan}}, \bibinfo {author} {\bibfnamefont {D.}~\bibnamefont {Wu}}, \bibinfo
  {author} {\bibfnamefont {H.}~\bibnamefont {Su}}, \bibinfo {author}
  {\bibfnamefont {H.}~\bibnamefont {Deng}}, \bibinfo {author} {\bibfnamefont
  {H.}~\bibnamefont {Rong}}, \bibinfo {author} {\bibfnamefont {Y.}~\bibnamefont
  {Li}}, \bibinfo {author} {\bibfnamefont {K.}~\bibnamefont {Zhang}}, \bibinfo
  {author} {\bibfnamefont {T.-H.}\ \bibnamefont {Chung}}, \bibinfo {author}
  {\bibfnamefont {F.}~\bibnamefont {Liang}}, \bibinfo {author} {\bibfnamefont
  {J.}~\bibnamefont {Lin}}, \bibinfo {author} {\bibfnamefont {Y.}~\bibnamefont
  {Xu}}, \bibinfo {author} {\bibfnamefont {C.}~\bibnamefont {Guo}}, \bibinfo
  {author} {\bibfnamefont {N.}~\bibnamefont {Li}}, \bibinfo {author}
  {\bibfnamefont {K.}~\bibnamefont {Yan}}, \bibinfo {author} {\bibfnamefont
  {F.-F.}\ \bibnamefont {Su}}, \bibinfo {author} {\bibfnamefont
  {G.}~\bibnamefont {Wu}}, \bibinfo {author} {\bibfnamefont {Y.-H.}\
  \bibnamefont {Huo}}, \bibinfo {author} {\bibfnamefont {C.-Z.}\ \bibnamefont
  {Peng}}, \bibinfo {author} {\bibfnamefont {C.-Y.}\ \bibnamefont {Lu}},
  \bibinfo {author} {\bibfnamefont {F.}~\bibnamefont {Mei}}, \bibinfo {author}
  {\bibfnamefont {S.}~\bibnamefont {Jia}}, \bibinfo {author} {\bibfnamefont
  {X.}~\bibnamefont {Zhu}},\ and\ \bibinfo {author} {\bibfnamefont {J.-W.}\
  \bibnamefont {Pan}},\ }\bibfield  {title} {\bibinfo {title} {Programmable
  higher-order nonequilibrium topological phases on a superconducting quantum
  processor},\ }\href {https://doi.org/10.1126/science.adp6802} {\bibfield
  {journal} {\bibinfo  {journal} {Science}\ }\textbf {\bibinfo {volume}
  {390}},\ \bibinfo {pages} {930} (\bibinfo {year} {2025})}\BibitemShut
  {NoStop}%
\bibitem [{\citenamefont {{Google Quantum AI and
  Collaborators}}(2025)}]{GoogleQuantumAI2025BelowThreshold}%
  \BibitemOpen
  \bibfield  {author} {\bibinfo {author} {\bibnamefont {{Google Quantum AI and
  Collaborators}}},\ }\bibfield  {title} {\bibinfo {title} {Quantum error
  correction below the surface code threshold},\ }\href
  {https://doi.org/10.1038/s41586-024-08449-y} {\bibfield  {journal} {\bibinfo
  {journal} {Nature}\ }\textbf {\bibinfo {volume} {638}},\ \bibinfo {pages}
  {920} (\bibinfo {year} {2025})},\ \bibinfo {note} {published online 9
  December 2024}\BibitemShut {NoStop}%
\bibitem [{\citenamefont {He}\ \emph {et~al.}(2025)\citenamefont {He},
  \citenamefont {Lin}, \citenamefont {Wang}, \citenamefont {Li}, \citenamefont
  {Bei}, \citenamefont {Cai}, \citenamefont {Cao}, \citenamefont {Chen},
  \citenamefont {Chen}, \citenamefont {Chen}, \citenamefont {Chen},
  \citenamefont {Chen}, \citenamefont {Chen}, \citenamefont {Chu},
  \citenamefont {Deng}, \citenamefont {Ding}, \citenamefont {Ding},
  \citenamefont {Fan}, \citenamefont {Fan}, \citenamefont {Fu}, \citenamefont
  {Gao}, \citenamefont {Gong}, \citenamefont {Gui}, \citenamefont {Guo},
  \citenamefont {Guo}, \citenamefont {Han}, \citenamefont {Hong}, \citenamefont
  {Hu}, \citenamefont {Huang}, \citenamefont {Huo}, \citenamefont {Jiang},
  \citenamefont {Jiang}, \citenamefont {Jiang}, \citenamefont {Jiang},
  \citenamefont {Jin}, \citenamefont {Li}, \citenamefont {Li}, \citenamefont
  {Li}, \citenamefont {Li}, \citenamefont {Li}, \citenamefont {Li},
  \citenamefont {Li}, \citenamefont {Li}, \citenamefont {Li}, \citenamefont
  {Liang}, \citenamefont {Liao}, \citenamefont {Lin}, \citenamefont {Liu},
  \citenamefont {Liu}, \citenamefont {Liu}, \citenamefont {Lou}, \citenamefont
  {Ma}, \citenamefont {Nan}, \citenamefont {Nie}, \citenamefont {Niu},
  \citenamefont {Peng}, \citenamefont {Qian}, \citenamefont {Rong},
  \citenamefont {Rong}, \citenamefont {Shen}, \citenamefont {Shen},
  \citenamefont {Su}, \citenamefont {Su}, \citenamefont {Sun}, \citenamefont
  {Sun}, \citenamefont {Sun}, \citenamefont {Sun}, \citenamefont {Tan},
  \citenamefont {Tan}, \citenamefont {Tu}, \citenamefont {Wang}, \citenamefont
  {Wang}, \citenamefont {Wang}, \citenamefont {Wang}, \citenamefont {Wang},
  \citenamefont {Wang}, \citenamefont {Wang}, \citenamefont {Wang},
  \citenamefont {Wei}, \citenamefont {Wu}, \citenamefont {Wu}, \citenamefont
  {Wu}, \citenamefont {Xu}, \citenamefont {Xue}, \citenamefont {Yan},
  \citenamefont {Yan}, \citenamefont {Yang}, \citenamefont {Yang},
  \citenamefont {Yang}, \citenamefont {Ye}, \citenamefont {Ye}, \citenamefont
  {Yi}, \citenamefont {Ying}, \citenamefont {Yu}, \citenamefont {Yu},
  \citenamefont {Zeng}, \citenamefont {Zha}, \citenamefont {Zhan},
  \citenamefont {Zhang}, \citenamefont {Zhang}, \citenamefont {Zhang},
  \citenamefont {Zhang}, \citenamefont {Zhang}, \citenamefont {Zhang},
  \citenamefont {Zhang}, \citenamefont {Zhao}, \citenamefont {Zhao},
  \citenamefont {Zhao}, \citenamefont {Zhao}, \citenamefont {Zheng},
  \citenamefont {Zhou}, \citenamefont {Zhou}, \citenamefont {Zhou},
  \citenamefont {Zhou}, \citenamefont {Zhu}, \citenamefont {Zhu}, \citenamefont
  {Zou}, \citenamefont {Zou}, \citenamefont {Zhang}, \citenamefont {Lu},
  \citenamefont {Peng}, \citenamefont {Chen}, \citenamefont {Zhu},\ and\
  \citenamefont {Pan}}]{rqkg-He2025AllMicrowaveLeakageSuppression}%
  \BibitemOpen
  \bibfield  {author} {\bibinfo {author} {\bibfnamefont {T.}~\bibnamefont
  {He}}, \bibinfo {author} {\bibfnamefont {W.}~\bibnamefont {Lin}}, \bibinfo
  {author} {\bibfnamefont {R.}~\bibnamefont {Wang}}, \bibinfo {author}
  {\bibfnamefont {Y.}~\bibnamefont {Li}}, \bibinfo {author} {\bibfnamefont
  {J.}~\bibnamefont {Bei}}, \bibinfo {author} {\bibfnamefont {J.}~\bibnamefont
  {Cai}}, \bibinfo {author} {\bibfnamefont {S.}~\bibnamefont {Cao}}, \bibinfo
  {author} {\bibfnamefont {D.}~\bibnamefont {Chen}}, \bibinfo {author}
  {\bibfnamefont {K.}~\bibnamefont {Chen}}, \bibinfo {author} {\bibfnamefont
  {X.}~\bibnamefont {Chen}}, \bibinfo {author} {\bibfnamefont {Z.}~\bibnamefont
  {Chen}}, \bibinfo {author} {\bibfnamefont {Z.}~\bibnamefont {Chen}}, \bibinfo
  {author} {\bibfnamefont {Z.}~\bibnamefont {Chen}}, \bibinfo {author}
  {\bibfnamefont {W.}~\bibnamefont {Chu}}, \bibinfo {author} {\bibfnamefont
  {H.}~\bibnamefont {Deng}}, \bibinfo {author} {\bibfnamefont {X.}~\bibnamefont
  {Ding}}, \bibinfo {author} {\bibfnamefont {Z.}~\bibnamefont {Ding}}, \bibinfo
  {author} {\bibfnamefont {B.}~\bibnamefont {Fan}}, \bibinfo {author}
  {\bibfnamefont {D.}~\bibnamefont {Fan}}, \bibinfo {author} {\bibfnamefont
  {Y.}~\bibnamefont {Fu}}, \bibinfo {author} {\bibfnamefont {D.}~\bibnamefont
  {Gao}}, \bibinfo {author} {\bibfnamefont {M.}~\bibnamefont {Gong}}, \bibinfo
  {author} {\bibfnamefont {J.}~\bibnamefont {Gui}}, \bibinfo {author}
  {\bibfnamefont {C.}~\bibnamefont {Guo}}, \bibinfo {author} {\bibfnamefont
  {S.}~\bibnamefont {Guo}}, \bibinfo {author} {\bibfnamefont {L.}~\bibnamefont
  {Han}}, \bibinfo {author} {\bibfnamefont {L.}~\bibnamefont {Hong}}, \bibinfo
  {author} {\bibfnamefont {Y.}~\bibnamefont {Hu}}, \bibinfo {author}
  {\bibfnamefont {H.-L.}\ \bibnamefont {Huang}}, \bibinfo {author}
  {\bibfnamefont {Y.-H.}\ \bibnamefont {Huo}}, \bibinfo {author} {\bibfnamefont
  {C.}~\bibnamefont {Jiang}}, \bibinfo {author} {\bibfnamefont
  {L.}~\bibnamefont {Jiang}}, \bibinfo {author} {\bibfnamefont
  {T.}~\bibnamefont {Jiang}}, \bibinfo {author} {\bibfnamefont
  {Z.}~\bibnamefont {Jiang}}, \bibinfo {author} {\bibfnamefont
  {H.}~\bibnamefont {Jin}}, \bibinfo {author} {\bibfnamefont {D.}~\bibnamefont
  {Li}}, \bibinfo {author} {\bibfnamefont {D.}~\bibnamefont {Li}}, \bibinfo
  {author} {\bibfnamefont {J.}~\bibnamefont {Li}}, \bibinfo {author}
  {\bibfnamefont {J.}~\bibnamefont {Li}}, \bibinfo {author} {\bibfnamefont
  {J.}~\bibnamefont {Li}}, \bibinfo {author} {\bibfnamefont {J.}~\bibnamefont
  {Li}}, \bibinfo {author} {\bibfnamefont {N.}~\bibnamefont {Li}}, \bibinfo
  {author} {\bibfnamefont {S.}~\bibnamefont {Li}}, \bibinfo {author}
  {\bibfnamefont {Y.}~\bibnamefont {Li}}, \bibinfo {author} {\bibfnamefont
  {F.}~\bibnamefont {Liang}}, \bibinfo {author} {\bibfnamefont
  {N.}~\bibnamefont {Liao}}, \bibinfo {author} {\bibfnamefont {J.}~\bibnamefont
  {Lin}}, \bibinfo {author} {\bibfnamefont {K.}~\bibnamefont {Liu}}, \bibinfo
  {author} {\bibfnamefont {M.}~\bibnamefont {Liu}}, \bibinfo {author}
  {\bibfnamefont {Y.}~\bibnamefont {Liu}}, \bibinfo {author} {\bibfnamefont
  {H.}~\bibnamefont {Lou}}, \bibinfo {author} {\bibfnamefont {Y.}~\bibnamefont
  {Ma}}, \bibinfo {author} {\bibfnamefont {K.}~\bibnamefont {Nan}}, \bibinfo
  {author} {\bibfnamefont {M.}~\bibnamefont {Nie}}, \bibinfo {author}
  {\bibfnamefont {L.}~\bibnamefont {Niu}}, \bibinfo {author} {\bibfnamefont
  {W.}~\bibnamefont {Peng}}, \bibinfo {author} {\bibfnamefont {H.}~\bibnamefont
  {Qian}}, \bibinfo {author} {\bibfnamefont {H.}~\bibnamefont {Rong}}, \bibinfo
  {author} {\bibfnamefont {T.}~\bibnamefont {Rong}}, \bibinfo {author}
  {\bibfnamefont {H.}~\bibnamefont {Shen}}, \bibinfo {author} {\bibfnamefont
  {Q.}~\bibnamefont {Shen}}, \bibinfo {author} {\bibfnamefont {H.}~\bibnamefont
  {Su}}, \bibinfo {author} {\bibfnamefont {F.}~\bibnamefont {Su}}, \bibinfo
  {author} {\bibfnamefont {C.}~\bibnamefont {Sun}}, \bibinfo {author}
  {\bibfnamefont {L.}~\bibnamefont {Sun}}, \bibinfo {author} {\bibfnamefont
  {T.}~\bibnamefont {Sun}}, \bibinfo {author} {\bibfnamefont {Y.}~\bibnamefont
  {Sun}}, \bibinfo {author} {\bibfnamefont {Y.}~\bibnamefont {Tan}}, \bibinfo
  {author} {\bibfnamefont {J.}~\bibnamefont {Tan}}, \bibinfo {author}
  {\bibfnamefont {W.}~\bibnamefont {Tu}}, \bibinfo {author} {\bibfnamefont
  {J.}~\bibnamefont {Wang}}, \bibinfo {author} {\bibfnamefont {B.}~\bibnamefont
  {Wang}}, \bibinfo {author} {\bibfnamefont {C.}~\bibnamefont {Wang}}, \bibinfo
  {author} {\bibfnamefont {C.}~\bibnamefont {Wang}}, \bibinfo {author}
  {\bibfnamefont {C.}~\bibnamefont {Wang}}, \bibinfo {author} {\bibfnamefont
  {J.}~\bibnamefont {Wang}}, \bibinfo {author} {\bibfnamefont {S.}~\bibnamefont
  {Wang}}, \bibinfo {author} {\bibfnamefont {X.}~\bibnamefont {Wang}}, \bibinfo
  {author} {\bibfnamefont {Z.}~\bibnamefont {Wei}}, \bibinfo {author}
  {\bibfnamefont {D.}~\bibnamefont {Wu}}, \bibinfo {author} {\bibfnamefont
  {G.}~\bibnamefont {Wu}}, \bibinfo {author} {\bibfnamefont {Y.}~\bibnamefont
  {Wu}}, \bibinfo {author} {\bibfnamefont {Y.}~\bibnamefont {Xu}}, \bibinfo
  {author} {\bibfnamefont {C.}~\bibnamefont {Xue}}, \bibinfo {author}
  {\bibfnamefont {K.}~\bibnamefont {Yan}}, \bibinfo {author} {\bibfnamefont
  {X.}~\bibnamefont {Yan}}, \bibinfo {author} {\bibfnamefont {W.}~\bibnamefont
  {Yang}}, \bibinfo {author} {\bibfnamefont {X.}~\bibnamefont {Yang}}, \bibinfo
  {author} {\bibfnamefont {Y.}~\bibnamefont {Yang}}, \bibinfo {author}
  {\bibfnamefont {Y.}~\bibnamefont {Ye}}, \bibinfo {author} {\bibfnamefont
  {Z.}~\bibnamefont {Ye}}, \bibinfo {author} {\bibfnamefont {Z.}~\bibnamefont
  {Yi}}, \bibinfo {author} {\bibfnamefont {C.}~\bibnamefont {Ying}}, \bibinfo
  {author} {\bibfnamefont {J.}~\bibnamefont {Yu}}, \bibinfo {author}
  {\bibfnamefont {Q.}~\bibnamefont {Yu}}, \bibinfo {author} {\bibfnamefont
  {X.}~\bibnamefont {Zeng}}, \bibinfo {author} {\bibfnamefont {C.}~\bibnamefont
  {Zha}}, \bibinfo {author} {\bibfnamefont {S.}~\bibnamefont {Zhan}}, \bibinfo
  {author} {\bibfnamefont {H.}~\bibnamefont {Zhang}}, \bibinfo {author}
  {\bibfnamefont {H.}~\bibnamefont {Zhang}}, \bibinfo {author} {\bibfnamefont
  {K.}~\bibnamefont {Zhang}}, \bibinfo {author} {\bibfnamefont
  {W.}~\bibnamefont {Zhang}}, \bibinfo {author} {\bibfnamefont
  {Y.}~\bibnamefont {Zhang}}, \bibinfo {author} {\bibfnamefont
  {Y.}~\bibnamefont {Zhang}}, \bibinfo {author} {\bibfnamefont
  {Z.}~\bibnamefont {Zhang}}, \bibinfo {author} {\bibfnamefont
  {G.}~\bibnamefont {Zhao}}, \bibinfo {author} {\bibfnamefont {X.}~\bibnamefont
  {Zhao}}, \bibinfo {author} {\bibfnamefont {Y.}~\bibnamefont {Zhao}}, \bibinfo
  {author} {\bibfnamefont {Z.}~\bibnamefont {Zhao}}, \bibinfo {author}
  {\bibfnamefont {L.}~\bibnamefont {Zheng}}, \bibinfo {author} {\bibfnamefont
  {F.}~\bibnamefont {Zhou}}, \bibinfo {author} {\bibfnamefont {L.}~\bibnamefont
  {Zhou}}, \bibinfo {author} {\bibfnamefont {N.}~\bibnamefont {Zhou}}, \bibinfo
  {author} {\bibfnamefont {N.}~\bibnamefont {Zhou}}, \bibinfo {author}
  {\bibfnamefont {C.}~\bibnamefont {Zhu}}, \bibinfo {author} {\bibfnamefont
  {Q.}~\bibnamefont {Zhu}}, \bibinfo {author} {\bibfnamefont {G.}~\bibnamefont
  {Zou}}, \bibinfo {author} {\bibfnamefont {H.}~\bibnamefont {Zou}}, \bibinfo
  {author} {\bibfnamefont {Q.}~\bibnamefont {Zhang}}, \bibinfo {author}
  {\bibfnamefont {C.-Y.}\ \bibnamefont {Lu}}, \bibinfo {author} {\bibfnamefont
  {C.-Z.}\ \bibnamefont {Peng}}, \bibinfo {author} {\bibfnamefont
  {F.}~\bibnamefont {Chen}}, \bibinfo {author} {\bibfnamefont {X.}~\bibnamefont
  {Zhu}},\ and\ \bibinfo {author} {\bibfnamefont {J.-W.}\ \bibnamefont {Pan}},\
  }\bibfield  {title} {\bibinfo {title} {Experimental quantum error correction
  below the surface code threshold via all-microwave leakage suppression},\
  }\href {https://doi.org/10.1103/rqkg-dw31} {\bibfield  {journal} {\bibinfo
  {journal} {Phys. Rev. Lett.}\ }\textbf {\bibinfo {volume} {135}},\ \bibinfo
  {pages} {260601} (\bibinfo {year} {2025})}\BibitemShut {NoStop}%
\bibitem [{\citenamefont {Lin}\ \emph {et~al.}(2026)\citenamefont {Lin},
  \citenamefont {Guo}, \citenamefont {Ma}, \citenamefont {Yi}, \citenamefont
  {Zhang}, \citenamefont {Bei}, \citenamefont {Cai}, \citenamefont {Cao},
  \citenamefont {Chen}, \citenamefont {Chen}, \citenamefont {Chen},
  \citenamefont {Chen}, \citenamefont {Chen}, \citenamefont {Chen},
  \citenamefont {Chen}, \citenamefont {Chen}, \citenamefont {Chu},
  \citenamefont {Deng}, \citenamefont {Ding}, \citenamefont {Ding},
  \citenamefont {Du}, \citenamefont {Fan}, \citenamefont {Fan}, \citenamefont
  {Fu}, \citenamefont {Gao}, \citenamefont {Gong}, \citenamefont {Gui},
  \citenamefont {Guo}, \citenamefont {Han}, \citenamefont {He}, \citenamefont
  {Hong}, \citenamefont {Hu}, \citenamefont {Huang}, \citenamefont {Huo},
  \citenamefont {Jiang}, \citenamefont {Jiang}, \citenamefont {Jiang},
  \citenamefont {Jiang}, \citenamefont {Jiang}, \citenamefont {Li},
  \citenamefont {Li}, \citenamefont {Li}, \citenamefont {Li}, \citenamefont
  {Li}, \citenamefont {Li}, \citenamefont {Li}, \citenamefont {Li},
  \citenamefont {Li}, \citenamefont {Li}, \citenamefont {Liang}, \citenamefont
  {Liao}, \citenamefont {Lin}, \citenamefont {Liu}, \citenamefont {Liu},
  \citenamefont {Lou}, \citenamefont {Nan}, \citenamefont {Nie}, \citenamefont
  {Niu}, \citenamefont {Peng}, \citenamefont {Qian}, \citenamefont {Rong},
  \citenamefont {Rong}, \citenamefont {Shao}, \citenamefont {Shen},
  \citenamefont {Shen}, \citenamefont {Song}, \citenamefont {Su}, \citenamefont
  {Su}, \citenamefont {Sun}, \citenamefont {Sun}, \citenamefont {Sun},
  \citenamefont {Sun}, \citenamefont {Sun}, \citenamefont {Tan}, \citenamefont
  {Tan}, \citenamefont {Tang}, \citenamefont {Tang}, \citenamefont {Tu},
  \citenamefont {Wang}, \citenamefont {Wang}, \citenamefont {Wang},
  \citenamefont {Wang}, \citenamefont {Wang}, \citenamefont {Wang},
  \citenamefont {Wang}, \citenamefont {Wang}, \citenamefont {Wang},
  \citenamefont {Wang}, \citenamefont {Wei}, \citenamefont {Wu}, \citenamefont
  {Wu}, \citenamefont {Xia}, \citenamefont {Xia}, \citenamefont {Xie},
  \citenamefont {Xie}, \citenamefont {Xiong}, \citenamefont {Xu}, \citenamefont
  {Xu}, \citenamefont {Xu}, \citenamefont {Xue}, \citenamefont {Yan},
  \citenamefont {Yan}, \citenamefont {Yang}, \citenamefont {Yang},
  \citenamefont {Yang}, \citenamefont {Yang}, \citenamefont {Ye}, \citenamefont
  {Ye}, \citenamefont {Yin}, \citenamefont {Ying}, \citenamefont {Yu},
  \citenamefont {Yu}, \citenamefont {Zha}, \citenamefont {Zhan}, \citenamefont
  {Zhang}, \citenamefont {Zhang}, \citenamefont {Zhang}, \citenamefont {Zhang},
  \citenamefont {Zhang}, \citenamefont {Zhang}, \citenamefont {Zhang},
  \citenamefont {Zhang}, \citenamefont {Zhang}, \citenamefont {Zhang},
  \citenamefont {Zhang}, \citenamefont {Zhang}, \citenamefont {Zhang},
  \citenamefont {Zhao}, \citenamefont {Zhao}, \citenamefont {Zhao},
  \citenamefont {Zhao}, \citenamefont {Zheng}, \citenamefont {Zhou},
  \citenamefont {Zhou}, \citenamefont {Zhou}, \citenamefont {Zhu},
  \citenamefont {Zhu}, \citenamefont {Zhu}, \citenamefont {Zou}, \citenamefont
  {Zou}, \citenamefont {Zhang}, \citenamefont {Lu}, \citenamefont {Chen},
  \citenamefont {Peng}, \citenamefont {Chen}, \citenamefont {Zhu},\ and\
  \citenamefont {Pan}}]{Lin2026SurfaceCodeLogicalOperations}%
  \BibitemOpen
  \bibfield  {author} {\bibinfo {author} {\bibfnamefont {W.}~\bibnamefont
  {Lin}}, \bibinfo {author} {\bibfnamefont {S.}~\bibnamefont {Guo}}, \bibinfo
  {author} {\bibfnamefont {Y.}~\bibnamefont {Ma}}, \bibinfo {author}
  {\bibfnamefont {Z.}~\bibnamefont {Yi}}, \bibinfo {author} {\bibfnamefont
  {K.}~\bibnamefont {Zhang}}, \bibinfo {author} {\bibfnamefont
  {J.}~\bibnamefont {Bei}}, \bibinfo {author} {\bibfnamefont {J.}~\bibnamefont
  {Cai}}, \bibinfo {author} {\bibfnamefont {S.}~\bibnamefont {Cao}}, \bibinfo
  {author} {\bibfnamefont {D.}~\bibnamefont {Chen}}, \bibinfo {author}
  {\bibfnamefont {G.}~\bibnamefont {Chen}}, \bibinfo {author} {\bibfnamefont
  {J.}~\bibnamefont {Chen}}, \bibinfo {author} {\bibfnamefont {K.}~\bibnamefont
  {Chen}}, \bibinfo {author} {\bibfnamefont {X.}~\bibnamefont {Chen}}, \bibinfo
  {author} {\bibfnamefont {Z.}~\bibnamefont {Chen}}, \bibinfo {author}
  {\bibfnamefont {Z.}~\bibnamefont {Chen}}, \bibinfo {author} {\bibfnamefont
  {Z.}~\bibnamefont {Chen}}, \bibinfo {author} {\bibfnamefont {W.}~\bibnamefont
  {Chu}}, \bibinfo {author} {\bibfnamefont {H.}~\bibnamefont {Deng}}, \bibinfo
  {author} {\bibfnamefont {X.}~\bibnamefont {Ding}}, \bibinfo {author}
  {\bibfnamefont {Z.}~\bibnamefont {Ding}}, \bibinfo {author} {\bibfnamefont
  {Y.}~\bibnamefont {Du}}, \bibinfo {author} {\bibfnamefont {B.}~\bibnamefont
  {Fan}}, \bibinfo {author} {\bibfnamefont {D.}~\bibnamefont {Fan}}, \bibinfo
  {author} {\bibfnamefont {Y.}~\bibnamefont {Fu}}, \bibinfo {author}
  {\bibfnamefont {D.}~\bibnamefont {Gao}}, \bibinfo {author} {\bibfnamefont
  {M.}~\bibnamefont {Gong}}, \bibinfo {author} {\bibfnamefont {J.}~\bibnamefont
  {Gui}}, \bibinfo {author} {\bibfnamefont {C.}~\bibnamefont {Guo}}, \bibinfo
  {author} {\bibfnamefont {L.}~\bibnamefont {Han}}, \bibinfo {author}
  {\bibfnamefont {T.}~\bibnamefont {He}}, \bibinfo {author} {\bibfnamefont
  {L.}~\bibnamefont {Hong}}, \bibinfo {author} {\bibfnamefont {Y.}~\bibnamefont
  {Hu}}, \bibinfo {author} {\bibfnamefont {H.-L.}\ \bibnamefont {Huang}},
  \bibinfo {author} {\bibfnamefont {Y.-H.}\ \bibnamefont {Huo}}, \bibinfo
  {author} {\bibfnamefont {C.}~\bibnamefont {Jiang}}, \bibinfo {author}
  {\bibfnamefont {L.}~\bibnamefont {Jiang}}, \bibinfo {author} {\bibfnamefont
  {T.}~\bibnamefont {Jiang}}, \bibinfo {author} {\bibfnamefont
  {W.}~\bibnamefont {Jiang}}, \bibinfo {author} {\bibfnamefont
  {Z.}~\bibnamefont {Jiang}}, \bibinfo {author} {\bibfnamefont
  {D.}~\bibnamefont {Li}}, \bibinfo {author} {\bibfnamefont {D.}~\bibnamefont
  {Li}}, \bibinfo {author} {\bibfnamefont {J.}~\bibnamefont {Li}}, \bibinfo
  {author} {\bibfnamefont {J.}~\bibnamefont {Li}}, \bibinfo {author}
  {\bibfnamefont {J.}~\bibnamefont {Li}}, \bibinfo {author} {\bibfnamefont
  {S.}~\bibnamefont {Li}}, \bibinfo {author} {\bibfnamefont {W.}~\bibnamefont
  {Li}}, \bibinfo {author} {\bibfnamefont {X.}~\bibnamefont {Li}}, \bibinfo
  {author} {\bibfnamefont {Y.}~\bibnamefont {Li}}, \bibinfo {author}
  {\bibfnamefont {Y.}~\bibnamefont {Li}}, \bibinfo {author} {\bibfnamefont
  {F.}~\bibnamefont {Liang}}, \bibinfo {author} {\bibfnamefont
  {N.}~\bibnamefont {Liao}}, \bibinfo {author} {\bibfnamefont {J.}~\bibnamefont
  {Lin}}, \bibinfo {author} {\bibfnamefont {M.}~\bibnamefont {Liu}}, \bibinfo
  {author} {\bibfnamefont {Y.}~\bibnamefont {Liu}}, \bibinfo {author}
  {\bibfnamefont {H.}~\bibnamefont {Lou}}, \bibinfo {author} {\bibfnamefont
  {K.}~\bibnamefont {Nan}}, \bibinfo {author} {\bibfnamefont {M.}~\bibnamefont
  {Nie}}, \bibinfo {author} {\bibfnamefont {L.}~\bibnamefont {Niu}}, \bibinfo
  {author} {\bibfnamefont {W.}~\bibnamefont {Peng}}, \bibinfo {author}
  {\bibfnamefont {H.}~\bibnamefont {Qian}}, \bibinfo {author} {\bibfnamefont
  {H.}~\bibnamefont {Rong}}, \bibinfo {author} {\bibfnamefont {T.}~\bibnamefont
  {Rong}}, \bibinfo {author} {\bibfnamefont {Y.}~\bibnamefont {Shao}}, \bibinfo
  {author} {\bibfnamefont {H.}~\bibnamefont {Shen}}, \bibinfo {author}
  {\bibfnamefont {Q.}~\bibnamefont {Shen}}, \bibinfo {author} {\bibfnamefont
  {G.}~\bibnamefont {Song}}, \bibinfo {author} {\bibfnamefont {H.}~\bibnamefont
  {Su}}, \bibinfo {author} {\bibfnamefont {F.}~\bibnamefont {Su}}, \bibinfo
  {author} {\bibfnamefont {C.}~\bibnamefont {Sun}}, \bibinfo {author}
  {\bibfnamefont {D.}~\bibnamefont {Sun}}, \bibinfo {author} {\bibfnamefont
  {L.}~\bibnamefont {Sun}}, \bibinfo {author} {\bibfnamefont {T.}~\bibnamefont
  {Sun}}, \bibinfo {author} {\bibfnamefont {Y.}~\bibnamefont {Sun}}, \bibinfo
  {author} {\bibfnamefont {Y.}~\bibnamefont {Tan}}, \bibinfo {author}
  {\bibfnamefont {J.}~\bibnamefont {Tan}}, \bibinfo {author} {\bibfnamefont
  {S.}~\bibnamefont {Tang}}, \bibinfo {author} {\bibfnamefont {Y.}~\bibnamefont
  {Tang}}, \bibinfo {author} {\bibfnamefont {W.}~\bibnamefont {Tu}}, \bibinfo
  {author} {\bibfnamefont {J.}~\bibnamefont {Wang}}, \bibinfo {author}
  {\bibfnamefont {B.}~\bibnamefont {Wang}}, \bibinfo {author} {\bibfnamefont
  {C.}~\bibnamefont {Wang}}, \bibinfo {author} {\bibfnamefont {C.}~\bibnamefont
  {Wang}}, \bibinfo {author} {\bibfnamefont {C.}~\bibnamefont {Wang}}, \bibinfo
  {author} {\bibfnamefont {J.}~\bibnamefont {Wang}}, \bibinfo {author}
  {\bibfnamefont {R.}~\bibnamefont {Wang}}, \bibinfo {author} {\bibfnamefont
  {S.}~\bibnamefont {Wang}}, \bibinfo {author} {\bibfnamefont {X.}~\bibnamefont
  {Wang}}, \bibinfo {author} {\bibfnamefont {Z.}~\bibnamefont {Wang}}, \bibinfo
  {author} {\bibfnamefont {Z.}~\bibnamefont {Wei}}, \bibinfo {author}
  {\bibfnamefont {G.}~\bibnamefont {Wu}}, \bibinfo {author} {\bibfnamefont
  {Y.}~\bibnamefont {Wu}}, \bibinfo {author} {\bibfnamefont {H.}~\bibnamefont
  {Xia}}, \bibinfo {author} {\bibfnamefont {S.}~\bibnamefont {Xia}}, \bibinfo
  {author} {\bibfnamefont {S.}~\bibnamefont {Xie}}, \bibinfo {author}
  {\bibfnamefont {Z.}~\bibnamefont {Xie}}, \bibinfo {author} {\bibfnamefont
  {L.}~\bibnamefont {Xiong}}, \bibinfo {author} {\bibfnamefont
  {J.}~\bibnamefont {Xu}}, \bibinfo {author} {\bibfnamefont {Y.}~\bibnamefont
  {Xu}}, \bibinfo {author} {\bibfnamefont {Y.}~\bibnamefont {Xu}}, \bibinfo
  {author} {\bibfnamefont {C.}~\bibnamefont {Xue}}, \bibinfo {author}
  {\bibfnamefont {K.}~\bibnamefont {Yan}}, \bibinfo {author} {\bibfnamefont
  {X.}~\bibnamefont {Yan}}, \bibinfo {author} {\bibfnamefont {W.}~\bibnamefont
  {Yang}}, \bibinfo {author} {\bibfnamefont {X.}~\bibnamefont {Yang}}, \bibinfo
  {author} {\bibfnamefont {Y.}~\bibnamefont {Yang}}, \bibinfo {author}
  {\bibfnamefont {Y.}~\bibnamefont {Yang}}, \bibinfo {author} {\bibfnamefont
  {Y.}~\bibnamefont {Ye}}, \bibinfo {author} {\bibfnamefont {Z.}~\bibnamefont
  {Ye}}, \bibinfo {author} {\bibfnamefont {J.}~\bibnamefont {Yin}}, \bibinfo
  {author} {\bibfnamefont {C.}~\bibnamefont {Ying}}, \bibinfo {author}
  {\bibfnamefont {J.}~\bibnamefont {Yu}}, \bibinfo {author} {\bibfnamefont
  {Q.}~\bibnamefont {Yu}}, \bibinfo {author} {\bibfnamefont {C.}~\bibnamefont
  {Zha}}, \bibinfo {author} {\bibfnamefont {S.}~\bibnamefont {Zhan}}, \bibinfo
  {author} {\bibfnamefont {C.}~\bibnamefont {Zhang}}, \bibinfo {author}
  {\bibfnamefont {F.}~\bibnamefont {Zhang}}, \bibinfo {author} {\bibfnamefont
  {H.}~\bibnamefont {Zhang}}, \bibinfo {author} {\bibfnamefont
  {H.}~\bibnamefont {Zhang}}, \bibinfo {author} {\bibfnamefont
  {H.}~\bibnamefont {Zhang}}, \bibinfo {author} {\bibfnamefont
  {K.}~\bibnamefont {Zhang}}, \bibinfo {author} {\bibfnamefont
  {Q.}~\bibnamefont {Zhang}}, \bibinfo {author} {\bibfnamefont
  {S.}~\bibnamefont {Zhang}}, \bibinfo {author} {\bibfnamefont
  {W.}~\bibnamefont {Zhang}}, \bibinfo {author} {\bibfnamefont
  {X.}~\bibnamefont {Zhang}}, \bibinfo {author} {\bibfnamefont
  {Y.}~\bibnamefont {Zhang}}, \bibinfo {author} {\bibfnamefont
  {Y.}~\bibnamefont {Zhang}}, \bibinfo {author} {\bibfnamefont
  {Z.}~\bibnamefont {Zhang}}, \bibinfo {author} {\bibfnamefont
  {G.}~\bibnamefont {Zhao}}, \bibinfo {author} {\bibfnamefont {X.}~\bibnamefont
  {Zhao}}, \bibinfo {author} {\bibfnamefont {Y.}~\bibnamefont {Zhao}}, \bibinfo
  {author} {\bibfnamefont {Z.}~\bibnamefont {Zhao}}, \bibinfo {author}
  {\bibfnamefont {L.}~\bibnamefont {Zheng}}, \bibinfo {author} {\bibfnamefont
  {F.}~\bibnamefont {Zhou}}, \bibinfo {author} {\bibfnamefont {L.}~\bibnamefont
  {Zhou}}, \bibinfo {author} {\bibfnamefont {N.}~\bibnamefont {Zhou}}, \bibinfo
  {author} {\bibfnamefont {C.}~\bibnamefont {Zhu}}, \bibinfo {author}
  {\bibfnamefont {Q.}~\bibnamefont {Zhu}}, \bibinfo {author} {\bibfnamefont
  {Y.}~\bibnamefont {Zhu}}, \bibinfo {author} {\bibfnamefont {G.}~\bibnamefont
  {Zou}}, \bibinfo {author} {\bibfnamefont {H.}~\bibnamefont {Zou}}, \bibinfo
  {author} {\bibfnamefont {Q.}~\bibnamefont {Zhang}}, \bibinfo {author}
  {\bibfnamefont {C.-Y.}\ \bibnamefont {Lu}}, \bibinfo {author} {\bibfnamefont
  {J.}~\bibnamefont {Chen}}, \bibinfo {author} {\bibfnamefont {C.-Z.}\
  \bibnamefont {Peng}}, \bibinfo {author} {\bibfnamefont {F.}~\bibnamefont
  {Chen}}, \bibinfo {author} {\bibfnamefont {X.}~\bibnamefont {Zhu}},\ and\
  \bibinfo {author} {\bibfnamefont {J.-W.}\ \bibnamefont {Pan}},\ }\href
  {https://doi.org/10.48550/arXiv.2607.01473} {\bibinfo {title} {Surface code
  logical operations on a superconducting quantum processor}} (\bibinfo {year}
  {2026}),\ \Eprint {https://arxiv.org/abs/2607.01473} {arXiv:2607.01473
  [quant-ph]} \BibitemShut {NoStop}%
\bibitem [{\citenamefont {Zhu}\ \emph {et~al.}(2026)\citenamefont {Zhu},
  \citenamefont {Zou}, \citenamefont {Jin}, \citenamefont {Mosharev},
  \citenamefont {Luo}, \citenamefont {Wu}, \citenamefont {Chen}, \citenamefont
  {Zhang}, \citenamefont {Gao}, \citenamefont {Wang}, \citenamefont {Zou},
  \citenamefont {Zhang}, \citenamefont {Shen}, \citenamefont {Bao},
  \citenamefont {Zhu}, \citenamefont {Zhong}, \citenamefont {Cui},
  \citenamefont {Han}, \citenamefont {He}, \citenamefont {Wang}, \citenamefont
  {Yang}, \citenamefont {Wang}, \citenamefont {Shen}, \citenamefont {Liu},
  \citenamefont {Song}, \citenamefont {Deng}, \citenamefont {Dong},
  \citenamefont {Zhang}, \citenamefont {Song}, \citenamefont {Wang},
  \citenamefont {Li}, \citenamefont {Guo}, \citenamefont {Yung},\ and\
  \citenamefont {Wang}}]{Zhu2026Combinatorial104Qubits}%
  \BibitemOpen
  \bibfield  {author} {\bibinfo {author} {\bibfnamefont {X.}~\bibnamefont
  {Zhu}}, \bibinfo {author} {\bibfnamefont {Z.}~\bibnamefont {Zou}}, \bibinfo
  {author} {\bibfnamefont {F.}~\bibnamefont {Jin}}, \bibinfo {author}
  {\bibfnamefont {P.}~\bibnamefont {Mosharev}}, \bibinfo {author}
  {\bibfnamefont {M.}~\bibnamefont {Luo}}, \bibinfo {author} {\bibfnamefont
  {Y.}~\bibnamefont {Wu}}, \bibinfo {author} {\bibfnamefont {J.}~\bibnamefont
  {Chen}}, \bibinfo {author} {\bibfnamefont {C.}~\bibnamefont {Zhang}},
  \bibinfo {author} {\bibfnamefont {Y.}~\bibnamefont {Gao}}, \bibinfo {author}
  {\bibfnamefont {N.}~\bibnamefont {Wang}}, \bibinfo {author} {\bibfnamefont
  {Y.}~\bibnamefont {Zou}}, \bibinfo {author} {\bibfnamefont {A.}~\bibnamefont
  {Zhang}}, \bibinfo {author} {\bibfnamefont {F.}~\bibnamefont {Shen}},
  \bibinfo {author} {\bibfnamefont {Z.}~\bibnamefont {Bao}}, \bibinfo {author}
  {\bibfnamefont {Z.}~\bibnamefont {Zhu}}, \bibinfo {author} {\bibfnamefont
  {J.}~\bibnamefont {Zhong}}, \bibinfo {author} {\bibfnamefont
  {Z.}~\bibnamefont {Cui}}, \bibinfo {author} {\bibfnamefont {Y.}~\bibnamefont
  {Han}}, \bibinfo {author} {\bibfnamefont {Y.}~\bibnamefont {He}}, \bibinfo
  {author} {\bibfnamefont {H.}~\bibnamefont {Wang}}, \bibinfo {author}
  {\bibfnamefont {J.-N.}\ \bibnamefont {Yang}}, \bibinfo {author}
  {\bibfnamefont {Y.}~\bibnamefont {Wang}}, \bibinfo {author} {\bibfnamefont
  {J.}~\bibnamefont {Shen}}, \bibinfo {author} {\bibfnamefont {G.}~\bibnamefont
  {Liu}}, \bibinfo {author} {\bibfnamefont {Z.}~\bibnamefont {Song}}, \bibinfo
  {author} {\bibfnamefont {J.}~\bibnamefont {Deng}}, \bibinfo {author}
  {\bibfnamefont {H.}~\bibnamefont {Dong}}, \bibinfo {author} {\bibfnamefont
  {P.}~\bibnamefont {Zhang}}, \bibinfo {author} {\bibfnamefont
  {C.}~\bibnamefont {Song}}, \bibinfo {author} {\bibfnamefont {Z.}~\bibnamefont
  {Wang}}, \bibinfo {author} {\bibfnamefont {H.}~\bibnamefont {Li}}, \bibinfo
  {author} {\bibfnamefont {Q.}~\bibnamefont {Guo}}, \bibinfo {author}
  {\bibfnamefont {M.-H.}\ \bibnamefont {Yung}},\ and\ \bibinfo {author}
  {\bibfnamefont {H.}~\bibnamefont {Wang}},\ }\bibfield  {title} {\bibinfo
  {title} {Combinatorial optimization enhanced by shallow quantum circuits with
  104 superconducting qubits},\ }\href {https://doi.org/10.1093/nsr/nwag124}
  {\bibfield  {journal} {\bibinfo  {journal} {National Science Review}\
  }\textbf {\bibinfo {volume} {13}},\ \bibinfo {pages} {nwag124} (\bibinfo
  {year} {2026})}\BibitemShut {NoStop}%
\bibitem [{\citenamefont {Wang}\ \emph
  {et~al.}(2026{\natexlab{a}})\citenamefont {Wang}, \citenamefont {Lu},
  \citenamefont {Zhang}, \citenamefont {Liu}, \citenamefont {Chen},
  \citenamefont {Wang}, \citenamefont {Wu}, \citenamefont {Xu}, \citenamefont
  {Zhu}, \citenamefont {Jin}, \citenamefont {Gao}, \citenamefont {Tan},
  \citenamefont {Cui}, \citenamefont {Wang}, \citenamefont {Zou}, \citenamefont
  {Zhang}, \citenamefont {Li}, \citenamefont {Shen}, \citenamefont {Zhong},
  \citenamefont {Bao}, \citenamefont {Zhu}, \citenamefont {Han}, \citenamefont
  {He}, \citenamefont {Shen}, \citenamefont {Wang}, \citenamefont {Yang},
  \citenamefont {Song}, \citenamefont {Deng}, \citenamefont {Dong},
  \citenamefont {Sun}, \citenamefont {Li}, \citenamefont {Ye}, \citenamefont
  {Jiang}, \citenamefont {Ma}, \citenamefont {Shen}, \citenamefont {Zhang},
  \citenamefont {Li}, \citenamefont {Guo}, \citenamefont {Wang}, \citenamefont
  {Song}, \citenamefont {Wang},\ and\ \citenamefont
  {Deng}}]{Wang2026LowOverheadQEC}%
  \BibitemOpen
  \bibfield  {author} {\bibinfo {author} {\bibfnamefont {K.}~\bibnamefont
  {Wang}}, \bibinfo {author} {\bibfnamefont {Z.}~\bibnamefont {Lu}}, \bibinfo
  {author} {\bibfnamefont {C.}~\bibnamefont {Zhang}}, \bibinfo {author}
  {\bibfnamefont {G.}~\bibnamefont {Liu}}, \bibinfo {author} {\bibfnamefont
  {J.}~\bibnamefont {Chen}}, \bibinfo {author} {\bibfnamefont {Y.}~\bibnamefont
  {Wang}}, \bibinfo {author} {\bibfnamefont {Y.}~\bibnamefont {Wu}}, \bibinfo
  {author} {\bibfnamefont {S.}~\bibnamefont {Xu}}, \bibinfo {author}
  {\bibfnamefont {X.}~\bibnamefont {Zhu}}, \bibinfo {author} {\bibfnamefont
  {F.}~\bibnamefont {Jin}}, \bibinfo {author} {\bibfnamefont {Y.}~\bibnamefont
  {Gao}}, \bibinfo {author} {\bibfnamefont {Z.}~\bibnamefont {Tan}}, \bibinfo
  {author} {\bibfnamefont {Z.}~\bibnamefont {Cui}}, \bibinfo {author}
  {\bibfnamefont {N.}~\bibnamefont {Wang}}, \bibinfo {author} {\bibfnamefont
  {Y.}~\bibnamefont {Zou}}, \bibinfo {author} {\bibfnamefont {A.}~\bibnamefont
  {Zhang}}, \bibinfo {author} {\bibfnamefont {T.}~\bibnamefont {Li}}, \bibinfo
  {author} {\bibfnamefont {F.}~\bibnamefont {Shen}}, \bibinfo {author}
  {\bibfnamefont {J.}~\bibnamefont {Zhong}}, \bibinfo {author} {\bibfnamefont
  {Z.}~\bibnamefont {Bao}}, \bibinfo {author} {\bibfnamefont {Z.}~\bibnamefont
  {Zhu}}, \bibinfo {author} {\bibfnamefont {Y.}~\bibnamefont {Han}}, \bibinfo
  {author} {\bibfnamefont {Y.}~\bibnamefont {He}}, \bibinfo {author}
  {\bibfnamefont {J.}~\bibnamefont {Shen}}, \bibinfo {author} {\bibfnamefont
  {H.}~\bibnamefont {Wang}}, \bibinfo {author} {\bibfnamefont {J.-N.}\
  \bibnamefont {Yang}}, \bibinfo {author} {\bibfnamefont {Z.}~\bibnamefont
  {Song}}, \bibinfo {author} {\bibfnamefont {J.}~\bibnamefont {Deng}}, \bibinfo
  {author} {\bibfnamefont {H.}~\bibnamefont {Dong}}, \bibinfo {author}
  {\bibfnamefont {Z.-Z.}\ \bibnamefont {Sun}}, \bibinfo {author} {\bibfnamefont
  {W.}~\bibnamefont {Li}}, \bibinfo {author} {\bibfnamefont {Q.}~\bibnamefont
  {Ye}}, \bibinfo {author} {\bibfnamefont {S.}~\bibnamefont {Jiang}}, \bibinfo
  {author} {\bibfnamefont {Y.}~\bibnamefont {Ma}}, \bibinfo {author}
  {\bibfnamefont {P.-X.}\ \bibnamefont {Shen}}, \bibinfo {author}
  {\bibfnamefont {P.}~\bibnamefont {Zhang}}, \bibinfo {author} {\bibfnamefont
  {H.}~\bibnamefont {Li}}, \bibinfo {author} {\bibfnamefont {Q.}~\bibnamefont
  {Guo}}, \bibinfo {author} {\bibfnamefont {Z.}~\bibnamefont {Wang}}, \bibinfo
  {author} {\bibfnamefont {C.}~\bibnamefont {Song}}, \bibinfo {author}
  {\bibfnamefont {H.}~\bibnamefont {Wang}},\ and\ \bibinfo {author}
  {\bibfnamefont {D.-L.}\ \bibnamefont {Deng}},\ }\bibfield  {title} {\bibinfo
  {title} {Demonstration of low-overhead quantum error correction codes},\
  }\bibfield  {journal} {\bibinfo  {journal} {Nature Physics}\ }\href
  {https://doi.org/10.1038/s41567-025-03157-4} {10.1038/s41567-025-03157-4}
  (\bibinfo {year} {2026}{\natexlab{a}})\BibitemShut {NoStop}%
\bibitem [{\citenamefont {Zhang}\ \emph {et~al.}(2026)\citenamefont {Zhang},
  \citenamefont {Li}, \citenamefont {Yang}, \citenamefont {Feng}, \citenamefont
  {Li}, \citenamefont {Chen}, \citenamefont {Liu}, \citenamefont {Xue},\ and\
  \citenamefont {Yu}}]{Zhang2026HighPrecisionCZCalibration}%
  \BibitemOpen
  \bibfield  {author} {\bibinfo {author} {\bibfnamefont {H.}~\bibnamefont
  {Zhang}}, \bibinfo {author} {\bibfnamefont {M.}~\bibnamefont {Li}}, \bibinfo
  {author} {\bibfnamefont {S.}~\bibnamefont {Yang}}, \bibinfo {author}
  {\bibfnamefont {Y.}~\bibnamefont {Feng}}, \bibinfo {author} {\bibfnamefont
  {Y.}~\bibnamefont {Li}}, \bibinfo {author} {\bibfnamefont {C.}~\bibnamefont
  {Chen}}, \bibinfo {author} {\bibfnamefont {P.}~\bibnamefont {Liu}}, \bibinfo
  {author} {\bibfnamefont {G.}~\bibnamefont {Xue}},\ and\ \bibinfo {author}
  {\bibfnamefont {H.}~\bibnamefont {Yu}},\ }\href
  {https://doi.org/10.48550/arXiv.2607.01422} {\bibinfo {title} {High-precision
  calibration workflow achieves above {99.9\%} {CZ} gate fidelity on a scalable
  superconducting processor}} (\bibinfo {year} {2026}),\ \Eprint
  {https://arxiv.org/abs/2607.01422} {arXiv:2607.01422 [quant-ph]} \BibitemShut
  {NoStop}%
\bibitem [{\citenamefont {Kim}\ \emph {et~al.}(2023)\citenamefont {Kim},
  \citenamefont {Eddins}, \citenamefont {Anand}, \citenamefont {Wei},
  \citenamefont {van~den Berg}, \citenamefont {Rosenblatt}, \citenamefont
  {Nayfeh}, \citenamefont {Wu}, \citenamefont {Zaletel}, \citenamefont
  {Temme},\ and\ \citenamefont {Kandala}}]{Kim2023QuantumUtility}%
  \BibitemOpen
  \bibfield  {author} {\bibinfo {author} {\bibfnamefont {Y.}~\bibnamefont
  {Kim}}, \bibinfo {author} {\bibfnamefont {A.}~\bibnamefont {Eddins}},
  \bibinfo {author} {\bibfnamefont {S.}~\bibnamefont {Anand}}, \bibinfo
  {author} {\bibfnamefont {K.~X.}\ \bibnamefont {Wei}}, \bibinfo {author}
  {\bibfnamefont {E.}~\bibnamefont {van~den Berg}}, \bibinfo {author}
  {\bibfnamefont {S.}~\bibnamefont {Rosenblatt}}, \bibinfo {author}
  {\bibfnamefont {H.}~\bibnamefont {Nayfeh}}, \bibinfo {author} {\bibfnamefont
  {Y.}~\bibnamefont {Wu}}, \bibinfo {author} {\bibfnamefont {M.}~\bibnamefont
  {Zaletel}}, \bibinfo {author} {\bibfnamefont {K.}~\bibnamefont {Temme}},\
  and\ \bibinfo {author} {\bibfnamefont {A.}~\bibnamefont {Kandala}},\
  }\bibfield  {title} {\bibinfo {title} {Evidence for the utility of quantum
  computing before fault tolerance},\ }\href
  {https://doi.org/10.1038/s41586-023-06096-3} {\bibfield  {journal} {\bibinfo
  {journal} {Nature}\ }\textbf {\bibinfo {volume} {618}},\ \bibinfo {pages}
  {500} (\bibinfo {year} {2023})}\BibitemShut {NoStop}%
\bibitem [{\citenamefont {Tan}\ \emph {et~al.}(2018)\citenamefont {Tan},
  \citenamefont {Zhang}, \citenamefont {Liu}, \citenamefont {Xue},
  \citenamefont {Yu}, \citenamefont {Zhu}, \citenamefont {Yan}, \citenamefont
  {Zhu},\ and\ \citenamefont {Yu}}]{PhysRevLett.120.130503}%
  \BibitemOpen
  \bibfield  {author} {\bibinfo {author} {\bibfnamefont {X.}~\bibnamefont
  {Tan}}, \bibinfo {author} {\bibfnamefont {D.-W.}\ \bibnamefont {Zhang}},
  \bibinfo {author} {\bibfnamefont {Q.}~\bibnamefont {Liu}}, \bibinfo {author}
  {\bibfnamefont {G.}~\bibnamefont {Xue}}, \bibinfo {author} {\bibfnamefont
  {H.-F.}\ \bibnamefont {Yu}}, \bibinfo {author} {\bibfnamefont {Y.-Q.}\
  \bibnamefont {Zhu}}, \bibinfo {author} {\bibfnamefont {H.}~\bibnamefont
  {Yan}}, \bibinfo {author} {\bibfnamefont {S.-L.}\ \bibnamefont {Zhu}},\ and\
  \bibinfo {author} {\bibfnamefont {Y.}~\bibnamefont {Yu}},\ }\bibfield
  {title} {\bibinfo {title} {Topological {Maxwell} metal bands in a
  superconducting qutrit},\ }\href
  {https://doi.org/10.1103/PhysRevLett.120.130503} {\bibfield  {journal}
  {\bibinfo  {journal} {Phys. Rev. Lett.}\ }\textbf {\bibinfo {volume} {120}},\
  \bibinfo {pages} {130503} (\bibinfo {year} {2018})}\BibitemShut {NoStop}%
\bibitem [{\citenamefont {Tan}\ \emph {et~al.}(2019)\citenamefont {Tan},
  \citenamefont {Zhang}, \citenamefont {Yang}, \citenamefont {Chu},
  \citenamefont {Zhu}, \citenamefont {Li}, \citenamefont {Yang}, \citenamefont
  {Song}, \citenamefont {Han}, \citenamefont {Li}, \citenamefont {Dong},
  \citenamefont {Yu}, \citenamefont {Yan}, \citenamefont {Zhu},\ and\
  \citenamefont {Yu}}]{PhysRevLett.122.210401}%
  \BibitemOpen
  \bibfield  {author} {\bibinfo {author} {\bibfnamefont {X.}~\bibnamefont
  {Tan}}, \bibinfo {author} {\bibfnamefont {D.-W.}\ \bibnamefont {Zhang}},
  \bibinfo {author} {\bibfnamefont {Z.}~\bibnamefont {Yang}}, \bibinfo {author}
  {\bibfnamefont {J.}~\bibnamefont {Chu}}, \bibinfo {author} {\bibfnamefont
  {Y.-Q.}\ \bibnamefont {Zhu}}, \bibinfo {author} {\bibfnamefont
  {D.}~\bibnamefont {Li}}, \bibinfo {author} {\bibfnamefont {X.}~\bibnamefont
  {Yang}}, \bibinfo {author} {\bibfnamefont {S.}~\bibnamefont {Song}}, \bibinfo
  {author} {\bibfnamefont {Z.}~\bibnamefont {Han}}, \bibinfo {author}
  {\bibfnamefont {Z.}~\bibnamefont {Li}}, \bibinfo {author} {\bibfnamefont
  {Y.}~\bibnamefont {Dong}}, \bibinfo {author} {\bibfnamefont {H.-F.}\
  \bibnamefont {Yu}}, \bibinfo {author} {\bibfnamefont {H.}~\bibnamefont
  {Yan}}, \bibinfo {author} {\bibfnamefont {S.-L.}\ \bibnamefont {Zhu}},\ and\
  \bibinfo {author} {\bibfnamefont {Y.}~\bibnamefont {Yu}},\ }\bibfield
  {title} {\bibinfo {title} {Experimental measurement of the quantum metric
  tensor and related topological phase transition with a superconducting
  qubit},\ }\href {https://doi.org/10.1103/PhysRevLett.122.210401} {\bibfield
  {journal} {\bibinfo  {journal} {Phys. Rev. Lett.}\ }\textbf {\bibinfo
  {volume} {122}},\ \bibinfo {pages} {210401} (\bibinfo {year}
  {2019})}\BibitemShut {NoStop}%
\bibitem [{\citenamefont {Shi}\ \emph {et~al.}(2023)\citenamefont {Shi},
  \citenamefont {Liu}, \citenamefont {Zhang}, \citenamefont {Xiang},
  \citenamefont {Huang}, \citenamefont {Liu}, \citenamefont {Wang},
  \citenamefont {Zhang}, \citenamefont {Deng}, \citenamefont {Liang},
  \citenamefont {Mei}, \citenamefont {Li}, \citenamefont {Li}, \citenamefont
  {Ma}, \citenamefont {Liu}, \citenamefont {Chen}, \citenamefont {Liu},
  \citenamefont {Tian}, \citenamefont {Song}, \citenamefont {Zhao},
  \citenamefont {Xu}, \citenamefont {Zheng}, \citenamefont {Nori},\ and\
  \citenamefont {Fan}}]{PhysRevLett.131.080401}%
  \BibitemOpen
  \bibfield  {author} {\bibinfo {author} {\bibfnamefont {Y.-H.}\ \bibnamefont
  {Shi}}, \bibinfo {author} {\bibfnamefont {Y.}~\bibnamefont {Liu}}, \bibinfo
  {author} {\bibfnamefont {Y.-R.}\ \bibnamefont {Zhang}}, \bibinfo {author}
  {\bibfnamefont {Z.}~\bibnamefont {Xiang}}, \bibinfo {author} {\bibfnamefont
  {K.}~\bibnamefont {Huang}}, \bibinfo {author} {\bibfnamefont
  {T.}~\bibnamefont {Liu}}, \bibinfo {author} {\bibfnamefont {Y.-Y.}\
  \bibnamefont {Wang}}, \bibinfo {author} {\bibfnamefont {J.-C.}\ \bibnamefont
  {Zhang}}, \bibinfo {author} {\bibfnamefont {C.-L.}\ \bibnamefont {Deng}},
  \bibinfo {author} {\bibfnamefont {G.-H.}\ \bibnamefont {Liang}}, \bibinfo
  {author} {\bibfnamefont {Z.-Y.}\ \bibnamefont {Mei}}, \bibinfo {author}
  {\bibfnamefont {H.}~\bibnamefont {Li}}, \bibinfo {author} {\bibfnamefont
  {T.-M.}\ \bibnamefont {Li}}, \bibinfo {author} {\bibfnamefont {W.-G.}\
  \bibnamefont {Ma}}, \bibinfo {author} {\bibfnamefont {H.-T.}\ \bibnamefont
  {Liu}}, \bibinfo {author} {\bibfnamefont {C.-T.}\ \bibnamefont {Chen}},
  \bibinfo {author} {\bibfnamefont {T.}~\bibnamefont {Liu}}, \bibinfo {author}
  {\bibfnamefont {Y.}~\bibnamefont {Tian}}, \bibinfo {author} {\bibfnamefont
  {X.}~\bibnamefont {Song}}, \bibinfo {author} {\bibfnamefont {S.~P.}\
  \bibnamefont {Zhao}}, \bibinfo {author} {\bibfnamefont {K.}~\bibnamefont
  {Xu}}, \bibinfo {author} {\bibfnamefont {D.}~\bibnamefont {Zheng}}, \bibinfo
  {author} {\bibfnamefont {F.}~\bibnamefont {Nori}},\ and\ \bibinfo {author}
  {\bibfnamefont {H.}~\bibnamefont {Fan}},\ }\bibfield  {title} {\bibinfo
  {title} {Quantum simulation of topological zero modes on a 41-qubit
  superconducting processor},\ }\href
  {https://doi.org/10.1103/PhysRevLett.131.080401} {\bibfield  {journal}
  {\bibinfo  {journal} {Phys. Rev. Lett.}\ }\textbf {\bibinfo {volume} {131}},\
  \bibinfo {pages} {080401} (\bibinfo {year} {2023})}\BibitemShut {NoStop}%
\bibitem [{\citenamefont {Yoshioka}\ \emph {et~al.}(2025)\citenamefont
  {Yoshioka}, \citenamefont {Amico}, \citenamefont {Kirby}, \citenamefont
  {Jurcevic}, \citenamefont {Dutt}, \citenamefont {Fuller}, \citenamefont
  {Garion}, \citenamefont {Haas}, \citenamefont {Hamamura}, \citenamefont
  {Ivrii}, \citenamefont {Majumdar}, \citenamefont {Minev}, \citenamefont
  {Motta}, \citenamefont {Pokharel}, \citenamefont {Rivero}, \citenamefont
  {Sharma}, \citenamefont {Wood}, \citenamefont {Javadi-Abhari},\ and\
  \citenamefont {Mezzacapo}}]{Yoshioka2025KrylovDiagonalization}%
  \BibitemOpen
  \bibfield  {author} {\bibinfo {author} {\bibfnamefont {N.}~\bibnamefont
  {Yoshioka}}, \bibinfo {author} {\bibfnamefont {M.}~\bibnamefont {Amico}},
  \bibinfo {author} {\bibfnamefont {W.}~\bibnamefont {Kirby}}, \bibinfo
  {author} {\bibfnamefont {P.}~\bibnamefont {Jurcevic}}, \bibinfo {author}
  {\bibfnamefont {A.}~\bibnamefont {Dutt}}, \bibinfo {author} {\bibfnamefont
  {B.}~\bibnamefont {Fuller}}, \bibinfo {author} {\bibfnamefont
  {S.}~\bibnamefont {Garion}}, \bibinfo {author} {\bibfnamefont
  {H.}~\bibnamefont {Haas}}, \bibinfo {author} {\bibfnamefont {I.}~\bibnamefont
  {Hamamura}}, \bibinfo {author} {\bibfnamefont {A.}~\bibnamefont {Ivrii}},
  \bibinfo {author} {\bibfnamefont {R.}~\bibnamefont {Majumdar}}, \bibinfo
  {author} {\bibfnamefont {Z.}~\bibnamefont {Minev}}, \bibinfo {author}
  {\bibfnamefont {M.}~\bibnamefont {Motta}}, \bibinfo {author} {\bibfnamefont
  {B.}~\bibnamefont {Pokharel}}, \bibinfo {author} {\bibfnamefont
  {P.}~\bibnamefont {Rivero}}, \bibinfo {author} {\bibfnamefont
  {K.}~\bibnamefont {Sharma}}, \bibinfo {author} {\bibfnamefont {C.~J.}\
  \bibnamefont {Wood}}, \bibinfo {author} {\bibfnamefont {A.}~\bibnamefont
  {Javadi-Abhari}},\ and\ \bibinfo {author} {\bibfnamefont {A.}~\bibnamefont
  {Mezzacapo}},\ }\bibfield  {title} {\bibinfo {title} {{Krylov diagonalization
  of large many-body Hamiltonians on a quantum processor}},\ }\href
  {https://doi.org/10.1038/s41467-025-59716-z} {\bibfield  {journal} {\bibinfo
  {journal} {Nature Communications}\ }\textbf {\bibinfo {volume} {16}},\
  \bibinfo {pages} {5014} (\bibinfo {year} {2025})}\BibitemShut {NoStop}%
\bibitem [{\citenamefont {McKay}\ \emph {et~al.}(2023)\citenamefont {McKay},
  \citenamefont {Hincks}, \citenamefont {Pritchett}, \citenamefont {Carroll},
  \citenamefont {Govia},\ and\ \citenamefont
  {Merkel}}]{McKay2023BenchmarkingScale}%
  \BibitemOpen
  \bibfield  {author} {\bibinfo {author} {\bibfnamefont {D.~C.}\ \bibnamefont
  {McKay}}, \bibinfo {author} {\bibfnamefont {I.}~\bibnamefont {Hincks}},
  \bibinfo {author} {\bibfnamefont {E.~J.}\ \bibnamefont {Pritchett}}, \bibinfo
  {author} {\bibfnamefont {M.}~\bibnamefont {Carroll}}, \bibinfo {author}
  {\bibfnamefont {L.~C.~G.}\ \bibnamefont {Govia}},\ and\ \bibinfo {author}
  {\bibfnamefont {S.~T.}\ \bibnamefont {Merkel}},\ }\href
  {https://doi.org/10.48550/arXiv.2311.05933} {\bibinfo {title} {Benchmarking
  quantum processor performance at scale}} (\bibinfo {year} {2023}),\ \Eprint
  {https://arxiv.org/abs/2311.05933} {arXiv:2311.05933 [quant-ph]} \BibitemShut
  {NoStop}%
\bibitem [{\citenamefont {Ni}\ \emph {et~al.}(2023)\citenamefont {Ni},
  \citenamefont {Li}, \citenamefont {Deng}, \citenamefont {Cai}, \citenamefont
  {Zhang}, \citenamefont {Wang}, \citenamefont {Yang}, \citenamefont {Yu},
  \citenamefont {Yan}, \citenamefont {Liu}, \citenamefont {Zou}, \citenamefont
  {Sun}, \citenamefont {Zheng}, \citenamefont {Xu},\ and\ \citenamefont
  {Yu}}]{Ni2023BreakEvenDiscreteVariable}%
  \BibitemOpen
  \bibfield  {author} {\bibinfo {author} {\bibfnamefont {Z.}~\bibnamefont
  {Ni}}, \bibinfo {author} {\bibfnamefont {S.}~\bibnamefont {Li}}, \bibinfo
  {author} {\bibfnamefont {X.}~\bibnamefont {Deng}}, \bibinfo {author}
  {\bibfnamefont {Y.}~\bibnamefont {Cai}}, \bibinfo {author} {\bibfnamefont
  {L.}~\bibnamefont {Zhang}}, \bibinfo {author} {\bibfnamefont
  {W.}~\bibnamefont {Wang}}, \bibinfo {author} {\bibfnamefont {Z.-B.}\
  \bibnamefont {Yang}}, \bibinfo {author} {\bibfnamefont {H.}~\bibnamefont
  {Yu}}, \bibinfo {author} {\bibfnamefont {F.}~\bibnamefont {Yan}}, \bibinfo
  {author} {\bibfnamefont {S.}~\bibnamefont {Liu}}, \bibinfo {author}
  {\bibfnamefont {C.-L.}\ \bibnamefont {Zou}}, \bibinfo {author} {\bibfnamefont
  {L.}~\bibnamefont {Sun}}, \bibinfo {author} {\bibfnamefont {S.-B.}\
  \bibnamefont {Zheng}}, \bibinfo {author} {\bibfnamefont {Y.}~\bibnamefont
  {Xu}},\ and\ \bibinfo {author} {\bibfnamefont {D.}~\bibnamefont {Yu}},\
  }\bibfield  {title} {\bibinfo {title} {Beating the break-even point with a
  discrete-variable-encoded logical qubit},\ }\href
  {https://doi.org/10.1038/s41586-023-05784-4} {\bibfield  {journal} {\bibinfo
  {journal} {Nature}\ }\textbf {\bibinfo {volume} {616}},\ \bibinfo {pages}
  {56} (\bibinfo {year} {2023})}\BibitemShut {NoStop}%
\bibitem [{\citenamefont {Liu}\ \emph {et~al.}(2023)\citenamefont {Liu},
  \citenamefont {Huang}, \citenamefont {Yan}, \citenamefont {Lu}, \citenamefont
  {Deng},\ and\ \citenamefont {Liu}}]{PhysRevApplied.19.054028}%
  \BibitemOpen
  \bibfield  {author} {\bibinfo {author} {\bibfnamefont {H.}~\bibnamefont
  {Liu}}, \bibinfo {author} {\bibfnamefont {X.}~\bibnamefont {Huang}}, \bibinfo
  {author} {\bibfnamefont {M.}~\bibnamefont {Yan}}, \bibinfo {author}
  {\bibfnamefont {J.}~\bibnamefont {Lu}}, \bibinfo {author} {\bibfnamefont
  {W.}~\bibnamefont {Deng}},\ and\ \bibinfo {author} {\bibfnamefont
  {Z.}~\bibnamefont {Liu}},\ }\bibfield  {title} {\bibinfo {title} {Acoustic
  topological metamaterials of large winding number},\ }\href
  {https://doi.org/10.1103/PhysRevApplied.19.054028} {\bibfield  {journal}
  {\bibinfo  {journal} {Phys. Rev. Appl.}\ }\textbf {\bibinfo {volume} {19}},\
  \bibinfo {pages} {054028} (\bibinfo {year} {2023})}\BibitemShut {NoStop}%
\bibitem [{\citenamefont {Su}\ \emph {et~al.}(1979)\citenamefont {Su},
  \citenamefont {Schrieffer},\ and\ \citenamefont
  {Heeger}}]{PhysRevLett.42.1698}%
  \BibitemOpen
  \bibfield  {author} {\bibinfo {author} {\bibfnamefont {W.~P.}\ \bibnamefont
  {Su}}, \bibinfo {author} {\bibfnamefont {J.~R.}\ \bibnamefont {Schrieffer}},\
  and\ \bibinfo {author} {\bibfnamefont {A.~J.}\ \bibnamefont {Heeger}},\
  }\bibfield  {title} {\bibinfo {title} {Solitons in polyacetylene},\ }\href
  {https://doi.org/10.1103/PhysRevLett.42.1698} {\bibfield  {journal} {\bibinfo
   {journal} {Phys. Rev. Lett.}\ }\textbf {\bibinfo {volume} {42}},\ \bibinfo
  {pages} {1698} (\bibinfo {year} {1979})}\BibitemShut {NoStop}%
\bibitem [{\citenamefont {Ryu}\ \emph {et~al.}(2010)\citenamefont {Ryu},
  \citenamefont {Schnyder}, \citenamefont {Furusaki},\ and\ \citenamefont
  {Ludwig}}]{Ryu2010}%
  \BibitemOpen
  \bibfield  {author} {\bibinfo {author} {\bibfnamefont {S.}~\bibnamefont
  {Ryu}}, \bibinfo {author} {\bibfnamefont {A.~P.}\ \bibnamefont {Schnyder}},
  \bibinfo {author} {\bibfnamefont {A.}~\bibnamefont {Furusaki}},\ and\
  \bibinfo {author} {\bibfnamefont {A.~W.~W.}\ \bibnamefont {Ludwig}},\
  }\bibfield  {title} {\bibinfo {title} {{Topological insulators and
  superconductors: tenfold way and dimensional hierarchy}},\ }\href
  {https://doi.org/10.1088/1367-2630/12/6/065010} {\bibfield  {journal}
  {\bibinfo  {journal} {New Journal of Physics}\ }\textbf {\bibinfo {volume}
  {12}},\ \bibinfo {pages} {065010} (\bibinfo {year} {2010})}\BibitemShut
  {NoStop}%
\bibitem [{\citenamefont {Mondragon-Shem}\ \emph {et~al.}(2014)\citenamefont
  {Mondragon-Shem}, \citenamefont {Hughes}, \citenamefont {Song},\ and\
  \citenamefont {Prodan}}]{PhysRevLett.113.046802}%
  \BibitemOpen
  \bibfield  {author} {\bibinfo {author} {\bibfnamefont {I.}~\bibnamefont
  {Mondragon-Shem}}, \bibinfo {author} {\bibfnamefont {T.~L.}\ \bibnamefont
  {Hughes}}, \bibinfo {author} {\bibfnamefont {J.}~\bibnamefont {Song}},\ and\
  \bibinfo {author} {\bibfnamefont {E.}~\bibnamefont {Prodan}},\ }\bibfield
  {title} {\bibinfo {title} {{Topological Criticality in the Chiral-Symmetric
  {AIII} Class at Strong Disorder}},\ }\href
  {https://doi.org/10.1103/PhysRevLett.113.046802} {\bibfield  {journal}
  {\bibinfo  {journal} {Phys. Rev. Lett.}\ }\textbf {\bibinfo {volume} {113}},\
  \bibinfo {pages} {046802} (\bibinfo {year} {2014})}\BibitemShut {NoStop}%
\bibitem [{\citenamefont {Sui}\ \emph {et~al.}(2026)\citenamefont {Sui},
  \citenamefont {Song}, \citenamefont {Huai}, \citenamefont {Li}, \citenamefont
  {Zong}, \citenamefont {Bu}, \citenamefont {Yang}, \citenamefont {Liu},
  \citenamefont {Jin}, \citenamefont {Chen}, \citenamefont {Zhang},
  \citenamefont {Wu}, \citenamefont {Zheng}, \citenamefont {Zhang},
  \citenamefont {v.~Chen},\ and\ \citenamefont {Yin}}]{Sui2026Platonic}%
  \BibitemOpen
  \bibfield  {author} {\bibinfo {author} {\bibfnamefont {L.}~\bibnamefont
  {Sui}}, \bibinfo {author} {\bibfnamefont {H.-H.}\ \bibnamefont {Song}},
  \bibinfo {author} {\bibfnamefont {S.}~\bibnamefont {Huai}}, \bibinfo {author}
  {\bibfnamefont {Y.}~\bibnamefont {Li}}, \bibinfo {author} {\bibfnamefont
  {Z.}~\bibnamefont {Zong}}, \bibinfo {author} {\bibfnamefont {K.}~\bibnamefont
  {Bu}}, \bibinfo {author} {\bibfnamefont {X.}~\bibnamefont {Yang}}, \bibinfo
  {author} {\bibfnamefont {X.}~\bibnamefont {Liu}}, \bibinfo {author}
  {\bibfnamefont {W.}~\bibnamefont {Jin}}, \bibinfo {author} {\bibfnamefont
  {B.}~\bibnamefont {Chen}}, \bibinfo {author} {\bibfnamefont {X.}~\bibnamefont
  {Zhang}}, \bibinfo {author} {\bibfnamefont {J.}~\bibnamefont {Wu}}, \bibinfo
  {author} {\bibfnamefont {Y.}~\bibnamefont {Zheng}}, \bibinfo {author}
  {\bibfnamefont {S.}~\bibnamefont {Zhang}}, \bibinfo {author} {\bibfnamefont
  {G.}~\bibnamefont {v.~Chen}},\ and\ \bibinfo {author} {\bibfnamefont
  {Y.}~\bibnamefont {Yin}},\ }\href {https://arxiv.org/abs/2606.16854}
  {\bibinfo {title} {3d {Ising} criticality with platonic lattice
  superconducting qubits}} (\bibinfo {year} {2026}),\ \Eprint
  {https://arxiv.org/abs/2606.16854} {arXiv:2606.16854 [quant-ph]} \BibitemShut
  {NoStop}%
\bibitem [{\citenamefont {Sete}\ \emph {et~al.}(2021)\citenamefont {Sete},
  \citenamefont {Chen}, \citenamefont {Manenti}, \citenamefont {Kulshreshtha},\
  and\ \citenamefont {Poletto}}]{Sete2021FloatingCoupler}%
  \BibitemOpen
  \bibfield  {author} {\bibinfo {author} {\bibfnamefont {E.~A.}\ \bibnamefont
  {Sete}}, \bibinfo {author} {\bibfnamefont {A.~Q.}\ \bibnamefont {Chen}},
  \bibinfo {author} {\bibfnamefont {R.}~\bibnamefont {Manenti}}, \bibinfo
  {author} {\bibfnamefont {S.}~\bibnamefont {Kulshreshtha}},\ and\ \bibinfo
  {author} {\bibfnamefont {S.}~\bibnamefont {Poletto}},\ }\bibfield  {title}
  {\bibinfo {title} {Floating tunable coupler for scalable quantum computing
  architectures},\ }\href {https://doi.org/10.1103/PhysRevApplied.15.064063}
  {\bibfield  {journal} {\bibinfo  {journal} {Phys. Rev. Appl.}\ }\textbf
  {\bibinfo {volume} {15}},\ \bibinfo {pages} {064063} (\bibinfo {year}
  {2021})}\BibitemShut {NoStop}%
\bibitem [{\citenamefont {Field}\ \emph {et~al.}(2024)\citenamefont {Field},
  \citenamefont {Chen}, \citenamefont {Scharmann}, \citenamefont {Sete},
  \citenamefont {Oruc}, \citenamefont {Vu}, \citenamefont {Kosenko},
  \citenamefont {Mutus}, \citenamefont {Poletto},\ and\ \citenamefont
  {Bestwick}}]{Field2024ModularCoupler}%
  \BibitemOpen
  \bibfield  {author} {\bibinfo {author} {\bibfnamefont {M.}~\bibnamefont
  {Field}}, \bibinfo {author} {\bibfnamefont {A.~Q.}\ \bibnamefont {Chen}},
  \bibinfo {author} {\bibfnamefont {B.}~\bibnamefont {Scharmann}}, \bibinfo
  {author} {\bibfnamefont {E.~A.}\ \bibnamefont {Sete}}, \bibinfo {author}
  {\bibfnamefont {F.}~\bibnamefont {Oruc}}, \bibinfo {author} {\bibfnamefont
  {K.}~\bibnamefont {Vu}}, \bibinfo {author} {\bibfnamefont {V.}~\bibnamefont
  {Kosenko}}, \bibinfo {author} {\bibfnamefont {J.~Y.}\ \bibnamefont {Mutus}},
  \bibinfo {author} {\bibfnamefont {S.}~\bibnamefont {Poletto}},\ and\ \bibinfo
  {author} {\bibfnamefont {A.}~\bibnamefont {Bestwick}},\ }\bibfield  {title}
  {\bibinfo {title} {Modular superconducting-qubit architecture with a
  multichip tunable coupler},\ }\href
  {https://doi.org/10.1103/PhysRevApplied.21.054063} {\bibfield  {journal}
  {\bibinfo  {journal} {Phys. Rev. Appl.}\ }\textbf {\bibinfo {volume} {21}},\
  \bibinfo {pages} {054063} (\bibinfo {year} {2024})}\BibitemShut {NoStop}%
\bibitem [{\citenamefont {Sete}\ \emph {et~al.}(2024)\citenamefont {Sete},
  \citenamefont {Tripathi}, \citenamefont {Valery}, \citenamefont {Lidar},\
  and\ \citenamefont {Mutus}}]{Sete2024ErrorBudget}%
  \BibitemOpen
  \bibfield  {author} {\bibinfo {author} {\bibfnamefont {E.~A.}\ \bibnamefont
  {Sete}}, \bibinfo {author} {\bibfnamefont {V.}~\bibnamefont {Tripathi}},
  \bibinfo {author} {\bibfnamefont {J.~A.}\ \bibnamefont {Valery}}, \bibinfo
  {author} {\bibfnamefont {D.}~\bibnamefont {Lidar}},\ and\ \bibinfo {author}
  {\bibfnamefont {J.~Y.}\ \bibnamefont {Mutus}},\ }\bibfield  {title} {\bibinfo
  {title} {Error budget of a parametric resonance entangling gate with a
  tunable coupler},\ }\href {https://doi.org/10.1103/PhysRevApplied.22.014059}
  {\bibfield  {journal} {\bibinfo  {journal} {Phys. Rev. Appl.}\ }\textbf
  {\bibinfo {volume} {22}},\ \bibinfo {pages} {014059} (\bibinfo {year}
  {2024})}\BibitemShut {NoStop}%
\bibitem [{\citenamefont {Bu}\ \emph {et~al.}(2025)\citenamefont {Bu},
  \citenamefont {Huai}, \citenamefont {Zhang}, \citenamefont {Li},
  \citenamefont {Li}, \citenamefont {Hu}, \citenamefont {Yang}, \citenamefont
  {Dai}, \citenamefont {Cai}, \citenamefont {Zheng},\ and\ \citenamefont
  {Zhang}}]{Bu2025TantalumAirbridges}%
  \BibitemOpen
  \bibfield  {author} {\bibinfo {author} {\bibfnamefont {K.}~\bibnamefont
  {Bu}}, \bibinfo {author} {\bibfnamefont {S.}~\bibnamefont {Huai}}, \bibinfo
  {author} {\bibfnamefont {Z.}~\bibnamefont {Zhang}}, \bibinfo {author}
  {\bibfnamefont {D.}~\bibnamefont {Li}}, \bibinfo {author} {\bibfnamefont
  {Y.}~\bibnamefont {Li}}, \bibinfo {author} {\bibfnamefont {J.}~\bibnamefont
  {Hu}}, \bibinfo {author} {\bibfnamefont {X.}~\bibnamefont {Yang}}, \bibinfo
  {author} {\bibfnamefont {M.}~\bibnamefont {Dai}}, \bibinfo {author}
  {\bibfnamefont {T.}~\bibnamefont {Cai}}, \bibinfo {author} {\bibfnamefont
  {Y.-C.}\ \bibnamefont {Zheng}},\ and\ \bibinfo {author} {\bibfnamefont
  {S.}~\bibnamefont {Zhang}},\ }\bibfield  {title} {\bibinfo {title} {Tantalum
  airbridges for scalable superconducting quantum processors},\ }\href
  {https://doi.org/10.1038/s41534-025-00972-8} {\bibfield  {journal} {\bibinfo
  {journal} {npj Quantum Information}\ }\textbf {\bibinfo {volume} {11}},\
  \bibinfo {pages} {17} (\bibinfo {year} {2025})}\BibitemShut {NoStop}%
\bibitem [{\citenamefont {Marxer}\ \emph {et~al.}(2023)\citenamefont {Marxer},
  \citenamefont {Veps\"{a}l\"{a}inen}, \citenamefont {Jolin}, \citenamefont
  {Tuorila}, \citenamefont {Landra}, \citenamefont {Ockeloen-Korppi},
  \citenamefont {Liu}, \citenamefont {Ahonen}, \citenamefont {Auer},
  \citenamefont {Belzane}, \citenamefont {Bergholm}, \citenamefont {Chan},
  \citenamefont {Chan}, \citenamefont {Hiltunen}, \citenamefont {Hotari},
  \citenamefont {Hyypp\"{a}}, \citenamefont {Ikonen}, \citenamefont {Janzso},
  \citenamefont {Koistinen}, \citenamefont {Kotilahti}, \citenamefont {Li},
  \citenamefont {Luus}, \citenamefont {Papic}, \citenamefont {Partanen},
  \citenamefont {R\"{a}bin\"{a}}, \citenamefont {Rosti}, \citenamefont
  {Savytskyi}, \citenamefont {Sepp\"{a}l\"{a}}, \citenamefont {Sevriuk},
  \citenamefont {Takala}, \citenamefont {Tarasinski}, \citenamefont {Thapa},
  \citenamefont {Tosto}, \citenamefont {Vorobeva}, \citenamefont {Yu},
  \citenamefont {Tan}, \citenamefont {Hassel}, \citenamefont
  {M\"{o}tt\"{o}nen},\ and\ \citenamefont {Heinsoo}}]{Marxer2023Long}%
  \BibitemOpen
  \bibfield  {author} {\bibinfo {author} {\bibfnamefont {F.}~\bibnamefont
  {Marxer}}, \bibinfo {author} {\bibfnamefont {A.}~\bibnamefont
  {Veps\"{a}l\"{a}inen}}, \bibinfo {author} {\bibfnamefont {S.~W.}\
  \bibnamefont {Jolin}}, \bibinfo {author} {\bibfnamefont {J.}~\bibnamefont
  {Tuorila}}, \bibinfo {author} {\bibfnamefont {A.}~\bibnamefont {Landra}},
  \bibinfo {author} {\bibfnamefont {C.}~\bibnamefont {Ockeloen-Korppi}},
  \bibinfo {author} {\bibfnamefont {W.}~\bibnamefont {Liu}}, \bibinfo {author}
  {\bibfnamefont {O.}~\bibnamefont {Ahonen}}, \bibinfo {author} {\bibfnamefont
  {A.}~\bibnamefont {Auer}}, \bibinfo {author} {\bibfnamefont {L.}~\bibnamefont
  {Belzane}}, \bibinfo {author} {\bibfnamefont {V.}~\bibnamefont {Bergholm}},
  \bibinfo {author} {\bibfnamefont {C.~F.}\ \bibnamefont {Chan}}, \bibinfo
  {author} {\bibfnamefont {K.~W.}\ \bibnamefont {Chan}}, \bibinfo {author}
  {\bibfnamefont {T.}~\bibnamefont {Hiltunen}}, \bibinfo {author}
  {\bibfnamefont {J.}~\bibnamefont {Hotari}}, \bibinfo {author} {\bibfnamefont
  {E.}~\bibnamefont {Hyypp\"{a}}}, \bibinfo {author} {\bibfnamefont
  {J.}~\bibnamefont {Ikonen}}, \bibinfo {author} {\bibfnamefont
  {D.}~\bibnamefont {Janzso}}, \bibinfo {author} {\bibfnamefont
  {M.}~\bibnamefont {Koistinen}}, \bibinfo {author} {\bibfnamefont
  {J.}~\bibnamefont {Kotilahti}}, \bibinfo {author} {\bibfnamefont
  {T.}~\bibnamefont {Li}}, \bibinfo {author} {\bibfnamefont {J.}~\bibnamefont
  {Luus}}, \bibinfo {author} {\bibfnamefont {M.}~\bibnamefont {Papic}},
  \bibinfo {author} {\bibfnamefont {M.}~\bibnamefont {Partanen}}, \bibinfo
  {author} {\bibfnamefont {J.}~\bibnamefont {R\"{a}bin\"{a}}}, \bibinfo
  {author} {\bibfnamefont {J.}~\bibnamefont {Rosti}}, \bibinfo {author}
  {\bibfnamefont {M.}~\bibnamefont {Savytskyi}}, \bibinfo {author}
  {\bibfnamefont {M.}~\bibnamefont {Sepp\"{a}l\"{a}}}, \bibinfo {author}
  {\bibfnamefont {V.}~\bibnamefont {Sevriuk}}, \bibinfo {author} {\bibfnamefont
  {E.}~\bibnamefont {Takala}}, \bibinfo {author} {\bibfnamefont
  {B.}~\bibnamefont {Tarasinski}}, \bibinfo {author} {\bibfnamefont {M.~J.}\
  \bibnamefont {Thapa}}, \bibinfo {author} {\bibfnamefont {F.}~\bibnamefont
  {Tosto}}, \bibinfo {author} {\bibfnamefont {N.}~\bibnamefont {Vorobeva}},
  \bibinfo {author} {\bibfnamefont {L.}~\bibnamefont {Yu}}, \bibinfo {author}
  {\bibfnamefont {K.~Y.}\ \bibnamefont {Tan}}, \bibinfo {author} {\bibfnamefont
  {J.}~\bibnamefont {Hassel}}, \bibinfo {author} {\bibfnamefont
  {M.}~\bibnamefont {M\"{o}tt\"{o}nen}},\ and\ \bibinfo {author} {\bibfnamefont
  {J.}~\bibnamefont {Heinsoo}},\ }\bibfield  {title} {\bibinfo {title}
  {Long-distance transmon coupler with cz-gate fidelity above 99.8\%},\ }\href
  {https://doi.org/10.1103/PRXQuantum.4.010314} {\bibfield  {journal} {\bibinfo
   {journal} {PRX Quantum}\ }\textbf {\bibinfo {volume} {4}},\ \bibinfo {pages}
  {010314} (\bibinfo {year} {2023})}\BibitemShut {NoStop}%
\bibitem [{\citenamefont {Bolgar}\ \emph {et~al.}(2025)\citenamefont {Bolgar},
  \citenamefont {Kalacheva}, \citenamefont {Lubsanov}, \citenamefont
  {Dmitriev}, \citenamefont {Alekseeva}, \citenamefont {Korostylev},\ and\
  \citenamefont {Astafiev}}]{Bolgar2025StiffenerAirbridges}%
  \BibitemOpen
  \bibfield  {author} {\bibinfo {author} {\bibfnamefont {A.~N.}\ \bibnamefont
  {Bolgar}}, \bibinfo {author} {\bibfnamefont {D.~A.}\ \bibnamefont
  {Kalacheva}}, \bibinfo {author} {\bibfnamefont {V.~B.}\ \bibnamefont
  {Lubsanov}}, \bibinfo {author} {\bibfnamefont {A.~Y.}\ \bibnamefont
  {Dmitriev}}, \bibinfo {author} {\bibfnamefont {E.~S.}\ \bibnamefont
  {Alekseeva}}, \bibinfo {author} {\bibfnamefont {E.~V.}\ \bibnamefont
  {Korostylev}},\ and\ \bibinfo {author} {\bibfnamefont {O.~V.}\ \bibnamefont
  {Astafiev}},\ }\bibfield  {title} {\bibinfo {title} {Highly stable aluminum
  air-bridges with stiffeners},\ }\href {https://doi.org/10.1063/5.0250195}
  {\bibfield  {journal} {\bibinfo  {journal} {Journal of Applied Physics}\
  }\textbf {\bibinfo {volume} {137}},\ \bibinfo {pages} {154401} (\bibinfo
  {year} {2025})}\BibitemShut {NoStop}%
\bibitem [{\citenamefont {Chen}(2026)}]{Chen2026InPrep}%
  \BibitemOpen
  \bibfield  {author} {\bibinfo {author} {\bibfnamefont {B.}~\bibnamefont
  {Chen}}} (\bibinfo {year} {2026}),\ \bibinfo {note} {in
  preparation}\BibitemShut {NoStop}%
\bibitem [{\citenamefont {Renger}\ \emph {et~al.}(2026)\citenamefont {Renger},
  \citenamefont {Verjauw}, \citenamefont {Wurz}, \citenamefont {Hosseinkhani},
  \citenamefont {Ockeloen-Korppi}, \citenamefont {Liu}, \citenamefont {Rath},
  \citenamefont {Thapa}, \citenamefont {Vigneau}, \citenamefont {Wybo},
  \citenamefont {Bergholm}, \citenamefont {Chan}, \citenamefont {Csatári},
  \citenamefont {Dahl}, \citenamefont {Davletkaliyev}, \citenamefont {Giri},
  \citenamefont {Gusenkova}, \citenamefont {Heimonen}, \citenamefont
  {Hiltunen}, \citenamefont {Hsu}, \citenamefont {Hyypp\"{a}}, \citenamefont
  {Ikonen}, \citenamefont {Jones}, \citenamefont {Khalid}, \citenamefont {Kim},
  \citenamefont {Koistinen}, \citenamefont {Komlev}, \citenamefont {Kotilahti},
  \citenamefont {Kukushkin}, \citenamefont {Lamprich}, \citenamefont {Landra},
  \citenamefont {Lee}, \citenamefont {Li}, \citenamefont {Liebermann},
  \citenamefont {Majumder}, \citenamefont {M\"{a}ntyl\"{a}}, \citenamefont
  {Marxer}, \citenamefont {van~de Griend}, \citenamefont {Milchakov},
  \citenamefont {Mrożek}, \citenamefont {Nath}, \citenamefont {Orell},
  \citenamefont {Papič}, \citenamefont {Partanen}, \citenamefont {Plyushch},
  \citenamefont {Pogorzalek}, \citenamefont {Ritvas}, \citenamefont {Romero},
  \citenamefont {Sampo}, \citenamefont {Sepp\"{a}l\"{a}}, \citenamefont
  {Selinmaa}, \citenamefont {Sundstr\"{o}m}, \citenamefont {Takmakov},
  \citenamefont {Tarasinski}, \citenamefont {Tuorila}, \citenamefont
  {Tyrkk\"{o}}, \citenamefont {V\"{a}limaa}, \citenamefont {Wesdorp},
  \citenamefont {Yang}, \citenamefont {Yu}, \citenamefont {Heinsoo},
  \citenamefont {Veps\"{a}l\"{a}inen}, \citenamefont {Kindel}, \citenamefont
  {Ku},\ and\ \citenamefont {Deppe}}]{Renger2026AllToAll}%
  \BibitemOpen
  \bibfield  {author} {\bibinfo {author} {\bibfnamefont {M.}~\bibnamefont
  {Renger}}, \bibinfo {author} {\bibfnamefont {J.}~\bibnamefont {Verjauw}},
  \bibinfo {author} {\bibfnamefont {N.}~\bibnamefont {Wurz}}, \bibinfo {author}
  {\bibfnamefont {A.}~\bibnamefont {Hosseinkhani}}, \bibinfo {author}
  {\bibfnamefont {C.}~\bibnamefont {Ockeloen-Korppi}}, \bibinfo {author}
  {\bibfnamefont {W.}~\bibnamefont {Liu}}, \bibinfo {author} {\bibfnamefont
  {A.}~\bibnamefont {Rath}}, \bibinfo {author} {\bibfnamefont {M.~J.}\
  \bibnamefont {Thapa}}, \bibinfo {author} {\bibfnamefont {F.}~\bibnamefont
  {Vigneau}}, \bibinfo {author} {\bibfnamefont {E.}~\bibnamefont {Wybo}},
  \bibinfo {author} {\bibfnamefont {V.}~\bibnamefont {Bergholm}}, \bibinfo
  {author} {\bibfnamefont {C.~F.}\ \bibnamefont {Chan}}, \bibinfo {author}
  {\bibfnamefont {B.}~\bibnamefont {Csatári}}, \bibinfo {author}
  {\bibfnamefont {S.}~\bibnamefont {Dahl}}, \bibinfo {author} {\bibfnamefont
  {R.}~\bibnamefont {Davletkaliyev}}, \bibinfo {author} {\bibfnamefont
  {R.}~\bibnamefont {Giri}}, \bibinfo {author} {\bibfnamefont {D.}~\bibnamefont
  {Gusenkova}}, \bibinfo {author} {\bibfnamefont {H.}~\bibnamefont {Heimonen}},
  \bibinfo {author} {\bibfnamefont {T.}~\bibnamefont {Hiltunen}}, \bibinfo
  {author} {\bibfnamefont {H.}~\bibnamefont {Hsu}}, \bibinfo {author}
  {\bibfnamefont {E.}~\bibnamefont {Hyypp\"{a}}}, \bibinfo {author}
  {\bibfnamefont {J.}~\bibnamefont {Ikonen}}, \bibinfo {author} {\bibfnamefont
  {T.}~\bibnamefont {Jones}}, \bibinfo {author} {\bibfnamefont
  {S.}~\bibnamefont {Khalid}}, \bibinfo {author} {\bibfnamefont {S.-G.}\
  \bibnamefont {Kim}}, \bibinfo {author} {\bibfnamefont {M.}~\bibnamefont
  {Koistinen}}, \bibinfo {author} {\bibfnamefont {A.}~\bibnamefont {Komlev}},
  \bibinfo {author} {\bibfnamefont {J.}~\bibnamefont {Kotilahti}}, \bibinfo
  {author} {\bibfnamefont {V.}~\bibnamefont {Kukushkin}}, \bibinfo {author}
  {\bibfnamefont {J.}~\bibnamefont {Lamprich}}, \bibinfo {author}
  {\bibfnamefont {A.}~\bibnamefont {Landra}}, \bibinfo {author} {\bibfnamefont
  {L.-H.}\ \bibnamefont {Lee}}, \bibinfo {author} {\bibfnamefont
  {T.}~\bibnamefont {Li}}, \bibinfo {author} {\bibfnamefont {P.}~\bibnamefont
  {Liebermann}}, \bibinfo {author} {\bibfnamefont {S.}~\bibnamefont
  {Majumder}}, \bibinfo {author} {\bibfnamefont {J.}~\bibnamefont
  {M\"{a}ntyl\"{a}}}, \bibinfo {author} {\bibfnamefont {F.}~\bibnamefont
  {Marxer}}, \bibinfo {author} {\bibfnamefont {A.~M.}\ \bibnamefont {van~de
  Griend}}, \bibinfo {author} {\bibfnamefont {V.}~\bibnamefont {Milchakov}},
  \bibinfo {author} {\bibfnamefont {J.}~\bibnamefont {Mrożek}}, \bibinfo
  {author} {\bibfnamefont {J.}~\bibnamefont {Nath}}, \bibinfo {author}
  {\bibfnamefont {T.}~\bibnamefont {Orell}}, \bibinfo {author} {\bibfnamefont
  {M.}~\bibnamefont {Papič}}, \bibinfo {author} {\bibfnamefont
  {M.}~\bibnamefont {Partanen}}, \bibinfo {author} {\bibfnamefont
  {A.}~\bibnamefont {Plyushch}}, \bibinfo {author} {\bibfnamefont
  {S.}~\bibnamefont {Pogorzalek}}, \bibinfo {author} {\bibfnamefont
  {J.}~\bibnamefont {Ritvas}}, \bibinfo {author} {\bibfnamefont {P.~F.}\
  \bibnamefont {Romero}}, \bibinfo {author} {\bibfnamefont {V.}~\bibnamefont
  {Sampo}}, \bibinfo {author} {\bibfnamefont {M.}~\bibnamefont
  {Sepp\"{a}l\"{a}}}, \bibinfo {author} {\bibfnamefont {V.}~\bibnamefont
  {Selinmaa}}, \bibinfo {author} {\bibfnamefont {L.}~\bibnamefont
  {Sundstr\"{o}m}}, \bibinfo {author} {\bibfnamefont {I.}~\bibnamefont
  {Takmakov}}, \bibinfo {author} {\bibfnamefont {B.}~\bibnamefont
  {Tarasinski}}, \bibinfo {author} {\bibfnamefont {J.}~\bibnamefont {Tuorila}},
  \bibinfo {author} {\bibfnamefont {O.}~\bibnamefont {Tyrkk\"{o}}}, \bibinfo
  {author} {\bibfnamefont {A.}~\bibnamefont {V\"{a}limaa}}, \bibinfo {author}
  {\bibfnamefont {J.}~\bibnamefont {Wesdorp}}, \bibinfo {author} {\bibfnamefont
  {P.}~\bibnamefont {Yang}}, \bibinfo {author} {\bibfnamefont {L.}~\bibnamefont
  {Yu}}, \bibinfo {author} {\bibfnamefont {J.}~\bibnamefont {Heinsoo}},
  \bibinfo {author} {\bibfnamefont {A.}~\bibnamefont {Veps\"{a}l\"{a}inen}},
  \bibinfo {author} {\bibfnamefont {W.}~\bibnamefont {Kindel}}, \bibinfo
  {author} {\bibfnamefont {H.-S.}\ \bibnamefont {Ku}},\ and\ \bibinfo {author}
  {\bibfnamefont {F.}~\bibnamefont {Deppe}},\ }\bibfield  {title} {\bibinfo
  {title} {Superconducting qubit-resonator quantum processor with effective
  all-to-all connectivity},\ }\href {https://doi.org/10.1103/8x68-cx1w}
  {\bibfield  {journal} {\bibinfo  {journal} {Phys. Rev. Res.}\ }\textbf
  {\bibinfo {volume} {8}},\ \bibinfo {pages} {013148} (\bibinfo {year}
  {2026})}\BibitemShut {NoStop}%
\bibitem [{\citenamefont {Song}\ \emph {et~al.}(2017)\citenamefont {Song},
  \citenamefont {Xu}, \citenamefont {Liu}, \citenamefont {Yang}, \citenamefont
  {Zheng}, \citenamefont {Deng}, \citenamefont {Xie}, \citenamefont {Huang},
  \citenamefont {Guo}, \citenamefont {Zhang}, \citenamefont {Zhang},
  \citenamefont {Xu}, \citenamefont {Zheng}, \citenamefont {Zhu}, \citenamefont
  {Wang}, \citenamefont {Chen}, \citenamefont {Lu}, \citenamefont {Han},\ and\
  \citenamefont {Pan}}]{Song2017TenQubit}%
  \BibitemOpen
  \bibfield  {author} {\bibinfo {author} {\bibfnamefont {C.}~\bibnamefont
  {Song}}, \bibinfo {author} {\bibfnamefont {K.}~\bibnamefont {Xu}}, \bibinfo
  {author} {\bibfnamefont {W.}~\bibnamefont {Liu}}, \bibinfo {author}
  {\bibfnamefont {C.-p.}\ \bibnamefont {Yang}}, \bibinfo {author}
  {\bibfnamefont {S.-B.}\ \bibnamefont {Zheng}}, \bibinfo {author}
  {\bibfnamefont {H.}~\bibnamefont {Deng}}, \bibinfo {author} {\bibfnamefont
  {Q.}~\bibnamefont {Xie}}, \bibinfo {author} {\bibfnamefont {K.}~\bibnamefont
  {Huang}}, \bibinfo {author} {\bibfnamefont {Q.}~\bibnamefont {Guo}}, \bibinfo
  {author} {\bibfnamefont {L.}~\bibnamefont {Zhang}}, \bibinfo {author}
  {\bibfnamefont {P.}~\bibnamefont {Zhang}}, \bibinfo {author} {\bibfnamefont
  {D.}~\bibnamefont {Xu}}, \bibinfo {author} {\bibfnamefont {D.}~\bibnamefont
  {Zheng}}, \bibinfo {author} {\bibfnamefont {X.}~\bibnamefont {Zhu}}, \bibinfo
  {author} {\bibfnamefont {H.}~\bibnamefont {Wang}}, \bibinfo {author}
  {\bibfnamefont {Y.-A.}\ \bibnamefont {Chen}}, \bibinfo {author}
  {\bibfnamefont {C.-Y.}\ \bibnamefont {Lu}}, \bibinfo {author} {\bibfnamefont
  {S.}~\bibnamefont {Han}},\ and\ \bibinfo {author} {\bibfnamefont {J.-W.}\
  \bibnamefont {Pan}},\ }\bibfield  {title} {\bibinfo {title} {10-qubit
  entanglement and parallel logic operations with a superconducting circuit},\
  }\href {https://doi.org/10.1103/PhysRevLett.119.180511} {\bibfield  {journal}
  {\bibinfo  {journal} {Phys. Rev. Lett.}\ }\textbf {\bibinfo {volume} {119}},\
  \bibinfo {pages} {180511} (\bibinfo {year} {2017})}\BibitemShut {NoStop}%
\bibitem [{\citenamefont {Song}\ \emph {et~al.}(2019)\citenamefont {Song},
  \citenamefont {Xu}, \citenamefont {Li}, \citenamefont {Zhang}, \citenamefont
  {Zhang}, \citenamefont {Liu}, \citenamefont {Guo}, \citenamefont {Wang},
  \citenamefont {Ren}, \citenamefont {Hao}, \citenamefont {Feng}, \citenamefont
  {Fan}, \citenamefont {Zheng}, \citenamefont {Wang}, \citenamefont {Wang},\
  and\ \citenamefont {Zhu}}]{Song2019TwentyQubit}%
  \BibitemOpen
  \bibfield  {author} {\bibinfo {author} {\bibfnamefont {C.}~\bibnamefont
  {Song}}, \bibinfo {author} {\bibfnamefont {K.}~\bibnamefont {Xu}}, \bibinfo
  {author} {\bibfnamefont {H.}~\bibnamefont {Li}}, \bibinfo {author}
  {\bibfnamefont {Y.-R.}\ \bibnamefont {Zhang}}, \bibinfo {author}
  {\bibfnamefont {X.}~\bibnamefont {Zhang}}, \bibinfo {author} {\bibfnamefont
  {W.}~\bibnamefont {Liu}}, \bibinfo {author} {\bibfnamefont {Q.}~\bibnamefont
  {Guo}}, \bibinfo {author} {\bibfnamefont {Z.}~\bibnamefont {Wang}}, \bibinfo
  {author} {\bibfnamefont {W.}~\bibnamefont {Ren}}, \bibinfo {author}
  {\bibfnamefont {J.}~\bibnamefont {Hao}}, \bibinfo {author} {\bibfnamefont
  {H.}~\bibnamefont {Feng}}, \bibinfo {author} {\bibfnamefont {H.}~\bibnamefont
  {Fan}}, \bibinfo {author} {\bibfnamefont {D.}~\bibnamefont {Zheng}}, \bibinfo
  {author} {\bibfnamefont {D.-W.}\ \bibnamefont {Wang}}, \bibinfo {author}
  {\bibfnamefont {H.}~\bibnamefont {Wang}},\ and\ \bibinfo {author}
  {\bibfnamefont {S.-Y.}\ \bibnamefont {Zhu}},\ }\bibfield  {title} {\bibinfo
  {title} {Generation of multicomponent atomic schr{\"o}dinger cat states of up
  to 20 qubits},\ }\href {https://doi.org/10.1126/science.aay0600} {\bibfield
  {journal} {\bibinfo  {journal} {Science}\ }\textbf {\bibinfo {volume}
  {365}},\ \bibinfo {pages} {574} (\bibinfo {year} {2019})}\BibitemShut
  {NoStop}%
\bibitem [{\citenamefont {Yan}\ \emph {et~al.}(2018)\citenamefont {Yan},
  \citenamefont {Krantz}, \citenamefont {Sung}, \citenamefont {Kjaergaard},
  \citenamefont {Campbell}, \citenamefont {Orlando}, \citenamefont
  {Gustavsson},\ and\ \citenamefont {Oliver}}]{Yan2018TunableCoupling}%
  \BibitemOpen
  \bibfield  {author} {\bibinfo {author} {\bibfnamefont {F.}~\bibnamefont
  {Yan}}, \bibinfo {author} {\bibfnamefont {P.}~\bibnamefont {Krantz}},
  \bibinfo {author} {\bibfnamefont {Y.}~\bibnamefont {Sung}}, \bibinfo {author}
  {\bibfnamefont {M.}~\bibnamefont {Kjaergaard}}, \bibinfo {author}
  {\bibfnamefont {D.~L.}\ \bibnamefont {Campbell}}, \bibinfo {author}
  {\bibfnamefont {T.~P.}\ \bibnamefont {Orlando}}, \bibinfo {author}
  {\bibfnamefont {S.}~\bibnamefont {Gustavsson}},\ and\ \bibinfo {author}
  {\bibfnamefont {W.~D.}\ \bibnamefont {Oliver}},\ }\bibfield  {title}
  {\bibinfo {title} {Tunable coupling scheme for implementing high-fidelity
  two-qubit gates},\ }\href {https://doi.org/10.1103/PhysRevApplied.10.054062}
  {\bibfield  {journal} {\bibinfo  {journal} {Phys. Rev. Appl.}\ }\textbf
  {\bibinfo {volume} {10}},\ \bibinfo {pages} {054062} (\bibinfo {year}
  {2018})}\BibitemShut {NoStop}%
\bibitem [{\citenamefont {Kandala}\ \emph {et~al.}(2021)\citenamefont
  {Kandala}, \citenamefont {Wei}, \citenamefont {Srinivasan}, \citenamefont
  {Magesan}, \citenamefont {Carnevale}, \citenamefont {Keefe}, \citenamefont
  {Klaus}, \citenamefont {Dial},\ and\ \citenamefont
  {McKay}}]{Kandala2021ZZSuppression}%
  \BibitemOpen
  \bibfield  {author} {\bibinfo {author} {\bibfnamefont {A.}~\bibnamefont
  {Kandala}}, \bibinfo {author} {\bibfnamefont {K.~X.}\ \bibnamefont {Wei}},
  \bibinfo {author} {\bibfnamefont {S.}~\bibnamefont {Srinivasan}}, \bibinfo
  {author} {\bibfnamefont {E.}~\bibnamefont {Magesan}}, \bibinfo {author}
  {\bibfnamefont {S.}~\bibnamefont {Carnevale}}, \bibinfo {author}
  {\bibfnamefont {G.~A.}\ \bibnamefont {Keefe}}, \bibinfo {author}
  {\bibfnamefont {D.}~\bibnamefont {Klaus}}, \bibinfo {author} {\bibfnamefont
  {O.}~\bibnamefont {Dial}},\ and\ \bibinfo {author} {\bibfnamefont {D.~C.}\
  \bibnamefont {McKay}},\ }\bibfield  {title} {\bibinfo {title} {Demonstration
  of a high-fidelity cnot gate for fixed-frequency transmons with engineered
  $zz$ suppression},\ }\href {https://doi.org/10.1103/PhysRevLett.127.130501}
  {\bibfield  {journal} {\bibinfo  {journal} {Phys. Rev. Lett.}\ }\textbf
  {\bibinfo {volume} {127}},\ \bibinfo {pages} {130501} (\bibinfo {year}
  {2021})}\BibitemShut {NoStop}%
\bibitem [{\citenamefont {Stehlik}\ \emph {et~al.}(2021)\citenamefont
  {Stehlik}, \citenamefont {Zajac}, \citenamefont {Underwood}, \citenamefont
  {Phung}, \citenamefont {Blair}, \citenamefont {Carnevale}, \citenamefont
  {Klaus}, \citenamefont {Keefe}, \citenamefont {Carniol}, \citenamefont
  {Kumph}, \citenamefont {Steffen},\ and\ \citenamefont
  {Dial}}]{Stehlik2021TunableCoupling}%
  \BibitemOpen
  \bibfield  {author} {\bibinfo {author} {\bibfnamefont {J.}~\bibnamefont
  {Stehlik}}, \bibinfo {author} {\bibfnamefont {D.~M.}\ \bibnamefont {Zajac}},
  \bibinfo {author} {\bibfnamefont {D.~L.}\ \bibnamefont {Underwood}}, \bibinfo
  {author} {\bibfnamefont {T.}~\bibnamefont {Phung}}, \bibinfo {author}
  {\bibfnamefont {J.}~\bibnamefont {Blair}}, \bibinfo {author} {\bibfnamefont
  {S.}~\bibnamefont {Carnevale}}, \bibinfo {author} {\bibfnamefont
  {D.}~\bibnamefont {Klaus}}, \bibinfo {author} {\bibfnamefont {G.~A.}\
  \bibnamefont {Keefe}}, \bibinfo {author} {\bibfnamefont {A.}~\bibnamefont
  {Carniol}}, \bibinfo {author} {\bibfnamefont {M.}~\bibnamefont {Kumph}},
  \bibinfo {author} {\bibfnamefont {M.}~\bibnamefont {Steffen}},\ and\ \bibinfo
  {author} {\bibfnamefont {O.~E.}\ \bibnamefont {Dial}},\ }\bibfield  {title}
  {\bibinfo {title} {Tunable coupling architecture for fixed-frequency transmon
  superconducting qubits},\ }\href
  {https://doi.org/10.1103/PhysRevLett.127.080505} {\bibfield  {journal}
  {\bibinfo  {journal} {Phys. Rev. Lett.}\ }\textbf {\bibinfo {volume} {127}},\
  \bibinfo {pages} {080505} (\bibinfo {year} {2021})}\BibitemShut {NoStop}%
\bibitem [{\citenamefont {Deng}\ \emph {et~al.}(2024)\citenamefont {Deng},
  \citenamefont {Liu}, \citenamefont {Zhang}, \citenamefont {Li}, \citenamefont
  {Liu}, \citenamefont {Chen}, \citenamefont {Liu}, \citenamefont {Lu},
  \citenamefont {Wang}, \citenamefont {Li}, \citenamefont {Fang}, \citenamefont
  {Zhou}, \citenamefont {Song}, \citenamefont {Xu}, \citenamefont {He},
  \citenamefont {Liu}, \citenamefont {Huang}, \citenamefont {Xiang},
  \citenamefont {Wang}, \citenamefont {Zheng}, \citenamefont {Xue},
  \citenamefont {Xu}, \citenamefont {Yu},\ and\ \citenamefont
  {Fan}}]{Deng2024HighOrderPumping}%
  \BibitemOpen
  \bibfield  {author} {\bibinfo {author} {\bibfnamefont {C.-L.}\ \bibnamefont
  {Deng}}, \bibinfo {author} {\bibfnamefont {Y.}~\bibnamefont {Liu}}, \bibinfo
  {author} {\bibfnamefont {Y.-R.}\ \bibnamefont {Zhang}}, \bibinfo {author}
  {\bibfnamefont {X.-G.}\ \bibnamefont {Li}}, \bibinfo {author} {\bibfnamefont
  {T.}~\bibnamefont {Liu}}, \bibinfo {author} {\bibfnamefont {C.-T.}\
  \bibnamefont {Chen}}, \bibinfo {author} {\bibfnamefont {T.}~\bibnamefont
  {Liu}}, \bibinfo {author} {\bibfnamefont {C.-W.}\ \bibnamefont {Lu}},
  \bibinfo {author} {\bibfnamefont {Y.-Y.}\ \bibnamefont {Wang}}, \bibinfo
  {author} {\bibfnamefont {T.-M.}\ \bibnamefont {Li}}, \bibinfo {author}
  {\bibfnamefont {C.-P.}\ \bibnamefont {Fang}}, \bibinfo {author}
  {\bibfnamefont {S.-Y.}\ \bibnamefont {Zhou}}, \bibinfo {author}
  {\bibfnamefont {J.-C.}\ \bibnamefont {Song}}, \bibinfo {author}
  {\bibfnamefont {Y.-S.}\ \bibnamefont {Xu}}, \bibinfo {author} {\bibfnamefont
  {Y.}~\bibnamefont {He}}, \bibinfo {author} {\bibfnamefont {Z.-H.}\
  \bibnamefont {Liu}}, \bibinfo {author} {\bibfnamefont {K.-X.}\ \bibnamefont
  {Huang}}, \bibinfo {author} {\bibfnamefont {Z.-C.}\ \bibnamefont {Xiang}},
  \bibinfo {author} {\bibfnamefont {J.-C.}\ \bibnamefont {Wang}}, \bibinfo
  {author} {\bibfnamefont {D.-N.}\ \bibnamefont {Zheng}}, \bibinfo {author}
  {\bibfnamefont {G.-M.}\ \bibnamefont {Xue}}, \bibinfo {author} {\bibfnamefont
  {K.}~\bibnamefont {Xu}}, \bibinfo {author} {\bibfnamefont {H.}~\bibnamefont
  {Yu}},\ and\ \bibinfo {author} {\bibfnamefont {H.}~\bibnamefont {Fan}},\
  }\bibfield  {title} {\bibinfo {title} {High-order topological pumping on a
  superconducting quantum processor},\ }\href
  {https://doi.org/10.1103/PhysRevLett.133.140402} {\bibfield  {journal}
  {\bibinfo  {journal} {Phys. Rev. Lett.}\ }\textbf {\bibinfo {volume} {133}},\
  \bibinfo {pages} {140402} (\bibinfo {year} {2024})}\BibitemShut {NoStop}%
\bibitem [{\citenamefont {Zhang}\ \emph {et~al.}(2024)\citenamefont {Zhang},
  \citenamefont {Gao}, \citenamefont {Xu}, \citenamefont {Wang}, \citenamefont
  {Dong}, \citenamefont {Guo}, \citenamefont {Deng}, \citenamefont {Zhang},
  \citenamefont {Chen}, \citenamefont {Xu}, \citenamefont {Wang}, \citenamefont
  {Wu}, \citenamefont {Zhang}, \citenamefont {Jin}, \citenamefont {Zhu},
  \citenamefont {Zhang}, \citenamefont {Zou}, \citenamefont {Tan},
  \citenamefont {Cui}, \citenamefont {Zhu}, \citenamefont {Shen}, \citenamefont
  {Li}, \citenamefont {Zhong}, \citenamefont {Bao}, \citenamefont {Zhao},
  \citenamefont {Hao}, \citenamefont {Li}, \citenamefont {Wang}, \citenamefont
  {Song}, \citenamefont {Guo}, \citenamefont {Wang},\ and\ \citenamefont
  {Poletti}}]{zhang2024emergence}%
  \BibitemOpen
  \bibfield  {author} {\bibinfo {author} {\bibfnamefont {P.}~\bibnamefont
  {Zhang}}, \bibinfo {author} {\bibfnamefont {Y.}~\bibnamefont {Gao}}, \bibinfo
  {author} {\bibfnamefont {X.}~\bibnamefont {Xu}}, \bibinfo {author}
  {\bibfnamefont {N.}~\bibnamefont {Wang}}, \bibinfo {author} {\bibfnamefont
  {H.}~\bibnamefont {Dong}}, \bibinfo {author} {\bibfnamefont {C.}~\bibnamefont
  {Guo}}, \bibinfo {author} {\bibfnamefont {J.}~\bibnamefont {Deng}}, \bibinfo
  {author} {\bibfnamefont {X.}~\bibnamefont {Zhang}}, \bibinfo {author}
  {\bibfnamefont {J.}~\bibnamefont {Chen}}, \bibinfo {author} {\bibfnamefont
  {S.}~\bibnamefont {Xu}}, \bibinfo {author} {\bibfnamefont {K.}~\bibnamefont
  {Wang}}, \bibinfo {author} {\bibfnamefont {Y.}~\bibnamefont {Wu}}, \bibinfo
  {author} {\bibfnamefont {C.}~\bibnamefont {Zhang}}, \bibinfo {author}
  {\bibfnamefont {F.}~\bibnamefont {Jin}}, \bibinfo {author} {\bibfnamefont
  {X.}~\bibnamefont {Zhu}}, \bibinfo {author} {\bibfnamefont {A.}~\bibnamefont
  {Zhang}}, \bibinfo {author} {\bibfnamefont {Y.}~\bibnamefont {Zou}}, \bibinfo
  {author} {\bibfnamefont {Z.}~\bibnamefont {Tan}}, \bibinfo {author}
  {\bibfnamefont {Z.}~\bibnamefont {Cui}}, \bibinfo {author} {\bibfnamefont
  {Z.}~\bibnamefont {Zhu}}, \bibinfo {author} {\bibfnamefont {F.}~\bibnamefont
  {Shen}}, \bibinfo {author} {\bibfnamefont {T.}~\bibnamefont {Li}}, \bibinfo
  {author} {\bibfnamefont {J.}~\bibnamefont {Zhong}}, \bibinfo {author}
  {\bibfnamefont {Z.}~\bibnamefont {Bao}}, \bibinfo {author} {\bibfnamefont
  {L.}~\bibnamefont {Zhao}}, \bibinfo {author} {\bibfnamefont {J.}~\bibnamefont
  {Hao}}, \bibinfo {author} {\bibfnamefont {H.}~\bibnamefont {Li}}, \bibinfo
  {author} {\bibfnamefont {Z.}~\bibnamefont {Wang}}, \bibinfo {author}
  {\bibfnamefont {C.}~\bibnamefont {Song}}, \bibinfo {author} {\bibfnamefont
  {Q.}~\bibnamefont {Guo}}, \bibinfo {author} {\bibfnamefont {H.}~\bibnamefont
  {Wang}},\ and\ \bibinfo {author} {\bibfnamefont {D.}~\bibnamefont
  {Poletti}},\ }\bibfield  {title} {\bibinfo {title} {Emergence of steady
  quantum transport in a superconducting processor},\ }\href
  {https://doi.org/10.1038/s41467-024-54332-9} {\bibfield  {journal} {\bibinfo
  {journal} {Nature Communications}\ }\textbf {\bibinfo {volume} {15}},\
  \bibinfo {pages} {10115} (\bibinfo {year} {2024})}\BibitemShut {NoStop}%
\bibitem [{\citenamefont {Wang}\ \emph
  {et~al.}(2026{\natexlab{b}})\citenamefont {Wang}, \citenamefont {Zhou},
  \citenamefont {Shi}, \citenamefont {Huang}, \citenamefont {Yang},
  \citenamefont {Zhang}, \citenamefont {Zhao}, \citenamefont {Xu},
  \citenamefont {Li}, \citenamefont {Zhao}, \citenamefont {Feng}, \citenamefont
  {Xue}, \citenamefont {Liu}, \citenamefont {Ma}, \citenamefont {Fang},
  \citenamefont {Liu}, \citenamefont {Wang}, \citenamefont {Xu}, \citenamefont
  {Yu}, \citenamefont {Fan},\ and\ \citenamefont
  {Zhao}}]{Wang2025TwoParticleCorrelation}%
  \BibitemOpen
  \bibfield  {author} {\bibinfo {author} {\bibfnamefont {Z.~T.}\ \bibnamefont
  {Wang}}, \bibinfo {author} {\bibfnamefont {S.-Y.}\ \bibnamefont {Zhou}},
  \bibinfo {author} {\bibfnamefont {Y.-H.}\ \bibnamefont {Shi}}, \bibinfo
  {author} {\bibfnamefont {K.}~\bibnamefont {Huang}}, \bibinfo {author}
  {\bibfnamefont {Z.~H.}\ \bibnamefont {Yang}}, \bibinfo {author}
  {\bibfnamefont {J.}~\bibnamefont {Zhang}}, \bibinfo {author} {\bibfnamefont
  {K.}~\bibnamefont {Zhao}}, \bibinfo {author} {\bibfnamefont {Y.}~\bibnamefont
  {Xu}}, \bibinfo {author} {\bibfnamefont {H.}~\bibnamefont {Li}}, \bibinfo
  {author} {\bibfnamefont {S.~K.}\ \bibnamefont {Zhao}}, \bibinfo {author}
  {\bibfnamefont {Y.}~\bibnamefont {Feng}}, \bibinfo {author} {\bibfnamefont
  {G.}~\bibnamefont {Xue}}, \bibinfo {author} {\bibfnamefont {Y.}~\bibnamefont
  {Liu}}, \bibinfo {author} {\bibfnamefont {W.-G.}\ \bibnamefont {Ma}},
  \bibinfo {author} {\bibfnamefont {C.-P.}\ \bibnamefont {Fang}}, \bibinfo
  {author} {\bibfnamefont {H.-T.}\ \bibnamefont {Liu}}, \bibinfo {author}
  {\bibfnamefont {Y.-Y.}\ \bibnamefont {Wang}}, \bibinfo {author}
  {\bibfnamefont {K.}~\bibnamefont {Xu}}, \bibinfo {author} {\bibfnamefont
  {H.}~\bibnamefont {Yu}}, \bibinfo {author} {\bibfnamefont {H.}~\bibnamefont
  {Fan}},\ and\ \bibinfo {author} {\bibfnamefont {S.~P.}\ \bibnamefont
  {Zhao}},\ }\href {https://arxiv.org/abs/2509.02180} {\bibinfo {title}
  {Observing quantum correlation dynamics in tunable superconducting
  bose-hubbard simulators}} (\bibinfo {year} {2026}{\natexlab{b}}),\ \Eprint
  {https://arxiv.org/abs/2509.02180} {arXiv:2509.02180 [quant-ph]} \BibitemShut
  {NoStop}%
\bibitem [{\citenamefont {Tao}\ \emph {et~al.}(2025)\citenamefont {Tao},
  \citenamefont {Huang}, \citenamefont {Niu}, \citenamefont {Zhang},
  \citenamefont {Ke}, \citenamefont {Gu}, \citenamefont {Lin}, \citenamefont
  {Qiu}, \citenamefont {Sun}, \citenamefont {Yang}, \citenamefont {Zhang},
  \citenamefont {Zhang}, \citenamefont {Zhou}, \citenamefont {Deng},
  \citenamefont {Hu}, \citenamefont {Hu}, \citenamefont {Li}, \citenamefont
  {Liu}, \citenamefont {Tan}, \citenamefont {Xu}, \citenamefont {Yan},
  \citenamefont {Chen}, \citenamefont {Lee}, \citenamefont {Zhong},
  \citenamefont {Liu},\ and\ \citenamefont {Yu}}]{Tao2025Emulating}%
  \BibitemOpen
  \bibfield  {author} {\bibinfo {author} {\bibfnamefont {Z.}~\bibnamefont
  {Tao}}, \bibinfo {author} {\bibfnamefont {W.}~\bibnamefont {Huang}}, \bibinfo
  {author} {\bibfnamefont {J.}~\bibnamefont {Niu}}, \bibinfo {author}
  {\bibfnamefont {L.}~\bibnamefont {Zhang}}, \bibinfo {author} {\bibfnamefont
  {Y.}~\bibnamefont {Ke}}, \bibinfo {author} {\bibfnamefont {X.}~\bibnamefont
  {Gu}}, \bibinfo {author} {\bibfnamefont {L.}~\bibnamefont {Lin}}, \bibinfo
  {author} {\bibfnamefont {J.}~\bibnamefont {Qiu}}, \bibinfo {author}
  {\bibfnamefont {X.}~\bibnamefont {Sun}}, \bibinfo {author} {\bibfnamefont
  {X.}~\bibnamefont {Yang}}, \bibinfo {author} {\bibfnamefont {J.}~\bibnamefont
  {Zhang}}, \bibinfo {author} {\bibfnamefont {J.}~\bibnamefont {Zhang}},
  \bibinfo {author} {\bibfnamefont {Y.}~\bibnamefont {Zhou}}, \bibinfo {author}
  {\bibfnamefont {X.}~\bibnamefont {Deng}}, \bibinfo {author} {\bibfnamefont
  {C.}~\bibnamefont {Hu}}, \bibinfo {author} {\bibfnamefont {L.}~\bibnamefont
  {Hu}}, \bibinfo {author} {\bibfnamefont {J.}~\bibnamefont {Li}}, \bibinfo
  {author} {\bibfnamefont {Y.}~\bibnamefont {Liu}}, \bibinfo {author}
  {\bibfnamefont {D.}~\bibnamefont {Tan}}, \bibinfo {author} {\bibfnamefont
  {Y.}~\bibnamefont {Xu}}, \bibinfo {author} {\bibfnamefont {T.}~\bibnamefont
  {Yan}}, \bibinfo {author} {\bibfnamefont {Y.}~\bibnamefont {Chen}}, \bibinfo
  {author} {\bibfnamefont {C.}~\bibnamefont {Lee}}, \bibinfo {author}
  {\bibfnamefont {Y.}~\bibnamefont {Zhong}}, \bibinfo {author} {\bibfnamefont
  {S.}~\bibnamefont {Liu}},\ and\ \bibinfo {author} {\bibfnamefont
  {D.}~\bibnamefont {Yu}},\ }\bibfield  {title} {\bibinfo {title} {Emulating
  {Thouless} pumping in the interacting {Rice-Mele} model using superconducting
  qutrits},\ }\href
  {https://doi.org/https://doi.org/10.15302/frontphys.2025.033202} {\bibfield
  {journal} {\bibinfo  {journal} {Frontiers of Physics}\ }\textbf {\bibinfo
  {volume} {20}},\ \bibinfo {pages} {033202} (\bibinfo {year}
  {2025})}\BibitemShut {NoStop}%
\bibitem [{\citenamefont {Roushan}\ \emph {et~al.}(2017)\citenamefont
  {Roushan}, \citenamefont {Neill}, \citenamefont {Tangpanitanon},
  \citenamefont {Bastidas}, \citenamefont {Megrant}, \citenamefont {Barends},
  \citenamefont {Chen}, \citenamefont {Chen}, \citenamefont {Chiaro},
  \citenamefont {Dunsworth}, \citenamefont {Fowler}, \citenamefont {Foxen},
  \citenamefont {Giustina}, \citenamefont {Jeffrey}, \citenamefont {Kelly},
  \citenamefont {Lucero}, \citenamefont {Mutus}, \citenamefont {Neeley},
  \citenamefont {Quintana}, \citenamefont {Sank}, \citenamefont {Vainsencher},
  \citenamefont {Wenner}, \citenamefont {White}, \citenamefont {Neven},
  \citenamefont {Angelakis},\ and\ \citenamefont
  {Martinis}}]{Roushan2017Spectroscopic}%
  \BibitemOpen
  \bibfield  {author} {\bibinfo {author} {\bibfnamefont {P.}~\bibnamefont
  {Roushan}}, \bibinfo {author} {\bibfnamefont {C.}~\bibnamefont {Neill}},
  \bibinfo {author} {\bibfnamefont {J.}~\bibnamefont {Tangpanitanon}}, \bibinfo
  {author} {\bibfnamefont {V.~M.}\ \bibnamefont {Bastidas}}, \bibinfo {author}
  {\bibfnamefont {A.}~\bibnamefont {Megrant}}, \bibinfo {author} {\bibfnamefont
  {R.}~\bibnamefont {Barends}}, \bibinfo {author} {\bibfnamefont
  {Y.}~\bibnamefont {Chen}}, \bibinfo {author} {\bibfnamefont {Z.}~\bibnamefont
  {Chen}}, \bibinfo {author} {\bibfnamefont {B.}~\bibnamefont {Chiaro}},
  \bibinfo {author} {\bibfnamefont {A.}~\bibnamefont {Dunsworth}}, \bibinfo
  {author} {\bibfnamefont {A.}~\bibnamefont {Fowler}}, \bibinfo {author}
  {\bibfnamefont {B.}~\bibnamefont {Foxen}}, \bibinfo {author} {\bibfnamefont
  {M.}~\bibnamefont {Giustina}}, \bibinfo {author} {\bibfnamefont
  {E.}~\bibnamefont {Jeffrey}}, \bibinfo {author} {\bibfnamefont
  {J.}~\bibnamefont {Kelly}}, \bibinfo {author} {\bibfnamefont
  {E.}~\bibnamefont {Lucero}}, \bibinfo {author} {\bibfnamefont
  {J.}~\bibnamefont {Mutus}}, \bibinfo {author} {\bibfnamefont
  {M.}~\bibnamefont {Neeley}}, \bibinfo {author} {\bibfnamefont
  {C.}~\bibnamefont {Quintana}}, \bibinfo {author} {\bibfnamefont
  {D.}~\bibnamefont {Sank}}, \bibinfo {author} {\bibfnamefont {A.}~\bibnamefont
  {Vainsencher}}, \bibinfo {author} {\bibfnamefont {J.}~\bibnamefont {Wenner}},
  \bibinfo {author} {\bibfnamefont {T.}~\bibnamefont {White}}, \bibinfo
  {author} {\bibfnamefont {H.}~\bibnamefont {Neven}}, \bibinfo {author}
  {\bibfnamefont {D.~G.}\ \bibnamefont {Angelakis}},\ and\ \bibinfo {author}
  {\bibfnamefont {J.}~\bibnamefont {Martinis}},\ }\bibfield  {title} {\bibinfo
  {title} {Spectroscopic signatures of localization with interacting photons in
  superconducting qubits},\ }\href {https://doi.org/10.1126/science.aao1401}
  {\bibfield  {journal} {\bibinfo  {journal} {Science}\ }\textbf {\bibinfo
  {volume} {358}},\ \bibinfo {pages} {1175} (\bibinfo {year}
  {2017})}\BibitemShut {NoStop}%
\bibitem [{\citenamefont {Rosenberg}\ \emph {et~al.}(2017)\citenamefont
  {Rosenberg}, \citenamefont {Kim}, \citenamefont {Das}, \citenamefont {Yost},
  \citenamefont {Gustavsson}, \citenamefont {Hover}, \citenamefont {Krantz},
  \citenamefont {Melville}, \citenamefont {Racz}, \citenamefont {Samach},
  \citenamefont {Weber}, \citenamefont {Yan}, \citenamefont {Yoder},
  \citenamefont {Kerman},\ and\ \citenamefont
  {Oliver}}]{Rosenberg2017FlipChip}%
  \BibitemOpen
  \bibfield  {author} {\bibinfo {author} {\bibfnamefont {D.}~\bibnamefont
  {Rosenberg}}, \bibinfo {author} {\bibfnamefont {D.}~\bibnamefont {Kim}},
  \bibinfo {author} {\bibfnamefont {R.}~\bibnamefont {Das}}, \bibinfo {author}
  {\bibfnamefont {D.}~\bibnamefont {Yost}}, \bibinfo {author} {\bibfnamefont
  {S.}~\bibnamefont {Gustavsson}}, \bibinfo {author} {\bibfnamefont
  {D.}~\bibnamefont {Hover}}, \bibinfo {author} {\bibfnamefont
  {P.}~\bibnamefont {Krantz}}, \bibinfo {author} {\bibfnamefont
  {A.}~\bibnamefont {Melville}}, \bibinfo {author} {\bibfnamefont
  {L.}~\bibnamefont {Racz}}, \bibinfo {author} {\bibfnamefont {G.~O.}\
  \bibnamefont {Samach}}, \bibinfo {author} {\bibfnamefont {S.~J.}\
  \bibnamefont {Weber}}, \bibinfo {author} {\bibfnamefont {F.}~\bibnamefont
  {Yan}}, \bibinfo {author} {\bibfnamefont {J.~L.}\ \bibnamefont {Yoder}},
  \bibinfo {author} {\bibfnamefont {A.~J.}\ \bibnamefont {Kerman}},\ and\
  \bibinfo {author} {\bibfnamefont {W.~D.}\ \bibnamefont {Oliver}},\ }\bibfield
   {title} {\bibinfo {title} {{3D} integrated superconducting qubits},\ }\href
  {https://doi.org/10.1038/s41534-017-0044-0} {\bibfield  {journal} {\bibinfo
  {journal} {npj Quantum Information}\ }\textbf {\bibinfo {volume} {3}},\
  \bibinfo {pages} {42} (\bibinfo {year} {2017})}\BibitemShut {NoStop}%
\bibitem [{\citenamefont {Foxen}\ \emph {et~al.}(2018)\citenamefont {Foxen},
  \citenamefont {Mutus}, \citenamefont {Lucero}, \citenamefont {Graff},
  \citenamefont {Megrant}, \citenamefont {Chen}, \citenamefont {Quintana},
  \citenamefont {Burkett}, \citenamefont {Kelly}, \citenamefont {Jeffrey},
  \citenamefont {Yang}, \citenamefont {Yu}, \citenamefont {Arya}, \citenamefont
  {Barends}, \citenamefont {Chen}, \citenamefont {Chiaro}, \citenamefont
  {Dunsworth}, \citenamefont {Fowler}, \citenamefont {Gidney}, \citenamefont
  {Giustina}, \citenamefont {Huang}, \citenamefont {Klimov}, \citenamefont
  {Neeley}, \citenamefont {Neill}, \citenamefont {Roushan}, \citenamefont
  {Sank}, \citenamefont {Vainsencher}, \citenamefont {Wenner}, \citenamefont
  {White},\ and\ \citenamefont {Martinis}}]{Foxen2018Interconnects}%
  \BibitemOpen
  \bibfield  {author} {\bibinfo {author} {\bibfnamefont {B.}~\bibnamefont
  {Foxen}}, \bibinfo {author} {\bibfnamefont {J.~Y.}\ \bibnamefont {Mutus}},
  \bibinfo {author} {\bibfnamefont {E.}~\bibnamefont {Lucero}}, \bibinfo
  {author} {\bibfnamefont {R.}~\bibnamefont {Graff}}, \bibinfo {author}
  {\bibfnamefont {A.}~\bibnamefont {Megrant}}, \bibinfo {author} {\bibfnamefont
  {Y.}~\bibnamefont {Chen}}, \bibinfo {author} {\bibfnamefont {C.}~\bibnamefont
  {Quintana}}, \bibinfo {author} {\bibfnamefont {B.}~\bibnamefont {Burkett}},
  \bibinfo {author} {\bibfnamefont {J.}~\bibnamefont {Kelly}}, \bibinfo
  {author} {\bibfnamefont {E.}~\bibnamefont {Jeffrey}}, \bibinfo {author}
  {\bibfnamefont {Y.}~\bibnamefont {Yang}}, \bibinfo {author} {\bibfnamefont
  {A.}~\bibnamefont {Yu}}, \bibinfo {author} {\bibfnamefont {K.}~\bibnamefont
  {Arya}}, \bibinfo {author} {\bibfnamefont {R.}~\bibnamefont {Barends}},
  \bibinfo {author} {\bibfnamefont {Z.}~\bibnamefont {Chen}}, \bibinfo {author}
  {\bibfnamefont {B.}~\bibnamefont {Chiaro}}, \bibinfo {author} {\bibfnamefont
  {A.}~\bibnamefont {Dunsworth}}, \bibinfo {author} {\bibfnamefont
  {A.}~\bibnamefont {Fowler}}, \bibinfo {author} {\bibfnamefont
  {C.}~\bibnamefont {Gidney}}, \bibinfo {author} {\bibfnamefont
  {M.}~\bibnamefont {Giustina}}, \bibinfo {author} {\bibfnamefont
  {T.}~\bibnamefont {Huang}}, \bibinfo {author} {\bibfnamefont
  {P.}~\bibnamefont {Klimov}}, \bibinfo {author} {\bibfnamefont
  {M.}~\bibnamefont {Neeley}}, \bibinfo {author} {\bibfnamefont
  {C.}~\bibnamefont {Neill}}, \bibinfo {author} {\bibfnamefont
  {P.}~\bibnamefont {Roushan}}, \bibinfo {author} {\bibfnamefont
  {D.}~\bibnamefont {Sank}}, \bibinfo {author} {\bibfnamefont {A.}~\bibnamefont
  {Vainsencher}}, \bibinfo {author} {\bibfnamefont {J.}~\bibnamefont {Wenner}},
  \bibinfo {author} {\bibfnamefont {T.~C.}\ \bibnamefont {White}},\ and\
  \bibinfo {author} {\bibfnamefont {J.~M.}\ \bibnamefont {Martinis}},\
  }\bibfield  {title} {\bibinfo {title} {Qubit compatible superconducting
  interconnects},\ }\href {https://doi.org/10.1088/2058-9565/aa94fc} {\bibfield
   {journal} {\bibinfo  {journal} {Quantum Science and Technology}\ }\textbf
  {\bibinfo {volume} {3}},\ \bibinfo {pages} {014005} (\bibinfo {year}
  {2018})}\BibitemShut {NoStop}%
\bibitem [{\citenamefont {Kosen}\ \emph {et~al.}(2022)\citenamefont {Kosen},
  \citenamefont {Li}, \citenamefont {Rommel}, \citenamefont {Shiri},
  \citenamefont {Warren}, \citenamefont {Gr\"{o}nberg}, \citenamefont
  {Salonen}, \citenamefont {Abad}, \citenamefont {Biznárová}, \citenamefont
  {Caputo}, \citenamefont {Chen}, \citenamefont {Grigoras}, \citenamefont
  {Johansson}, \citenamefont {Kockum}, \citenamefont {Križan}, \citenamefont
  {Lozano}, \citenamefont {Norris}, \citenamefont {Osman}, \citenamefont
  {Fernández-Pendás}, \citenamefont {Ronzani}, \citenamefont {Roudsari},
  \citenamefont {Simbierowicz}, \citenamefont {Tancredi}, \citenamefont
  {Wallraff}, \citenamefont {Eichler}, \citenamefont {Govenius},\ and\
  \citenamefont {Bylander}}]{Kosen2022FlipChip}%
  \BibitemOpen
  \bibfield  {author} {\bibinfo {author} {\bibfnamefont {S.}~\bibnamefont
  {Kosen}}, \bibinfo {author} {\bibfnamefont {H.-X.}\ \bibnamefont {Li}},
  \bibinfo {author} {\bibfnamefont {M.}~\bibnamefont {Rommel}}, \bibinfo
  {author} {\bibfnamefont {D.}~\bibnamefont {Shiri}}, \bibinfo {author}
  {\bibfnamefont {C.}~\bibnamefont {Warren}}, \bibinfo {author} {\bibfnamefont
  {L.}~\bibnamefont {Gr\"{o}nberg}}, \bibinfo {author} {\bibfnamefont
  {J.}~\bibnamefont {Salonen}}, \bibinfo {author} {\bibfnamefont
  {T.}~\bibnamefont {Abad}}, \bibinfo {author} {\bibfnamefont {J.}~\bibnamefont
  {Biznárová}}, \bibinfo {author} {\bibfnamefont {M.}~\bibnamefont {Caputo}},
  \bibinfo {author} {\bibfnamefont {L.}~\bibnamefont {Chen}}, \bibinfo {author}
  {\bibfnamefont {K.}~\bibnamefont {Grigoras}}, \bibinfo {author}
  {\bibfnamefont {G.}~\bibnamefont {Johansson}}, \bibinfo {author}
  {\bibfnamefont {A.~F.}\ \bibnamefont {Kockum}}, \bibinfo {author}
  {\bibfnamefont {C.}~\bibnamefont {Križan}}, \bibinfo {author} {\bibfnamefont
  {D.~P.}\ \bibnamefont {Lozano}}, \bibinfo {author} {\bibfnamefont {G.~J.}\
  \bibnamefont {Norris}}, \bibinfo {author} {\bibfnamefont {A.}~\bibnamefont
  {Osman}}, \bibinfo {author} {\bibfnamefont {J.}~\bibnamefont
  {Fernández-Pendás}}, \bibinfo {author} {\bibfnamefont {A.}~\bibnamefont
  {Ronzani}}, \bibinfo {author} {\bibfnamefont {A.~F.}\ \bibnamefont
  {Roudsari}}, \bibinfo {author} {\bibfnamefont {S.}~\bibnamefont
  {Simbierowicz}}, \bibinfo {author} {\bibfnamefont {G.}~\bibnamefont
  {Tancredi}}, \bibinfo {author} {\bibfnamefont {A.}~\bibnamefont {Wallraff}},
  \bibinfo {author} {\bibfnamefont {C.}~\bibnamefont {Eichler}}, \bibinfo
  {author} {\bibfnamefont {J.}~\bibnamefont {Govenius}},\ and\ \bibinfo
  {author} {\bibfnamefont {J.}~\bibnamefont {Bylander}},\ }\bibfield  {title}
  {\bibinfo {title} {Building blocks of a flip-chip integrated superconducting
  quantum processor},\ }\href {https://doi.org/10.1088/2058-9565/ac734b}
  {\bibfield  {journal} {\bibinfo  {journal} {Quantum Science and Technology}\
  }\textbf {\bibinfo {volume} {7}},\ \bibinfo {pages} {035018} (\bibinfo {year}
  {2022})}\BibitemShut {NoStop}%
\bibitem [{\citenamefont {Manenti}\ \emph {et~al.}(2021)\citenamefont
  {Manenti}, \citenamefont {Sete}, \citenamefont {Chen}, \citenamefont
  {Kulshreshtha}, \citenamefont {Yeh}, \citenamefont {Oruc}, \citenamefont
  {Bestwick}, \citenamefont {Field}, \citenamefont {Jackson},\ and\
  \citenamefont {Poletto}}]{Manenti2021CombinedXYZ}%
  \BibitemOpen
  \bibfield  {author} {\bibinfo {author} {\bibfnamefont {R.}~\bibnamefont
  {Manenti}}, \bibinfo {author} {\bibfnamefont {E.~A.}\ \bibnamefont {Sete}},
  \bibinfo {author} {\bibfnamefont {A.~Q.}\ \bibnamefont {Chen}}, \bibinfo
  {author} {\bibfnamefont {S.}~\bibnamefont {Kulshreshtha}}, \bibinfo {author}
  {\bibfnamefont {J.-H.}\ \bibnamefont {Yeh}}, \bibinfo {author} {\bibfnamefont
  {F.}~\bibnamefont {Oruc}}, \bibinfo {author} {\bibfnamefont {A.}~\bibnamefont
  {Bestwick}}, \bibinfo {author} {\bibfnamefont {M.}~\bibnamefont {Field}},
  \bibinfo {author} {\bibfnamefont {K.}~\bibnamefont {Jackson}},\ and\ \bibinfo
  {author} {\bibfnamefont {S.}~\bibnamefont {Poletto}},\ }\bibfield  {title}
  {\bibinfo {title} {Full control of superconducting qubits with combined
  on-chip microwave and flux lines},\ }\href
  {https://doi.org/10.1063/5.0065517} {\bibfield  {journal} {\bibinfo
  {journal} {Applied Physics Letters}\ }\textbf {\bibinfo {volume} {119}},\
  \bibinfo {pages} {144001} (\bibinfo {year} {2021})}\BibitemShut {NoStop}%
\bibitem [{\citenamefont {Place}\ \emph {et~al.}(2021)\citenamefont {Place},
  \citenamefont {Rodgers}, \citenamefont {Mundada}, \citenamefont {Smitham},
  \citenamefont {Fitzpatrick}, \citenamefont {Leng}, \citenamefont {Premkumar},
  \citenamefont {Bryon}, \citenamefont {Vrajitoarea}, \citenamefont {Sussman},
  \citenamefont {Cheng}, \citenamefont {Madhavan}, \citenamefont {Babla},
  \citenamefont {Le}, \citenamefont {Gang}, \citenamefont {J\"{a}ck},
  \citenamefont {Gyenis}, \citenamefont {Yao}, \citenamefont {Cava},
  \citenamefont {de~Leon},\ and\ \citenamefont
  {Houck}}]{Place2021TantalumTransmon}%
  \BibitemOpen
  \bibfield  {author} {\bibinfo {author} {\bibfnamefont {A.~P.~M.}\
  \bibnamefont {Place}}, \bibinfo {author} {\bibfnamefont {L.~V.~H.}\
  \bibnamefont {Rodgers}}, \bibinfo {author} {\bibfnamefont {P.}~\bibnamefont
  {Mundada}}, \bibinfo {author} {\bibfnamefont {B.~M.}\ \bibnamefont
  {Smitham}}, \bibinfo {author} {\bibfnamefont {M.}~\bibnamefont
  {Fitzpatrick}}, \bibinfo {author} {\bibfnamefont {Z.}~\bibnamefont {Leng}},
  \bibinfo {author} {\bibfnamefont {A.}~\bibnamefont {Premkumar}}, \bibinfo
  {author} {\bibfnamefont {J.}~\bibnamefont {Bryon}}, \bibinfo {author}
  {\bibfnamefont {A.}~\bibnamefont {Vrajitoarea}}, \bibinfo {author}
  {\bibfnamefont {S.}~\bibnamefont {Sussman}}, \bibinfo {author} {\bibfnamefont
  {G.}~\bibnamefont {Cheng}}, \bibinfo {author} {\bibfnamefont
  {T.}~\bibnamefont {Madhavan}}, \bibinfo {author} {\bibfnamefont {H.~K.}\
  \bibnamefont {Babla}}, \bibinfo {author} {\bibfnamefont {X.~H.}\ \bibnamefont
  {Le}}, \bibinfo {author} {\bibfnamefont {Y.}~\bibnamefont {Gang}}, \bibinfo
  {author} {\bibfnamefont {B.}~\bibnamefont {J\"{a}ck}}, \bibinfo {author}
  {\bibfnamefont {A.}~\bibnamefont {Gyenis}}, \bibinfo {author} {\bibfnamefont
  {N.}~\bibnamefont {Yao}}, \bibinfo {author} {\bibfnamefont {R.~J.}\
  \bibnamefont {Cava}}, \bibinfo {author} {\bibfnamefont {N.~P.}\ \bibnamefont
  {de~Leon}},\ and\ \bibinfo {author} {\bibfnamefont {A.~A.}\ \bibnamefont
  {Houck}},\ }\bibfield  {title} {\bibinfo {title} {New material platform for
  superconducting transmon qubits with coherence times exceeding 0.3
  milliseconds},\ }\href {https://doi.org/10.1038/s41467-021-22030-5}
  {\bibfield  {journal} {\bibinfo  {journal} {Nature Communications}\ }\textbf
  {\bibinfo {volume} {12}},\ \bibinfo {pages} {1779} (\bibinfo {year}
  {2021})}\BibitemShut {NoStop}%
\bibitem [{\citenamefont {Wang}\ \emph {et~al.}(2022)\citenamefont {Wang},
  \citenamefont {Li}, \citenamefont {Xu}, \citenamefont {Li}, \citenamefont
  {Wang}, \citenamefont {Yang}, \citenamefont {Mi}, \citenamefont {Liang},
  \citenamefont {Su}, \citenamefont {Yang}, \citenamefont {Wang}, \citenamefont
  {Wang}, \citenamefont {Li}, \citenamefont {Chen}, \citenamefont {Li},
  \citenamefont {Linghu}, \citenamefont {Han}, \citenamefont {Zhang},
  \citenamefont {Feng}, \citenamefont {Song}, \citenamefont {Ma}, \citenamefont
  {Zhang}, \citenamefont {Wang}, \citenamefont {Zhao}, \citenamefont {Liu},
  \citenamefont {Xue}, \citenamefont {Jin},\ and\ \citenamefont
  {Yu}}]{Wang2022TantalumLifetime}%
  \BibitemOpen
  \bibfield  {author} {\bibinfo {author} {\bibfnamefont {C.}~\bibnamefont
  {Wang}}, \bibinfo {author} {\bibfnamefont {X.}~\bibnamefont {Li}}, \bibinfo
  {author} {\bibfnamefont {H.}~\bibnamefont {Xu}}, \bibinfo {author}
  {\bibfnamefont {Z.}~\bibnamefont {Li}}, \bibinfo {author} {\bibfnamefont
  {J.}~\bibnamefont {Wang}}, \bibinfo {author} {\bibfnamefont {Z.}~\bibnamefont
  {Yang}}, \bibinfo {author} {\bibfnamefont {Z.}~\bibnamefont {Mi}}, \bibinfo
  {author} {\bibfnamefont {X.}~\bibnamefont {Liang}}, \bibinfo {author}
  {\bibfnamefont {T.}~\bibnamefont {Su}}, \bibinfo {author} {\bibfnamefont
  {C.}~\bibnamefont {Yang}}, \bibinfo {author} {\bibfnamefont {G.}~\bibnamefont
  {Wang}}, \bibinfo {author} {\bibfnamefont {W.}~\bibnamefont {Wang}}, \bibinfo
  {author} {\bibfnamefont {Y.}~\bibnamefont {Li}}, \bibinfo {author}
  {\bibfnamefont {M.}~\bibnamefont {Chen}}, \bibinfo {author} {\bibfnamefont
  {C.}~\bibnamefont {Li}}, \bibinfo {author} {\bibfnamefont {K.}~\bibnamefont
  {Linghu}}, \bibinfo {author} {\bibfnamefont {J.}~\bibnamefont {Han}},
  \bibinfo {author} {\bibfnamefont {Y.}~\bibnamefont {Zhang}}, \bibinfo
  {author} {\bibfnamefont {Y.}~\bibnamefont {Feng}}, \bibinfo {author}
  {\bibfnamefont {Y.}~\bibnamefont {Song}}, \bibinfo {author} {\bibfnamefont
  {T.}~\bibnamefont {Ma}}, \bibinfo {author} {\bibfnamefont {J.}~\bibnamefont
  {Zhang}}, \bibinfo {author} {\bibfnamefont {R.}~\bibnamefont {Wang}},
  \bibinfo {author} {\bibfnamefont {P.}~\bibnamefont {Zhao}}, \bibinfo {author}
  {\bibfnamefont {W.}~\bibnamefont {Liu}}, \bibinfo {author} {\bibfnamefont
  {G.}~\bibnamefont {Xue}}, \bibinfo {author} {\bibfnamefont {Y.}~\bibnamefont
  {Jin}},\ and\ \bibinfo {author} {\bibfnamefont {H.}~\bibnamefont {Yu}},\
  }\bibfield  {title} {\bibinfo {title} {Towards practical quantum computers:
  transmon qubit with a lifetime approaching 0.5 milliseconds},\ }\href
  {https://doi.org/10.1038/s41534-021-00510-2} {\bibfield  {journal} {\bibinfo
  {journal} {npj Quantum Information}\ }\textbf {\bibinfo {volume} {8}},\
  \bibinfo {pages} {3} (\bibinfo {year} {2022})}\BibitemShut {NoStop}%
\bibitem [{\citenamefont {Chen}\ \emph {et~al.}(2014)\citenamefont {Chen},
  \citenamefont {Megrant}, \citenamefont {Kelly}, \citenamefont {Barends},
  \citenamefont {Bochmann}, \citenamefont {Chen}, \citenamefont {Chiaro},
  \citenamefont {Dunsworth}, \citenamefont {Jeffrey}, \citenamefont {Mutus},
  \citenamefont {O’Malley}, \citenamefont {Neill}, \citenamefont {Roushan},
  \citenamefont {Sank}, \citenamefont {Vainsencher}, \citenamefont {Wenner},
  \citenamefont {White}, \citenamefont {Cleland},\ and\ \citenamefont
  {Martinis}}]{Chen2014Airbridges}%
  \BibitemOpen
  \bibfield  {author} {\bibinfo {author} {\bibfnamefont {Z.}~\bibnamefont
  {Chen}}, \bibinfo {author} {\bibfnamefont {A.}~\bibnamefont {Megrant}},
  \bibinfo {author} {\bibfnamefont {J.}~\bibnamefont {Kelly}}, \bibinfo
  {author} {\bibfnamefont {R.}~\bibnamefont {Barends}}, \bibinfo {author}
  {\bibfnamefont {J.}~\bibnamefont {Bochmann}}, \bibinfo {author}
  {\bibfnamefont {Y.}~\bibnamefont {Chen}}, \bibinfo {author} {\bibfnamefont
  {B.}~\bibnamefont {Chiaro}}, \bibinfo {author} {\bibfnamefont
  {A.}~\bibnamefont {Dunsworth}}, \bibinfo {author} {\bibfnamefont
  {E.}~\bibnamefont {Jeffrey}}, \bibinfo {author} {\bibfnamefont {J.~Y.}\
  \bibnamefont {Mutus}}, \bibinfo {author} {\bibfnamefont {P.~J.~J.}\
  \bibnamefont {O’Malley}}, \bibinfo {author} {\bibfnamefont
  {C.}~\bibnamefont {Neill}}, \bibinfo {author} {\bibfnamefont
  {P.}~\bibnamefont {Roushan}}, \bibinfo {author} {\bibfnamefont
  {D.}~\bibnamefont {Sank}}, \bibinfo {author} {\bibfnamefont {A.}~\bibnamefont
  {Vainsencher}}, \bibinfo {author} {\bibfnamefont {J.}~\bibnamefont {Wenner}},
  \bibinfo {author} {\bibfnamefont {T.~C.}\ \bibnamefont {White}}, \bibinfo
  {author} {\bibfnamefont {A.~N.}\ \bibnamefont {Cleland}},\ and\ \bibinfo
  {author} {\bibfnamefont {J.~M.}\ \bibnamefont {Martinis}},\ }\bibfield
  {title} {\bibinfo {title} {Fabrication and characterization of aluminum
  airbridges for superconducting microwave circuits},\ }\href
  {https://doi.org/10.1063/1.4863745} {\bibfield  {journal} {\bibinfo
  {journal} {Applied Physics Letters}\ }\textbf {\bibinfo {volume} {104}},\
  \bibinfo {pages} {052602} (\bibinfo {year} {2014})}\BibitemShut {NoStop}%
\bibitem [{\citenamefont {Dunsworth}\ \emph {et~al.}(2018)\citenamefont
  {Dunsworth}, \citenamefont {Megrant}, \citenamefont {Quintana}, \citenamefont
  {Chen}, \citenamefont {Kelly}, \citenamefont {Barends}, \citenamefont
  {Chiaro}, \citenamefont {Foxen}, \citenamefont {Jeffrey}, \citenamefont
  {Kreikebaum}, \citenamefont {Lucero}, \citenamefont {Mutus}, \citenamefont
  {Neeley}, \citenamefont {Neill}, \citenamefont {O'Malley}, \citenamefont
  {Roushan}, \citenamefont {Sank}, \citenamefont {Vainsencher}, \citenamefont
  {Wenner}, \citenamefont {White}, \citenamefont {Cleland},\ and\ \citenamefont
  {Martinis}}]{Dunsworth2018MultiLayerWiring}%
  \BibitemOpen
  \bibfield  {author} {\bibinfo {author} {\bibfnamefont {A.}~\bibnamefont
  {Dunsworth}}, \bibinfo {author} {\bibfnamefont {A.}~\bibnamefont {Megrant}},
  \bibinfo {author} {\bibfnamefont {C.}~\bibnamefont {Quintana}}, \bibinfo
  {author} {\bibfnamefont {Z.}~\bibnamefont {Chen}}, \bibinfo {author}
  {\bibfnamefont {J.}~\bibnamefont {Kelly}}, \bibinfo {author} {\bibfnamefont
  {R.}~\bibnamefont {Barends}}, \bibinfo {author} {\bibfnamefont
  {B.}~\bibnamefont {Chiaro}}, \bibinfo {author} {\bibfnamefont
  {B.}~\bibnamefont {Foxen}}, \bibinfo {author} {\bibfnamefont
  {E.}~\bibnamefont {Jeffrey}}, \bibinfo {author} {\bibfnamefont
  {J.}~\bibnamefont {Kreikebaum}}, \bibinfo {author} {\bibfnamefont
  {E.}~\bibnamefont {Lucero}}, \bibinfo {author} {\bibfnamefont {J.~Y.}\
  \bibnamefont {Mutus}}, \bibinfo {author} {\bibfnamefont {M.}~\bibnamefont
  {Neeley}}, \bibinfo {author} {\bibfnamefont {C.}~\bibnamefont {Neill}},
  \bibinfo {author} {\bibfnamefont {P.~J.~J.}\ \bibnamefont {O'Malley}},
  \bibinfo {author} {\bibfnamefont {P.}~\bibnamefont {Roushan}}, \bibinfo
  {author} {\bibfnamefont {D.}~\bibnamefont {Sank}}, \bibinfo {author}
  {\bibfnamefont {A.}~\bibnamefont {Vainsencher}}, \bibinfo {author}
  {\bibfnamefont {J.}~\bibnamefont {Wenner}}, \bibinfo {author} {\bibfnamefont
  {T.~C.}\ \bibnamefont {White}}, \bibinfo {author} {\bibfnamefont {A.~N.}\
  \bibnamefont {Cleland}},\ and\ \bibinfo {author} {\bibfnamefont {J.~M.}\
  \bibnamefont {Martinis}},\ }\bibfield  {title} {\bibinfo {title} {A method
  for building low loss multi-layer wiring for superconducting microwave
  devices},\ }\href {https://doi.org/10.1063/1.5014033} {\bibfield  {journal}
  {\bibinfo  {journal} {Applied Physics Letters}\ }\textbf {\bibinfo {volume}
  {112}},\ \bibinfo {pages} {063502} (\bibinfo {year} {2018})}\BibitemShut
  {NoStop}%
\bibitem [{\citenamefont {Girgis}\ \emph {et~al.}(2006)\citenamefont {Girgis},
  \citenamefont {Liu},\ and\ \citenamefont
  {Benkhedar}}]{Girgis2006MetallicAirBridges}%
  \BibitemOpen
  \bibfield  {author} {\bibinfo {author} {\bibfnamefont {E.}~\bibnamefont
  {Girgis}}, \bibinfo {author} {\bibfnamefont {J.}~\bibnamefont {Liu}},\ and\
  \bibinfo {author} {\bibfnamefont {M.}~\bibnamefont {Benkhedar}},\ }\bibfield
  {title} {\bibinfo {title} {Fabrication of metallic air bridges using
  multiple-dose electron beam lithography},\ }\href
  {https://doi.org/10.1063/1.2203210} {\bibfield  {journal} {\bibinfo
  {journal} {Applied Physics Letters}\ }\textbf {\bibinfo {volume} {88}},\
  \bibinfo {pages} {202103} (\bibinfo {year} {2006})}\BibitemShut {NoStop}%
\bibitem [{\citenamefont {Stavenga}\ \emph {et~al.}(2023)\citenamefont
  {Stavenga}, \citenamefont {Khan}, \citenamefont {Liu}, \citenamefont
  {Krogstrup},\ and\ \citenamefont
  {DiCarlo}}]{Stavenga2023GrayscaleAirbridges}%
  \BibitemOpen
  \bibfield  {author} {\bibinfo {author} {\bibfnamefont {T.}~\bibnamefont
  {Stavenga}}, \bibinfo {author} {\bibfnamefont {S.~A.}\ \bibnamefont {Khan}},
  \bibinfo {author} {\bibfnamefont {Y.}~\bibnamefont {Liu}}, \bibinfo {author}
  {\bibfnamefont {P.}~\bibnamefont {Krogstrup}},\ and\ \bibinfo {author}
  {\bibfnamefont {L.}~\bibnamefont {DiCarlo}},\ }\bibfield  {title} {\bibinfo
  {title} {Lower-temperature fabrication of airbridges by grayscale lithography
  to increase yield of nanowire transmons in circuit qed quantum processors},\
  }\href {https://doi.org/10.1063/5.0146814} {\bibfield  {journal} {\bibinfo
  {journal} {Applied Physics Letters}\ }\textbf {\bibinfo {volume} {123}},\
  \bibinfo {pages} {024004} (\bibinfo {year} {2023})}\BibitemShut {NoStop}%
\bibitem [{\citenamefont {Alegria}\ \emph {et~al.}(2025)\citenamefont
  {Alegria}, \citenamefont {Abelson}, \citenamefont {Kim}, \citenamefont {Im},
  \citenamefont {Voyles}, \citenamefont {Lordi}, \citenamefont {DuBois},\ and\
  \citenamefont {Rosen}}]{Alegria2025TaAirbridge}%
  \BibitemOpen
  \bibfield  {author} {\bibinfo {author} {\bibfnamefont {L.~D.}\ \bibnamefont
  {Alegria}}, \bibinfo {author} {\bibfnamefont {A.}~\bibnamefont {Abelson}},
  \bibinfo {author} {\bibfnamefont {E.}~\bibnamefont {Kim}}, \bibinfo {author}
  {\bibfnamefont {S.}~\bibnamefont {Im}}, \bibinfo {author} {\bibfnamefont
  {P.~M.}\ \bibnamefont {Voyles}}, \bibinfo {author} {\bibfnamefont
  {V.}~\bibnamefont {Lordi}}, \bibinfo {author} {\bibfnamefont {J.~L.}\
  \bibnamefont {DuBois}},\ and\ \bibinfo {author} {\bibfnamefont {Y.~J.}\
  \bibnamefont {Rosen}},\ }\bibfield  {title} {\bibinfo {title} {Growth and
  structure of alpha-ta films for quantum circuit integration},\ }\href
  {https://doi.org/10.1063/5.0245710} {\bibfield  {journal} {\bibinfo
  {journal} {Journal of Applied Physics}\ }\textbf {\bibinfo {volume} {137}},\
  \bibinfo {pages} {044402} (\bibinfo {year} {2025})}\BibitemShut {NoStop}%
\bibitem [{\citenamefont {Sun}\ \emph {et~al.}(2022)\citenamefont {Sun},
  \citenamefont {Ding}, \citenamefont {Xia}, \citenamefont {Wang},
  \citenamefont {Xu}, \citenamefont {Song}, \citenamefont {Lan}, \citenamefont
  {Zhao},\ and\ \citenamefont {Yu}}]{Sun2022GradientExposure}%
  \BibitemOpen
  \bibfield  {author} {\bibinfo {author} {\bibfnamefont {Y.}~\bibnamefont
  {Sun}}, \bibinfo {author} {\bibfnamefont {J.}~\bibnamefont {Ding}}, \bibinfo
  {author} {\bibfnamefont {X.}~\bibnamefont {Xia}}, \bibinfo {author}
  {\bibfnamefont {X.}~\bibnamefont {Wang}}, \bibinfo {author} {\bibfnamefont
  {J.}~\bibnamefont {Xu}}, \bibinfo {author} {\bibfnamefont {S.}~\bibnamefont
  {Song}}, \bibinfo {author} {\bibfnamefont {D.}~\bibnamefont {Lan}}, \bibinfo
  {author} {\bibfnamefont {J.}~\bibnamefont {Zhao}},\ and\ \bibinfo {author}
  {\bibfnamefont {Y.}~\bibnamefont {Yu}},\ }\bibfield  {title} {\bibinfo
  {title} {Fabrication of airbridges with gradient exposure},\ }\href
  {https://doi.org/10.1063/5.0102555} {\bibfield  {journal} {\bibinfo
  {journal} {Applied Physics Letters}\ }\textbf {\bibinfo {volume} {121}},\
  \bibinfo {pages} {074001} (\bibinfo {year} {2022})}\BibitemShut {NoStop}%
\bibitem [{\citenamefont {Malevannaya}\ \emph {et~al.}(2025)\citenamefont
  {Malevannaya}, \citenamefont {Polozov}, \citenamefont {Ivanov}, \citenamefont
  {Matanin}, \citenamefont {Smirnov}, \citenamefont {Echeistov}, \citenamefont
  {Moskalev}, \citenamefont {Mikhalin}, \citenamefont {Shirokov}, \citenamefont
  {Panfilov}, \citenamefont {Ryzhikov}, \citenamefont {Andriyash},\ and\
  \citenamefont {Rodionov}}]{Malevannaya2025Shielding}%
  \BibitemOpen
  \bibfield  {author} {\bibinfo {author} {\bibfnamefont {E.~I.}\ \bibnamefont
  {Malevannaya}}, \bibinfo {author} {\bibfnamefont {V.~I.}\ \bibnamefont
  {Polozov}}, \bibinfo {author} {\bibfnamefont {A.~I.}\ \bibnamefont {Ivanov}},
  \bibinfo {author} {\bibfnamefont {A.~R.}\ \bibnamefont {Matanin}}, \bibinfo
  {author} {\bibfnamefont {N.~S.}\ \bibnamefont {Smirnov}}, \bibinfo {author}
  {\bibfnamefont {V.~V.}\ \bibnamefont {Echeistov}}, \bibinfo {author}
  {\bibfnamefont {D.~O.}\ \bibnamefont {Moskalev}}, \bibinfo {author}
  {\bibfnamefont {D.~A.}\ \bibnamefont {Mikhalin}}, \bibinfo {author}
  {\bibfnamefont {D.~E.}\ \bibnamefont {Shirokov}}, \bibinfo {author}
  {\bibfnamefont {Y.~V.}\ \bibnamefont {Panfilov}}, \bibinfo {author}
  {\bibfnamefont {I.~A.}\ \bibnamefont {Ryzhikov}}, \bibinfo {author}
  {\bibfnamefont {A.~V.}\ \bibnamefont {Andriyash}},\ and\ \bibinfo {author}
  {\bibfnamefont {I.~A.}\ \bibnamefont {Rodionov}},\ }\bibfield  {title}
  {\bibinfo {title} {An engineering guide to superconducting quantum circuit
  shielding},\ }\href {https://doi.org/10.1063/5.0250262} {\bibfield  {journal}
  {\bibinfo  {journal} {Applied Physics Reviews}\ }\textbf {\bibinfo {volume}
  {12}},\ \bibinfo {pages} {031334} (\bibinfo {year} {2025})}\BibitemShut
  {NoStop}%
\bibitem [{\citenamefont {Krantz}\ \emph {et~al.}(2019)\citenamefont {Krantz},
  \citenamefont {Kjaergaard}, \citenamefont {Yan}, \citenamefont {Orlando},
  \citenamefont {Gustavsson},\ and\ \citenamefont {Oliver}}]{Krantz2019Guide}%
  \BibitemOpen
  \bibfield  {author} {\bibinfo {author} {\bibfnamefont {P.}~\bibnamefont
  {Krantz}}, \bibinfo {author} {\bibfnamefont {M.}~\bibnamefont {Kjaergaard}},
  \bibinfo {author} {\bibfnamefont {F.}~\bibnamefont {Yan}}, \bibinfo {author}
  {\bibfnamefont {T.~P.}\ \bibnamefont {Orlando}}, \bibinfo {author}
  {\bibfnamefont {S.}~\bibnamefont {Gustavsson}},\ and\ \bibinfo {author}
  {\bibfnamefont {W.~D.}\ \bibnamefont {Oliver}},\ }\bibfield  {title}
  {\bibinfo {title} {A quantum engineer's guide to superconducting qubits},\
  }\href {https://doi.org/10.1063/1.5089550} {\bibfield  {journal} {\bibinfo
  {journal} {Applied Physics Reviews}\ }\textbf {\bibinfo {volume} {6}},\
  \bibinfo {pages} {021318} (\bibinfo {year} {2019})}\BibitemShut {NoStop}%
\end{thebibliography}%


\begin{thebibliography}{9}%
\makeatletter
\providecommand \@ifxundefined [1]{%
 \@ifx{#1\undefined}
}%
\providecommand \@ifnum [1]{%
 \ifnum #1\expandafter \@firstoftwo
 \else \expandafter \@secondoftwo
 \fi
}%
\providecommand \@ifx [1]{%
 \ifx #1\expandafter \@firstoftwo
 \else \expandafter \@secondoftwo
 \fi
}%
\providecommand \natexlab [1]{#1}%
\providecommand \enquote  [1]{``#1''}%
\providecommand \bibnamefont  [1]{#1}%
\providecommand \bibfnamefont [1]{#1}%
\providecommand \citenamefont [1]{#1}%
\providecommand \href@noop [0]{\@secondoftwo}%
\providecommand \href [0]{\begingroup \@sanitize@url \@href}%
\providecommand \@href[1]{\@@startlink{#1}\@@href}%
\providecommand \@@href[1]{\endgroup#1\@@endlink}%
\providecommand \@sanitize@url [0]{\catcode `\\12\catcode `\$12\catcode
  `\&12\catcode `\#12\catcode `\^12\catcode `\_12\catcode `\%12\relax}%
\providecommand \@@startlink[1]{}%
\providecommand \@@endlink[0]{}%
\providecommand \url  [0]{\begingroup\@sanitize@url \@url }%
\providecommand \@url [1]{\endgroup\@href {#1}{\urlprefix }}%
\providecommand \urlprefix  [0]{URL }%
\providecommand \Eprint [0]{\href }%
\providecommand \doibase [0]{https://doi.org/}%
\providecommand \selectlanguage [0]{\@gobble}%
\providecommand \bibinfo  [0]{\@secondoftwo}%
\providecommand \bibfield  [0]{\@secondoftwo}%
\providecommand \translation [1]{[#1]}%
\providecommand \BibitemOpen [0]{}%
\providecommand \bibitemStop [0]{}%
\providecommand \bibitemNoStop [0]{.\EOS\space}%
\providecommand \EOS [0]{\spacefactor3000\relax}%
\providecommand \BibitemShut  [1]{\csname bibitem#1\endcsname}%
\let\auto@bib@innerbib\@empty
\bibitem [{\citenamefont {Huai}\ \emph {et~al.}(2024)\citenamefont {Huai},
  \citenamefont {Bu}, \citenamefont {Gu}, \citenamefont {Zhang}, \citenamefont
  {An}, \citenamefont {Yang}, \citenamefont {Li}, \citenamefont {Cai},\ and\
  \citenamefont {Zheng}}]{huai2024fast}%
  \BibitemOpen
  \bibfield  {author} {\bibinfo {author} {\bibfnamefont {S.}~\bibnamefont
  {Huai}}, \bibinfo {author} {\bibfnamefont {K.}~\bibnamefont {Bu}}, \bibinfo
  {author} {\bibfnamefont {X.}~\bibnamefont {Gu}}, \bibinfo {author}
  {\bibfnamefont {Z.}~\bibnamefont {Zhang}}, \bibinfo {author} {\bibfnamefont
  {S.}~\bibnamefont {An}}, \bibinfo {author} {\bibfnamefont {X.}~\bibnamefont
  {Yang}}, \bibinfo {author} {\bibfnamefont {Y.}~\bibnamefont {Li}}, \bibinfo
  {author} {\bibfnamefont {T.}~\bibnamefont {Cai}},\ and\ \bibinfo {author}
  {\bibfnamefont {Y.}~\bibnamefont {Zheng}},\ }\bibfield  {title} {\bibinfo
  {title} {Fast joint parity measurement via collective interactions induced by
  stimulated emission},\ }\href {https://doi.org/10.1038/s41467-024-47379-1}
  {\bibfield  {journal} {\bibinfo  {journal} {Nature Communications}\ }\textbf
  {\bibinfo {volume} {15}},\ \bibinfo {pages} {3045} (\bibinfo {year}
  {2024})}\BibitemShut {NoStop}%
\bibitem [{\citenamefont {Wenner}\ \emph {et~al.}(2011)\citenamefont {Wenner},
  \citenamefont {Barends}, \citenamefont {Bialczak}, \citenamefont {Chen},
  \citenamefont {Kelly}, \citenamefont {Lucero}, \citenamefont {Mariantoni},
  \citenamefont {Megrant}, \citenamefont {O'Malley}, \citenamefont {Sank},
  \citenamefont {Vainsencher}, \citenamefont {Wang}, \citenamefont {White},
  \citenamefont {Yin}, \citenamefont {Zhao}, \citenamefont {Cleland},\ and\
  \citenamefont {Martinis}}]{wenner2011surface}%
  \BibitemOpen
  \bibfield  {author} {\bibinfo {author} {\bibfnamefont {J.}~\bibnamefont
  {Wenner}}, \bibinfo {author} {\bibfnamefont {R.}~\bibnamefont {Barends}},
  \bibinfo {author} {\bibfnamefont {R.~C.}\ \bibnamefont {Bialczak}}, \bibinfo
  {author} {\bibfnamefont {Y.}~\bibnamefont {Chen}}, \bibinfo {author}
  {\bibfnamefont {J.}~\bibnamefont {Kelly}}, \bibinfo {author} {\bibfnamefont
  {E.}~\bibnamefont {Lucero}}, \bibinfo {author} {\bibfnamefont
  {M.}~\bibnamefont {Mariantoni}}, \bibinfo {author} {\bibfnamefont
  {A.}~\bibnamefont {Megrant}}, \bibinfo {author} {\bibfnamefont {P.~J.~J.}\
  \bibnamefont {O'Malley}}, \bibinfo {author} {\bibfnamefont {D.}~\bibnamefont
  {Sank}}, \bibinfo {author} {\bibfnamefont {A.}~\bibnamefont {Vainsencher}},
  \bibinfo {author} {\bibfnamefont {H.}~\bibnamefont {Wang}}, \bibinfo {author}
  {\bibfnamefont {T.~C.}\ \bibnamefont {White}}, \bibinfo {author}
  {\bibfnamefont {Y.}~\bibnamefont {Yin}}, \bibinfo {author} {\bibfnamefont
  {J.}~\bibnamefont {Zhao}}, \bibinfo {author} {\bibfnamefont {A.~N.}\
  \bibnamefont {Cleland}},\ and\ \bibinfo {author} {\bibfnamefont {J.~M.}\
  \bibnamefont {Martinis}},\ }\bibfield  {title} {\bibinfo {title} {Surface
  loss simulations of superconducting coplanar waveguide resonators},\ }\href
  {https://doi.org/10.1063/1.3637047} {\bibfield  {journal} {\bibinfo
  {journal} {Applied Physics Letters}\ }\textbf {\bibinfo {volume} {99}},\
  \bibinfo {pages} {113513} (\bibinfo {year} {2011})}\BibitemShut {NoStop}%
\bibitem [{\citenamefont {Wang}\ \emph {et~al.}(2015)\citenamefont {Wang},
  \citenamefont {Axline}, \citenamefont {Gao}, \citenamefont {Brecht},
  \citenamefont {Chu}, \citenamefont {Frunzio}, \citenamefont {Devoret},\ and\
  \citenamefont {Schoelkopf}}]{wang2015surface}%
  \BibitemOpen
  \bibfield  {author} {\bibinfo {author} {\bibfnamefont {C.}~\bibnamefont
  {Wang}}, \bibinfo {author} {\bibfnamefont {C.}~\bibnamefont {Axline}},
  \bibinfo {author} {\bibfnamefont {Y.~Y.}\ \bibnamefont {Gao}}, \bibinfo
  {author} {\bibfnamefont {T.}~\bibnamefont {Brecht}}, \bibinfo {author}
  {\bibfnamefont {Y.}~\bibnamefont {Chu}}, \bibinfo {author} {\bibfnamefont
  {L.}~\bibnamefont {Frunzio}}, \bibinfo {author} {\bibfnamefont {M.~H.}\
  \bibnamefont {Devoret}},\ and\ \bibinfo {author} {\bibfnamefont {R.~J.}\
  \bibnamefont {Schoelkopf}},\ }\bibfield  {title} {\bibinfo {title} {Surface
  participation and dielectric loss in superconducting qubits},\ }\href
  {https://doi.org/10.1063/1.4934486} {\bibfield  {journal} {\bibinfo
  {journal} {Applied Physics Letters}\ }\textbf {\bibinfo {volume} {107}},\
  \bibinfo {pages} {162601} (\bibinfo {year} {2015})}\BibitemShut {NoStop}%
\bibitem [{\citenamefont {Woods}\ \emph {et~al.}(2019)\citenamefont {Woods},
  \citenamefont {Calusine}, \citenamefont {Melville}, \citenamefont {Sevi},
  \citenamefont {Golden}, \citenamefont {Kim}, \citenamefont {Rosenberg},
  \citenamefont {Yoder},\ and\ \citenamefont {Oliver}}]{woods2019determining}%
  \BibitemOpen
  \bibfield  {author} {\bibinfo {author} {\bibfnamefont {W.}~\bibnamefont
  {Woods}}, \bibinfo {author} {\bibfnamefont {G.}~\bibnamefont {Calusine}},
  \bibinfo {author} {\bibfnamefont {A.}~\bibnamefont {Melville}}, \bibinfo
  {author} {\bibfnamefont {A.}~\bibnamefont {Sevi}}, \bibinfo {author}
  {\bibfnamefont {E.}~\bibnamefont {Golden}}, \bibinfo {author} {\bibfnamefont
  {D.~K.}\ \bibnamefont {Kim}}, \bibinfo {author} {\bibfnamefont
  {D.}~\bibnamefont {Rosenberg}}, \bibinfo {author} {\bibfnamefont {J.~L.}\
  \bibnamefont {Yoder}},\ and\ \bibinfo {author} {\bibfnamefont {W.~D.}\
  \bibnamefont {Oliver}},\ }\bibfield  {title} {\bibinfo {title} {Determining
  interface dielectric losses in superconducting coplanar-waveguide
  resonators},\ }\href {https://doi.org/10.1103/PhysRevApplied.12.014012}
  {\bibfield  {journal} {\bibinfo  {journal} {Physical Review Applied}\
  }\textbf {\bibinfo {volume} {12}},\ \bibinfo {pages} {014012} (\bibinfo
  {year} {2019})}\BibitemShut {NoStop}%
\bibitem [{\citenamefont {Gu}\ \emph {et~al.}(2026)\citenamefont {Gu},
  \citenamefont {Liang}, \citenamefont {Wang}, \citenamefont {Xiao},
  \citenamefont {Deng}, \citenamefont {Liu}, \citenamefont {Li}, \citenamefont
  {Xu}, \citenamefont {Xiang},\ and\ \citenamefont {Fan}}]{gu2026connectivity}%
  \BibitemOpen
  \bibfield  {author} {\bibinfo {author} {\bibfnamefont {X.-Y.}\ \bibnamefont
  {Gu}}, \bibinfo {author} {\bibfnamefont {G.-H.}\ \bibnamefont {Liang}},
  \bibinfo {author} {\bibfnamefont {M.-C.}\ \bibnamefont {Wang}}, \bibinfo
  {author} {\bibfnamefont {Y.}~\bibnamefont {Xiao}}, \bibinfo {author}
  {\bibfnamefont {C.-L.}\ \bibnamefont {Deng}}, \bibinfo {author}
  {\bibfnamefont {Z.-H.}\ \bibnamefont {Liu}}, \bibinfo {author} {\bibfnamefont
  {T.-M.}\ \bibnamefont {Li}}, \bibinfo {author} {\bibfnamefont
  {K.}~\bibnamefont {Xu}}, \bibinfo {author} {\bibfnamefont {Z.}~\bibnamefont
  {Xiang}},\ and\ \bibinfo {author} {\bibfnamefont {H.}~\bibnamefont {Fan}},\
  }\href {https://doi.org/10.48550/arXiv.2607.10743} {\bibinfo {title}
  {Connectivity-induced surface-loss penalty in superconducting qubit-coupler
  lattices}} (\bibinfo {year} {2026}),\ \Eprint
  {https://arxiv.org/abs/2607.10743} {arXiv:2607.10743 [quant-ph]} \BibitemShut
  {NoStop}%
\bibitem [{\citenamefont {Minev}\ \emph {et~al.}(2021)\citenamefont {Minev},
  \citenamefont {Leghtas}, \citenamefont {Mundhada}, \citenamefont
  {Christakis}, \citenamefont {Pop},\ and\ \citenamefont
  {Devoret}}]{minev2021epr}%
  \BibitemOpen
  \bibfield  {author} {\bibinfo {author} {\bibfnamefont {Z.~K.}\ \bibnamefont
  {Minev}}, \bibinfo {author} {\bibfnamefont {Z.}~\bibnamefont {Leghtas}},
  \bibinfo {author} {\bibfnamefont {S.~O.}\ \bibnamefont {Mundhada}}, \bibinfo
  {author} {\bibfnamefont {L.}~\bibnamefont {Christakis}}, \bibinfo {author}
  {\bibfnamefont {I.~M.}\ \bibnamefont {Pop}},\ and\ \bibinfo {author}
  {\bibfnamefont {M.~H.}\ \bibnamefont {Devoret}},\ }\bibfield  {title}
  {\bibinfo {title} {Energy-participation quantization of josephson circuits},\
  }\href {https://doi.org/10.1038/s41534-021-00461-8} {\bibfield  {journal}
  {\bibinfo  {journal} {npj Quantum Information}\ }\textbf {\bibinfo {volume}
  {7}},\ \bibinfo {pages} {131} (\bibinfo {year} {2021})}\BibitemShut {NoStop}%
\bibitem [{\citenamefont {Zhang}\ \emph {et~al.}(2024)\citenamefont {Zhang},
  \citenamefont {Gao}, \citenamefont {Xu}, \citenamefont {Wang}, \citenamefont
  {Dong}, \citenamefont {Guo}, \citenamefont {Deng}, \citenamefont {Zhang},
  \citenamefont {Chen}, \citenamefont {Xu}, \citenamefont {Wang}, \citenamefont
  {Wu}, \citenamefont {Zhang}, \citenamefont {Jin}, \citenamefont {Zhu},
  \citenamefont {Zhang}, \citenamefont {Zou}, \citenamefont {Tan},
  \citenamefont {Cui}, \citenamefont {Zhu}, \citenamefont {Shen}, \citenamefont
  {Li}, \citenamefont {Zhong}, \citenamefont {Bao}, \citenamefont {Zhao},
  \citenamefont {Hao}, \citenamefont {Li}, \citenamefont {Wang}, \citenamefont
  {Song}, \citenamefont {Guo}, \citenamefont {Wang},\ and\ \citenamefont
  {Poletti}}]{zhang2024emergence}%
  \BibitemOpen
  \bibfield  {author} {\bibinfo {author} {\bibfnamefont {P.}~\bibnamefont
  {Zhang}}, \bibinfo {author} {\bibfnamefont {Y.}~\bibnamefont {Gao}}, \bibinfo
  {author} {\bibfnamefont {X.}~\bibnamefont {Xu}}, \bibinfo {author}
  {\bibfnamefont {N.}~\bibnamefont {Wang}}, \bibinfo {author} {\bibfnamefont
  {H.}~\bibnamefont {Dong}}, \bibinfo {author} {\bibfnamefont {C.}~\bibnamefont
  {Guo}}, \bibinfo {author} {\bibfnamefont {J.}~\bibnamefont {Deng}}, \bibinfo
  {author} {\bibfnamefont {X.}~\bibnamefont {Zhang}}, \bibinfo {author}
  {\bibfnamefont {J.}~\bibnamefont {Chen}}, \bibinfo {author} {\bibfnamefont
  {S.}~\bibnamefont {Xu}}, \bibinfo {author} {\bibfnamefont {K.}~\bibnamefont
  {Wang}}, \bibinfo {author} {\bibfnamefont {Y.}~\bibnamefont {Wu}}, \bibinfo
  {author} {\bibfnamefont {C.}~\bibnamefont {Zhang}}, \bibinfo {author}
  {\bibfnamefont {F.}~\bibnamefont {Jin}}, \bibinfo {author} {\bibfnamefont
  {X.}~\bibnamefont {Zhu}}, \bibinfo {author} {\bibfnamefont {A.}~\bibnamefont
  {Zhang}}, \bibinfo {author} {\bibfnamefont {Y.}~\bibnamefont {Zou}}, \bibinfo
  {author} {\bibfnamefont {Z.}~\bibnamefont {Tan}}, \bibinfo {author}
  {\bibfnamefont {Z.}~\bibnamefont {Cui}}, \bibinfo {author} {\bibfnamefont
  {Z.}~\bibnamefont {Zhu}}, \bibinfo {author} {\bibfnamefont {F.}~\bibnamefont
  {Shen}}, \bibinfo {author} {\bibfnamefont {T.}~\bibnamefont {Li}}, \bibinfo
  {author} {\bibfnamefont {J.}~\bibnamefont {Zhong}}, \bibinfo {author}
  {\bibfnamefont {Z.}~\bibnamefont {Bao}}, \bibinfo {author} {\bibfnamefont
  {L.}~\bibnamefont {Zhao}}, \bibinfo {author} {\bibfnamefont {J.}~\bibnamefont
  {Hao}}, \bibinfo {author} {\bibfnamefont {H.}~\bibnamefont {Li}}, \bibinfo
  {author} {\bibfnamefont {Z.}~\bibnamefont {Wang}}, \bibinfo {author}
  {\bibfnamefont {C.}~\bibnamefont {Song}}, \bibinfo {author} {\bibfnamefont
  {Q.}~\bibnamefont {Guo}}, \bibinfo {author} {\bibfnamefont {H.}~\bibnamefont
  {Wang}},\ and\ \bibinfo {author} {\bibfnamefont {D.}~\bibnamefont
  {Poletti}},\ }\bibfield  {title} {\bibinfo {title} {Emergence of steady
  quantum transport in a superconducting processor},\ }\href
  {https://doi.org/10.1038/s41467-024-54332-9} {\bibfield  {journal} {\bibinfo
  {journal} {Nature Communications}\ }\textbf {\bibinfo {volume} {15}},\
  \bibinfo {pages} {10115} (\bibinfo {year} {2024})}\BibitemShut {NoStop}%
\bibitem [{\citenamefont {Koch}\ \emph {et~al.}(2007)\citenamefont {Koch},
  \citenamefont {Yu}, \citenamefont {Gambetta}, \citenamefont {Houck},
  \citenamefont {Schuster}, \citenamefont {Majer}, \citenamefont {Blais},
  \citenamefont {Devoret}, \citenamefont {Girvin},\ and\ \citenamefont
  {Schoelkopf}}]{koch2007transmon}%
  \BibitemOpen
  \bibfield  {author} {\bibinfo {author} {\bibfnamefont {J.}~\bibnamefont
  {Koch}}, \bibinfo {author} {\bibfnamefont {T.~M.}\ \bibnamefont {Yu}},
  \bibinfo {author} {\bibfnamefont {J.}~\bibnamefont {Gambetta}}, \bibinfo
  {author} {\bibfnamefont {A.~A.}\ \bibnamefont {Houck}}, \bibinfo {author}
  {\bibfnamefont {D.~I.}\ \bibnamefont {Schuster}}, \bibinfo {author}
  {\bibfnamefont {J.}~\bibnamefont {Majer}}, \bibinfo {author} {\bibfnamefont
  {A.}~\bibnamefont {Blais}}, \bibinfo {author} {\bibfnamefont {M.~H.}\
  \bibnamefont {Devoret}}, \bibinfo {author} {\bibfnamefont {S.~M.}\
  \bibnamefont {Girvin}},\ and\ \bibinfo {author} {\bibfnamefont {R.~J.}\
  \bibnamefont {Schoelkopf}},\ }\bibfield  {title} {\bibinfo {title}
  {Charge-insensitive qubit design derived from the cooper pair box},\ }\href
  {https://doi.org/10.1103/PhysRevA.76.042319} {\bibfield  {journal} {\bibinfo
  {journal} {Physical Review A}\ }\textbf {\bibinfo {volume} {76}},\ \bibinfo
  {pages} {042319} (\bibinfo {year} {2007})}\BibitemShut {NoStop}%
\bibitem [{\citenamefont {Wang}\ \emph {et~al.}(2024)\citenamefont {Wang},
  \citenamefont {Wu}, \citenamefont {Wang}, \citenamefont {Ma}, \citenamefont
  {Zhang}, \citenamefont {Chen}, \citenamefont {Deng}, \citenamefont {Gao},
  \citenamefont {Hu}, \citenamefont {Ma}, \citenamefont {Song}, \citenamefont
  {Xia}, \citenamefont {Ying}, \citenamefont {Zhan}, \citenamefont {Zhao},\
  and\ \citenamefont {Deng}}]{wang2024efficient}%
  \BibitemOpen
  \bibfield  {author} {\bibinfo {author} {\bibfnamefont {T.}~\bibnamefont
  {Wang}}, \bibinfo {author} {\bibfnamefont {F.}~\bibnamefont {Wu}}, \bibinfo
  {author} {\bibfnamefont {F.}~\bibnamefont {Wang}}, \bibinfo {author}
  {\bibfnamefont {X.}~\bibnamefont {Ma}}, \bibinfo {author} {\bibfnamefont
  {G.}~\bibnamefont {Zhang}}, \bibinfo {author} {\bibfnamefont
  {J.}~\bibnamefont {Chen}}, \bibinfo {author} {\bibfnamefont {H.}~\bibnamefont
  {Deng}}, \bibinfo {author} {\bibfnamefont {R.}~\bibnamefont {Gao}}, \bibinfo
  {author} {\bibfnamefont {R.}~\bibnamefont {Hu}}, \bibinfo {author}
  {\bibfnamefont {L.}~\bibnamefont {Ma}}, \bibinfo {author} {\bibfnamefont
  {Z.}~\bibnamefont {Song}}, \bibinfo {author} {\bibfnamefont {T.}~\bibnamefont
  {Xia}}, \bibinfo {author} {\bibfnamefont {M.}~\bibnamefont {Ying}}, \bibinfo
  {author} {\bibfnamefont {H.}~\bibnamefont {Zhan}}, \bibinfo {author}
  {\bibfnamefont {H.-H.}\ \bibnamefont {Zhao}},\ and\ \bibinfo {author}
  {\bibfnamefont {C.}~\bibnamefont {Deng}},\ }\bibfield  {title} {\bibinfo
  {title} {Efficient initialization of fluxonium qubits based on auxiliary
  energy levels},\ }\href {https://doi.org/10.1103/PhysRevLett.132.230601}
  {\bibfield  {journal} {\bibinfo  {journal} {Physical Review Letters}\
  }\textbf {\bibinfo {volume} {132}},\ \bibinfo {pages} {230601} (\bibinfo
  {year} {2024})}\BibitemShut {NoStop}%
\end{thebibliography}%

\bigskip

\textbf{Data Availability.} The data that support the findings of this study are available from the corresponding authors Z.X.Z., S.Y.Z. and C.H.L. upon request.
\bigskip

\textbf{Code Availability.} The code that supports the simulations of this study are available from the corresponding authors Z.X.Z., S.Y.Z. and C.H.L. upon request.
\bigskip

\textbf{Acknowledgments.} We thank Guanglei Xi, Qiaonian Yu, Hualiang Zhang and Fuming Liu for supporting room-temperature electronics. We thank Xiaopei Yang for providing Josephson parametric amplifiers. The authors from Shenzhen University (L. Lin and C.H. Lee) acknowledge the support from the Quantum Science and Technology - National Science and Technology Major Project (2025ZD0300800), the National Natural Science Foundation of China (92476201 and 12505025), and the Guangdong Provincial Quantum Science Strategic Initiative (GDZX2305006 and GDZX2405002).
\bigskip

\textbf{Author Contributions.}
L.L. developed the theory with the discussion of C.H.L. and conceived the experiment. S.N.H. designed the device with the help of Z.X.Z. and X.P.Y., T.Q.C.. K.L.B. developed the fabrication method and fabricated the devices with assistance from S.N.H., T.Q.C., Z.X.Z. and Y.L.. S.N.H., Z.X.Z., T.Q.C. and Y.Y. established the measurement setup. S.N.H. performed the experimental measurements with support from Z.X.Z., Z.W.Z. and T.Q.C.. L.L. performed the numerical simulations aided by S.N.H., Z.X.Z. and T.Q.C.. S.N.H. and L.L. composed the manuscript draft. S.N.H., L.L. and C.H.L. finalized the manuscript with feedback from all authors. Z.X.Z., S.Y.Z. and C.H.L. supervised the project. All authors contributed to the discussion of the results and the development of the manuscript.
\bigskip

\textbf{Additional information.}
The authors declare no competing financial interests. Correspondence and requests for materials should be addressed to Z.X.Z., S.Y.Z. and C.H.L..

\end{document}


\title{Supplementary Materials: Exploring topology via momentum-selective tomography}

\author{Sainan Huai}
\thanks{These authors contributed equally to this work.}
\affiliation{Tencent Quantum Laboratory, Tencent, Shenzhen 518057, China}

\author{Ling Lin}
\thanks{These authors contributed equally to this work.}
\affiliation{Institute of Quantum Precision Measurement, State Key Laboratory of Radio Frequency Heterogeneous Integration, College of Physics and Optoelectronic Engineering, Shenzhen University, Shenzhen 518060, China}

\author{Kunliang Bu}
\thanks{These authors contributed equally to this work.}
\affiliation{Tencent Quantum Laboratory, Tencent, Shenzhen 518057, China}

\author{Tianqi Cai}
\thanks{These authors contributed equally to this work.}
\affiliation{Tencent Quantum Laboratory, Tencent, Shenzhen 518057, China}

\author{Zhiwen Zong}
\affiliation{Tencent Quantum Laboratory, Tencent, Shenzhen 518057, China}

\author{Xiaopei Yang}
\affiliation{Tencent Quantum Laboratory, Tencent, Shenzhen 518057, China}

\author{Yuan Li}
\affiliation{Tencent Quantum Laboratory, Tencent, Shenzhen 518057, China}

\author{Yulei Ye}
\affiliation{Tencent Quantum Laboratory, Tencent, Shenzhen 518057, China}

\author{Zhenxing Zhang}
\email{zzxht3@gmail.com}
\affiliation{Tencent Quantum Laboratory, Tencent, Shenzhen 518057, China}

\author{Shengyu Zhang}
\email{shengyzhang@tencent.com}
\affiliation{Tencent Quantum Laboratory, Tencent, Shenzhen 518057, China}

\author{Chaohong Lee}
\email{chleecn@szu.edu.cn}
\affiliation{Institute of Quantum Precision Measurement, State Key Laboratory of Radio Frequency Heterogeneous Integration, College of Physics and Optoelectronic Engineering, Shenzhen University, Shenzhen 518060, China}

\maketitle

\tableofcontents

\newpage

\section{Theoretical model}

\subsection{Selective coupling to a single Bloch eigenstate}

Here we show that the central site can be accurately coupled to a certain eigenstate with appropriate coupling modulation. For convenience, we diagonalize the kinetic part of the Hamiltonian (Eq.(1) in main text) via the unitary transformation
\begin{equation}
{\hat{d} _{n,k}^\dag}  = \sum\limits_\sigma  {u_{n,k}^\sigma \hat c_{\sigma ,k}^\dag } ,
\end{equation}
with the orthonormal relations
\begin{eqnarray}
\label{eqn:u_k_orthonormal}
\sum\limits_n {{{\left( {u_{n,k}^\sigma } \right)}^*}u_{n,k}^{\sigma '}}  = {\delta _{\sigma ,\sigma '}},  \\ \nonumber
\sum\limits_\sigma  {{{\left( {u_{n,k}^\sigma } \right)}^*}u_{n',k}^\sigma }  = {\delta _{n,n'}}.
\end{eqnarray}
The inverse transformation reads
\begin{equation}
\hat c_{\sigma ,k}^\dag  = \sum\limits_n {{{\left( {u_{n,k}^\sigma } \right)}^*}{\hat{d} _{n,k}}^\dag }.
\end{equation}
Under this transformation, the Hamiltonian becomes:
\begin{eqnarray}
\hat H &=& \sum\limits_{n,k} {{E_{n,k}}{\hat{d} _{n,k}}^\dag {\hat{d} _{n,k}}}    + \sum\limits_{k,\sigma } {\left[ {\;{\eta _\sigma }\left( k \right)\hat c_0^\dag \sum\limits_n {u_{n,k}^\sigma {\hat{d} _{n,k}} + {\rm{H}}.{\rm{c}}.} } \right]}  \nonumber \\
& =& \sum\limits_{n,k} {{E_{n,k}}{\hat{d} _{n,k}}^\dag {\hat{d} _{n,k}}}  + \sum\limits_{k,n,\sigma } {\left[ {\;u_{n,k}^\sigma {\eta _\sigma }\left( k \right)\hat c_0^\dag {\hat{d} _{n,k}} + {\rm{H}}.{\rm{c}}.} \right]}   \nonumber \\
&=& \sum\limits_{n,k} {{E_{n,k}}{\hat{d} _{n,k}}^\dag {\hat{d} _{n,k}}}  + \sum\limits_{k,n} {\left( {\;{\lambda _{n,k}}\hat c_0^\dag {\hat{d} _{n,k}} + {\rm{H}}.{\rm{c}}.} \right)} ,
\end{eqnarray}
where we have defined ${\lambda _{n,k}} = \sum\nolimits_\sigma  {u_{n,k}^\sigma {\eta _\sigma }\left( k \right)}$ as the coupling strength between the central site and the bare eigenstate of the chain. This formula offers the possibility to use the central site as a probe to couple to a certain target momentum and band state. To this end, we now assume that we want to address a specific target band $m$, and the modulated coupling strength satisfies ${\eta _\sigma }\left( x \right) = {\eta _0}{(u_{m ,q}^\sigma)^*}{e^{ - iqx}}$ in real space. By leveraging the orthonormal relation in Eq.~\eqref{eqn:u_k_orthonormal}, this leads to a precise momentum selection ${\eta _\sigma }\left( k \right) = {\tilde\eta _0}{(u_{m ,q}^\sigma)^*}{\delta _{k,q}}$ (with $\tilde\eta_0$ absorbing the Fourier transform prefactor) and furthermore we have the band selection $\sum\nolimits_\sigma  {u_{n,k}^\sigma {\eta _\sigma }\left( k \right)}  = {\tilde\eta _0}{\delta _{k,q}}{\delta _{n,m}}$. Consequently, the central site couples exclusively to the chain's eigenstates characterized by specific quantum numbers $k=q$ and $n=m$. One can excite the central site to excite the desired eigenstate on the chain.

Implementing such a precise spatial modulation, however, remains experimentally challenging. Targeting a specific eigenstate demands exact knowledge of the Bloch state $u_{n,k}$ in advance, creating a tautological paradox for unknown topological systems. To circumvent this fundamental limitation, our protocol in the main text exploits the intrinsic chiral symmetry of the lattice. Rather than engineering complex modulations, we employ a drastically simplified, real-valued coupling profile. Protected by chiral symmetry, this reduced hardware overhead is sufficient to isolate the zero-energy mode, allowing us to deterministically extract the topological invariant without requiring any prior knowledge of the full Bloch manifold.

Nevertheless, despite the stringent experimental prerequisites of this exact mapping, the theoretical derivation above reveals a fundamentally powerful feature of the star-like architecture: the capacity for deterministic, momentum-resolved eigenstate manipulation. While the main text focuses on a simplified measurement protocol, the full state-selective coupling demonstrated here provides a highly versatile toolkit for future explorations of topological matter. As quantum control technologies continue to advance, this unique ability to directly address and coherently manipulate specific bulk eigenstates could open new frontiers in preparing exotic topological phases, engineering non-trivial momentum-space dynamics, and advancing topologically protected quantum information processing.

\section{Device overview}\label{SecS2}

As illustrated in Fig.~2(b) of the main text, the nine qubits are arranged in a $3\times3$ square lattice, with $Q_0$ positioned at the center surrounded by an outer ring of $Q_1\text{--}Q_8$. Within this heterogeneous coupling network, in addition to standard nearest-neighbor interactions, cross-diagonal and longer-range tunable inter-qubit couplings are integrated, thereby yielding a physical connectivity of 8 for the central qubit $Q_0$. This high-connectivity topology provides the essential hardware foundation for constructing and observing the complex topological states described in the main text.

\begin{figure}[tb]
\includegraphics{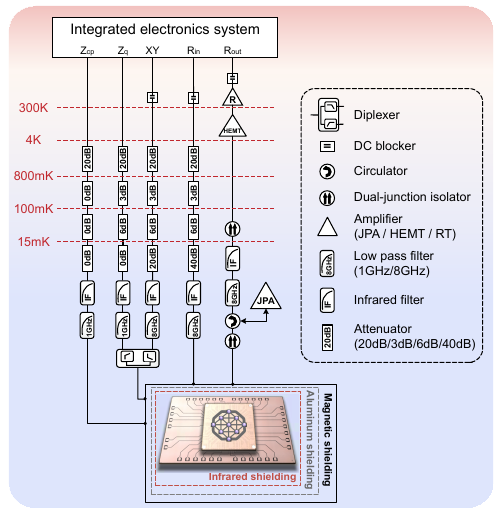}
\caption{\textbf{Measurement setup.} Wiring diagram across the dilution-refrigerator temperature stages ($300\,\mathrm{K}\rightarrow 15\,\mathrm{mK}$). The qubit XY/Z, coupler Z, and readout ($R_\mathrm{in}/R_\mathrm{out}$) channels are each routed through stage-by-stage attenuation and filtering, with the readout signal amplified before digitization. Thermal stages and component specifications are labeled in the figure.}
\label{fig:FigS1}
\end{figure}

The complete wiring topology and the cryogenic isolation environment of the device are shown in Fig.~\ref{fig:FigS1}. The signal chain comprises two primary branches: coherent control and quantum state readout. For qubit XY manipulation, baseband I/Q envelopes generated by room-temperature arbitrary waveform generators (AWGs) are up-converted using IQ mixers and microwave local oscillator (LO) sources. Z control of both qubits and couplers, in contrast, is driven directly at baseband by independent high-bandwidth AWG channels. To suppress thermal noise reaching the device, all input coaxial lines are equipped with attenuators stepped sequentially across each temperature stage from $300\,\mathrm{K}$ down to $15\,\mathrm{mK}$. At the MC stage, the signals sequentially pass through Eccosorb infrared (IR) filters and low-pass filters prior to entering the sample holder. The weak readout signal reflected from the chip is enhanced via a three-stage amplification chain. Finally, the output radio-frequency (RF) signal is down-converted by an IQ mixer and digitized by a high-speed analog-to-digital converter (ADC). Digital I/Q demodulation and single-shot state discrimination are subsequently performed on the master controller.

\subsection{Single-qubit parameters}\label{sec:single-qubit}

\begin{figure*}[tb]
\includegraphics{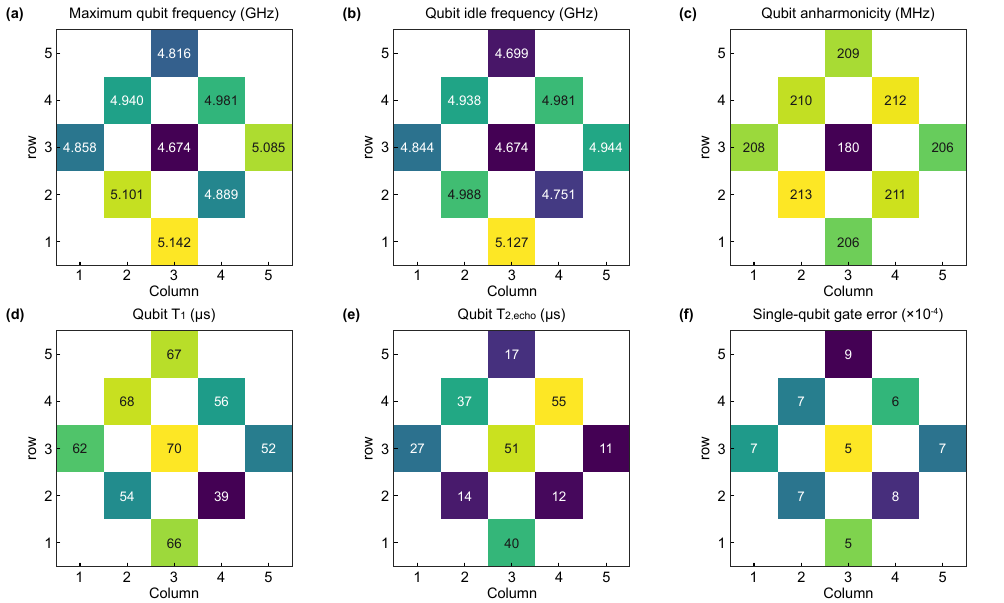}
\caption{\textbf{Single-qubit parameters of chip L9.} Key single-qubit metrics of the nine transmon qubits ($Q_0$--$Q_8$) measured with all couplers biased at their idle points. (a) Maximum qubit frequency (sweet spot). (b) Qubit idle frequency. (c) Qubit anharmonicity. (d) Qubit energy relaxation time $T_1$. (e) Hahn-echo dephasing time $T_2^{E}$. (f) Pauli error of single-qubit gates. Parameters in (c)-(f) are measured at the qubit idle frequency in (b).}
\label{fig:FigS2}
\end{figure*}

We systematically characterized the nine transmon qubits following the calibration process described in Ref.~\cite{huai2024fast}. Figure~\ref{fig:FigS2} summarises the key single-qubit performance metrics and their spatial distribution across the $3\times3$ array. During the characterization, all tunable couplers were biased at their idle points to suppress residual coupling, thereby reducing spurious frequency shifts and residual $ZZ$ interactions induced by spectator qubits. At the idle operating point $f_{10}$, we measured the energy relaxation time $T_1$ [Fig.~\ref{fig:FigS2}(d)] and the Hahn-echo decoherence time $T_2^{\mathrm{echo}}$ [Fig.~\ref{fig:FigS2}(e)] for each qubit. The device exhibits an average $T_1$ of $59.3\,\mathrm{\mu s}$ and $T_2^{\mathrm{echo}}$ of $29.3\,\mathrm{\mu s}$, with a mean single-qubit gate fidelity of $99.93\%$. These metrics indicate that the dense interconnects and long airbridges do not appreciably degrade the baseline coherence of the platform. One of the major factors is a surface-participation-ratio (SPR)-guided layout~\cite{wenner2011surface,wang2015surface,woods2019determining,gu2026connectivity,minev2021epr}, which keeps the electric-field participation in lossy interfaces low despite the increased connectivity.

\subsection{Two-qubit parameters and couplers}\label{sec:two-qubit}

The quantum processor integrates 24 frequency-tunable floating transmon couplers to construct the complete network of qubit interconnects. Each coupler is capacitively coupled to two neighbouring qubits and provides a flux-tunable effective interaction between them. Based on the qubit separation, these coupling links are quantitatively categorized into three types: $s=1$ (nearest-neighbor coupling), $s=\sqrt{2}$ (diagonal coupling), and $s=2$ (next-nearest-neighbor coupling).

\begin{figure*}[tb]
\includegraphics{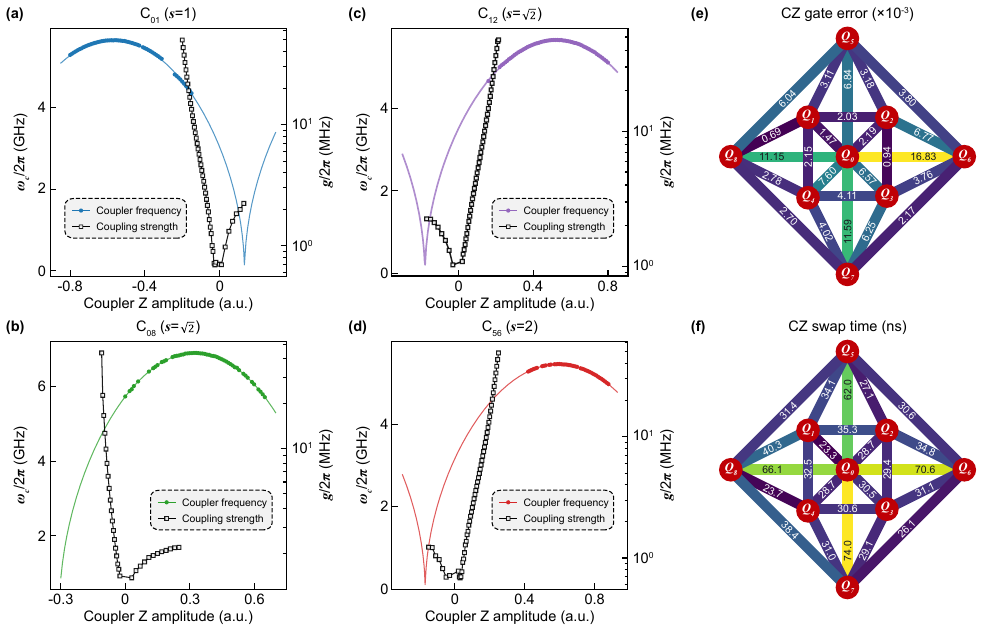}
\caption{\textbf{Coupler characterization and two-qubit gate performance.} (a)-(d) Coupler eigenfrequency $\omega_c/2\pi$ (colored dots, left axis) and the resulting effective $XY$ coupling strength $g/2\pi$ (black squares, right axis) as functions of the coupler Z-pulse amplitude. Data are shown for four representative couplers with different qubit separations $s$ (in units of the lattice constant): $C_{01}(s=1)$, $C_{08} (s=\sqrt{2})$, $C_{12} (s=\sqrt{2})$ and $C_{56}  (s=2)$. (e),(f) Spatial distribution of CZ-gate randomized benchmarking (RB) errors (e) and CZ swap duration (f) across all 24 couplers, with qubits arranged according to their connectivity. The array-averaged CZ-gate fidelity is $99.5\%$ and the average swap duration is ${\sim}\,37\,\mathrm{ns}$.}
\label{fig:FigS3}
\end{figure*}

To suppress potential crosstalk between cross-diagonal couplers within the highly connected lattice, we introduce two floating-coupler
configurations, i.e., symmetric and asymmetric, at the device stage. By exploiting the versatile pad configurations inherent to the floating-coupler design alongside the distinct operating frequency ranges of the two geometries, such crosstalk is substantially mitigated at the fundamental architectural level. Figure~\ref{fig:FigS3} illustrates the coupler eigenspectra and the response of the effective $XY$ coupling strength to the coupler flux Z bias. At the designated OFF bias points, the effective $XY$ exchange interaction between a given qubit pair is strongly suppressed, with a residual coupling rate of $\sim10~\mathrm{kHz}$ that is negligible over the timescales of our experimental protocols. This ensures a well-isolated, local Hamiltonian during both single-qubit operations and many-body evolution.

Building upon this coupling architecture, we calibrated and characterized two-qubit controlled-Z (CZ) gates on all 24 couplers to further evaluate the control quality of the physical channels. Gate fidelities were quantitatively extracted using randomized benchmarking (RB). The 24 CZ gates reach an average fidelity of $99.5\%$, with an average swap duration of approximately $37\,\mathrm{ns}$. The complete experimental results are summarized in Fig.~\ref{fig:FigS3}. These two-qubit control metrics demonstrate that the coupler architecture can reliably deliver high-fidelity operations, even within a complex interconnected lattice containing diverse physical spans and configurations.

\section{Hamiltonian calibration for the extended SSH model}\label{sec:calibration}

Simulating the extended Su--Schrieffer--Heeger (SSH) model and observing its topological dynamics on a superconducting processor requires reliable initialization of the target Hamiltonian. Although topological phases are intrinsically robust against specific local perturbations, establishing a pristine Hamiltonian baseline in an interacting superconducting many-body system remains a prerequisite for verifying topological protection mechanisms. To this end, we implement a two-stage coarse-to-fine calibration protocol~\cite{zhang2024emergence}, whose core is a progressive approximation of the measurement environment. In the coarse stage, the nominal Z-pulse amplitude (ZPA) values of the lattice-site frequencies and couplers are determined in nearly isolated subsystems. In the fine stage, the system is brought close to the realistic many-body Hamiltonian to align the diagonal terms (site eigenfrequencies) and off-diagonal terms (intra-cell hopping $t_1$, inter-cell hopping $t_2$, cross-cell couplings $\gamma_1$, $\gamma_2$, and radial coupling $\eta_B$), thereby refining the nominal values to their actual operating points.

\subsection{Coarse calibration of frequencies and couplers}\label{sec:coarse}

Following the device characterization framework outlined in Sec.~\ref{sec:single-qubit}, we first establish the coarse calibration curves for each control channel with all neighbouring couplers biased at their OFF points. To map the frequency response of each qubit as a function of the $Z$-pulse amplitude (ZPA), we employ the Ramsey interferometry sequence shown in Fig.~\ref{fig:FigS4}(a), in which two $\pi/2$ pulses applied on the $XY$ drive line bracket a free-evolution window. During this window, a rectangular detuning pulse of amplitude ZPA and duration $T$ is applied to the $Z$ line, and the resulting detuning drives the qubit to accumulate a dynamical phase [Fig.~\ref{fig:FigS4}(b)]. For a given ZPA, the qubit frequency is extracted from the linear slope of this phase versus $T$; sweeping the ZPA then reconstructs the full frequency response. Taking the central qubit $Q_0$ as an example, the qubit frequencies measured at various ZPAs (blue circles) are well described (red solid line) by the standard transmon flux-modulation relation~\cite{koch2007transmon}. The response curve is symmetric about $\mathrm{ZPA}=0$, corresponding to the qubit sweet spot. The target frequency $\omega_I$ required for the extended SSH model is marked by the grey horizontal dashed line, and its intersection with the fitted curve yields the nominal ZPA needed to initialize the qubit at the specific operating point.

For each of the 24 frequency-tunable couplers, we track the single-excitation population oscillations between the adjacent qubit pairs, yielding a coarse mapping between the coupler ZPA and the effective coupling strength $g$. At this stage, each coupler is characterized within an isolated two-qubit subsystem, from which we extract the nominal ZPAs that realize the target intra-cell ($t_1$), inter-cell ($t_2$), cross-cell ($\gamma$), and radial ($\eta$) couplings, together with the OFF biases for the remaining idle couplers.

\subsection{Precise alignment of local chemical potentials}\label{sec:fine}

\subsubsection{Selection of the operating frequency \texorpdfstring{$\omega_I$}{omega\_I}}

\begin{figure*}[tb]
\includegraphics{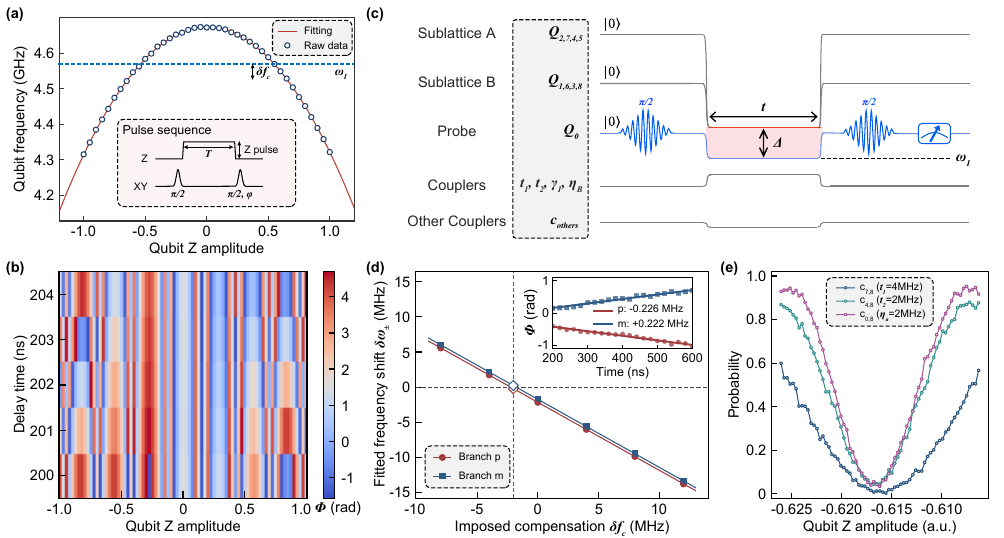}
\caption{\textbf{Coarse-to-fine calibration of qubit operating frequencies.} (a) Qubit frequency versus Z-pulse amplitude (ZPA) for the central qubit $Q_0$, measured with the Ramsey phase-based spectroscopy sequence shown in the inset. The measured data (blue circles) are fitted to the transmon flux-modulation relation (red line). The target operating frequency $\omega_I$ (grey dashed line) sets the nominal ZPA in the isolated single-qubit configuration; $\delta f_c$ denotes the fine correction subsequently applied in the many-body configuration. (b) Raw phase accumulation used to extract (a), measured as a function of ZPA and free-evolution delay $T$. (c) Pulse sequence for the fine calibration, exemplified for target qubit $Q_0$. The qubit is prepared on the equator by an $X/2$ pulse and biased to $\omega_I$ with all couplers active, while spectator qubits (sublattices A and B) are detuned to $\omega_I\pm\Delta$; the accumulated phase is then read out by tomography after an evolution time $t$. (d) Fitted frequency shifts $\delta\omega_\pm$ for the two spectator configurations (branches $p$ and $m$) versus the imposed compensation $\delta f_c$; the optimal correction $\delta f_c^{*}$ (red dashed line) is identified as the point where $\delta\omega_\pm$ become symmetric about zero, i.e. equal in magnitude and opposite in sign. Inset, measured linear phase accumulation $\varphi(t)$ and extraction of $\delta\omega_\pm$ at $\delta f_c^{*}$. (e) Multi-channel in-situ verification at node $Q_8$: transition probability $P_1$ versus $Q_8$ ZPA for three active links $C_{08}$, $C_{48}$ and $C_{78}$ (coupling strengths as labelled), which share a common minimum at $\mathrm{ZPA}=-0.6175$~a.u.}
\label{fig:FigS4}
\end{figure*}

The selection of the global resonant operating point $\omega_I$ is constrained by a physical trade-off between intrinsic device parameters and the fine calibration protocol. First, within the current ``window-blossom'' topological array, the central hub qubit $Q_0$ has the lowest sweet-spot frequency of all qubits, at $4.674\,\mathrm{GHz}$ (Fig.~\ref{fig:FigS2}(b)). Since transmon qubits can only be tuned downward from their respective sweet spots, the target interacting point must lie below this value to ensure all qubits can be brought into resonance. Second, in the subsequent fine calibration, a large symmetric detuning ($\omega_I \pm \Delta$, with $\Delta/2\pi = 100\,\mathrm{MHz}$) is used to effectively isolate the target qubit from the spectators, pushing the system into the deep dispersive regime. Given a maximum topological hopping rate of $J_{\max}\approx 4\,\mathrm{MHz}$ (yielding a detuning ratio of $\Delta/J_{\max}\approx 25$), the global resonant frequency $\omega_I$ should lie below $4.574\,\mathrm{GHz}$. Subject to these constraints, we anchor the global resonant baseline of the system at $\omega_I/2\pi = 4.57\,\mathrm{GHz}$.

\subsubsection{Transition from the nominal to fine-tuned ZPA}

The response curve in Fig.~\ref{fig:FigS4}(a) is measured and fitted with all adjacent couplers turned off and every qubit parked at its idle frequency, i.e., in a single-qubit configuration free of inter-qubit coupling. Consequently, the ZPA obtained by directly solving this curve for $\omega_I$ generally deviates from the optimal operating point required under the actual working Hamiltonian. To align the local chemical potentials across all lattice sites---that is, to bring all qubit frequencies into resonance during the many-body dynamics---we need to fine-tune each qubit under a control configuration that closely approximates the target operating Hamiltonian.

We adapt the Ramsey fine-tuning protocol of Ref.~\cite{wang2024efficient} to the present ``window-blossom'' topological array to calibrate the Z-pulse amplitude (ZPA) corresponding to each qubit's resonance frequency. For each target qubit, a $\pi/2$ pulse first prepares its state on the equator of the Bloch sphere. The target is then biased at $\omega_I$ while all couplers are activated to their respective target operating frequencies, so that it carries the same coupling load as in the full protocol. Concurrently, the remaining spectator qubits are fast-biased to $\omega_I\pm\Delta$ in two distinct measurement configurations. The total frequency shift $\delta\omega_\pm$ extracted from this accumulated dynamical phase comprises two components: a static background frequency offset $\delta\omega_0$ to be eliminated, and a residual dispersive shift $\delta\omega_\pm^{\mathrm{dis}}\approx\mp\sum J_{i,j}^2/\Delta$ induced by the active couplers. Averaging the two configurations cancels the antisymmetric dispersive shift induced by the $\pm\Delta$ spectator detuning, isolating the constant background offset.

Feeding $\delta\omega_0$ back into the qubit ZPA--frequency response function is equivalent to correcting the nominal target qubit frequency to $\omega_I - \delta f_c$ [Fig.~\ref{fig:FigS4}(a)] when solving for the exact ZPA. Sweeping $\delta f_c$ yields the two tracking $\delta\omega_\pm$ curves shown in Fig.~\ref{fig:FigS4}(d); the point $\delta f_c^{*}$ at which they lie symmetrically about zero defines the final correction (red dashed line in Fig.~\ref{fig:FigS4}(d)). A verification measurement of $\delta\omega_\pm$ at $\delta f_c^{*}$ reveals that the two offsets become equal in magnitude and opposite in sign [inset of Fig.~\ref{fig:FigS4}(d)]. Their absolute values closely match the theoretical dispersive estimate of $|\delta\omega_\pm^{\mathrm{dis}}|\approx 0.18\,\mathrm{MHz}$, confirming the elimination of the background error. This fine-tuning protocol ultimately establishes the precise operating bias baseline for each qubit at $\omega_I$.

\subsubsection{Multi-channel in-situ verification} \label{sec:fine_step3}

The preceding Ramsey-phase calibration fixes the bias baseline of each qubit. As a complementary in-situ check, we verify these baselines under a more realistic, resonant configuration: a coupled pair is brought simultaneously to $\omega_I$--with the remaining spectator qubits held at $\omega_I\pm\Delta$--and we confirm that the associated channels share a common operating point. Consider node $Q_8$ in the extended SSH model, which shares physical edges with $Q_0$, $Q_4$, $Q_5$, and $Q_7$, with coupling strengths $|J_{0,8}|/2\pi = 2\,\mathrm{MHz}$, $|J_{4,8}|/2\pi = 2\,\mathrm{MHz}$, and $|J_{7,8}|/2\pi = 4\,\mathrm{MHz}$; the $(5,8)$ channel is left idle. We prepare $Q_8$ in the excited state and bias it to $\omega_I$ together with one coupled neighbor at a time. For each channel we fix the evolution time at its half-period $t_k = \pi/(2|J_{k,8}|)$ ($k\in\{0,4,7\}$) and sweep the ZPA of $Q_8$. Overlaying the responses of all active channels, we find that at a $Q_8$ ZPA of $-0.6175$~a.u. their transition probabilities reach a common minimum, forming a shared V-shaped valley, whereas the idle $(5,8)$ link stays essentially flat (not shown). Locking $Q_8$ at this minimum aligns the hub node with all of its evolution branches at once.

\subsection{Calibration of the topological hopping network}\label{sec:hopping}

After aligning the local chemical potentials, the core of topological phase construction shifts to the fine-mapping of multi-scale hopping rates ($J_{i,j}$). As in the multi-channel in-situ verification (Sec.~\ref{sec:fine_step3}), we bias both qubits of the target pair to $\omega_I$, keep the remaining qubits at $\omega_I\pm\Delta$, and activate the target coupler to its operating frequency. Here we instead monitor the single-excitation transfer $\ket{10}\leftrightarrow\ket{01}$ under coupler ZPA modulation. This exchange completes at $t_m = (m+1/2)\,\pi/|J_{i,j}|$, and, as in Ref.~\cite{zhang2024emergence}, we can choose several half-periods to amplify the sensitivity of the population to the coupler bias. For our target range of $|J_{i,j}|/2\pi = 1\text{--}4\,\mathrm{MHz}$ we take $m=2$--$4$, fix the evolution time at this $t_m$ and continuously sweep the target coupler ZPA; the peak of the $\ket{01}$ state population identifies the optimal ZPA for the targeted hopping rate $J_{i,j}$. It is worth noting that tuning the coupler operating frequency inevitably shifts the target qubit frequencies due to parasitic coupling; that is, the diagonal-term calibration (qubit frequencies) and the off-diagonal-term calibration (coupling strengths) are weakly coupled. This interdependency can be resolved through multiple iterations between the two calibrations, driving the ZPAs of the entire array toward a self-consistent operating point. This iterative protocol establishes a static Hamiltonian baseline for subsequent many-body dynamical evolution.

\begin{figure}[tb]
\includegraphics[width = 0.5\columnwidth ]{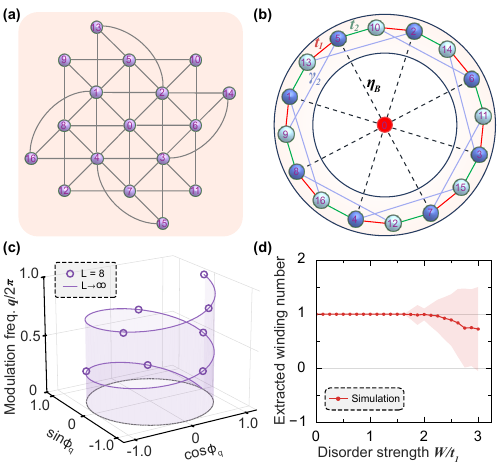}
\caption{\textbf{Scalability to $L=8$ and robustness against disorder.} 
(a) Schematic physical layout of the proposed 17-qubit superconducting processor ($16+1$ configuration). Circles denote transmon qubits and lines tunable floating couplers. 
(b) Topological mapping onto an extended SSH ring with $L=8$ unit cells under periodic boundary conditions, formed by all qubits and a subset of the couplers in (a). Intra- and inter-cell hoppings are represented by red and green bonds, while long-range couplings are indicated by purple lines. The central auxiliary qubit (red sphere) connects to the target sublattice sites via spatially modulated star-like couplings (dashed lines). (c) Extracted $L=8$ winding trajectory (open rings) overlaid on the analytic $L\to\infty$ (solid line), showing the $4\pi$ phase winding of the $\nu=+2$ phase. (d) Numerically extracted winding number $\nu=+1$ versus bond disorder strength $W/t_{1}$, averaged over 100 disorder realizations; the shaded band marks $\pm1\sigma$ across realizations. Other parameters are the same as Fig.~4 in the main text.
}
\label{fig:FigS5}
\end{figure}

\section{Scaling toward larger systems and higher winding phases}
To demonstrate the scalability and future potential of our experimental architecture, we present an extended layout scheme capable of hosting a larger 1D topological lattice with $L=8$ unit cells (comprising $16+1 = 17$ transmon qubits in total), as schematically illustrated in Fig.~\ref{fig:FigS5}. While the current proof-of-principle experiment is implemented on a compact $L=4$ platform, our hardware, characterized by 2.5D flip-chip integration, multi-functional floating couplers, and stiffener-reinforced $200~\mathrm{\mu m}$ Ta airbridges, is inherently scalable to larger quantum networks. As shown in Fig.~\ref{fig:FigS5}(a), the physical layout configures 17 qubits in a highly connected 2D grid, where long-distance crossover structures (curved gray lines) enable the extended long-range hoppings ($\gamma_{1,2}$) without routing bottlenecks. Fig.~\ref{fig:FigS5}(b) illustrates the corresponding topological mapping: 16 outer qubits form an 8-cell extended SSH ring with periodic boundary conditions, while the central probe qubit (red sphere) maintains star-like, spatially modulated couplings to the targeted sublattices. 

In this expanded $L=8$ configuration, the Brillouin zone is discretized into 8 distinct momentum modes $q = 2n\pi/8$ ($n=1, \dots, 8$). 
It can be numerically demonstrated that the enlarged
system supports a direct and unambiguous resolution of $4\pi$ winding of the $\nu=+2$ phase [Fig.~\ref{fig:FigS5}(c)]: the eight momentum-resolved points lie on the analytic $(\cos\phi_q,\sin\phi_q,q/2\pi)$ helix in the $L\to\infty$ limit, sharpening the mapping of the topological winding angle $\phi_q$ and providing a route toward exploring topological sectors with larger winding numbers. Furthermore, we verify the robustness of the protocol at $L=8$ under chiral-preserving bond disorder on the inter- and intra-cell hopping.
Fig.~\ref{fig:FigS5}(d) shows that the extracted winding number remains pinned at $+1$ up to $W/t_1 \sim 1.75$ (100 random realizations; shaded band, $\pm1\sigma$), before deviating at larger disorder. 
This robustness, together with the reliable sampling resolution, demonstrates the scalability of the proposed protocol toward larger processors.
Although experimental realization of this 17-qubit system remains a future prospect, this concrete blueprint underscores that our star-like band tomography framework is readily extendable to larger-scale processors, offering a practical pathway toward probing complex quantum topological phases with higher precision.

\makeatletter
\renewcommand\@biblabel[1]{#1.}
\makeatother
\bibliography{Citation_sp}